\documentstyle[11pt,subfigure,epsfig,rotating,longtable]{article}
\def\q2{$Q^2 \, $}

\def\rx{$R(x,Q^2) \;$}
 
\title{\begin{center}  \sc
Measurement of the proton \\ and
deuteron  structure functions, \\
$F_{2}^p$ and $F_{2}^d$, and of the ratio $\sigma_L/\sigma_T$  \\
\end{center}}

\author{}
\date{}
\begin{document}
\begin{center}
\maketitle
\end{center}

\begin{center} 
THE NEW MUON COLLABORATION (NMC) \\
\vspace{0.6cm}
{\footnotesize{\sl{
Bielefeld~University$^{1+}$,
Freiburg~University$^{2+}$,
Max-Planck~Institut f\"{u}r Kernphysik, Heidelberg$^{3+}$,
Heidelberg~University$^{4+}$,
Mainz~University$^{5+}$, Mons~University$^6$,
Neuch\^{a}tel~University$^7$,
NIKHEF$^{8++}$,
Saclay DAPNIA/SPP$^{9**}$,
University~of~California, Santa~Cruz$^{10}$,
Paul~Scherrer~Institut$^{11}$,
Torino~University and INFN~Torino$^{12}$,
Uppsala~University$^{13}$,
Soltan~Institute~for~Nuclear~Studies, Warsaw$^{14*}$,
Warsaw~University$^{15*}$}}}\\
\vspace{0.6cm}
{\small{M.~Arneodo$^{12a)}$,
A.~Arvidson$^{13}$,
B.~Bade{\l }ek$^{13,15}$,
M.~Ballintijn$^{8}$,
G.~Baum$^1$,
J.~Beaufays$^{8}$,
I.G.~Bird$^{3,8b)}$,
P.~Bj\"{o}rkholm$^{13}$,
M.~Botje$^{11c)}$,
C.~Broggini$^{7d)}$,
W.~Br\"{u}ckner$^3$,
A.~Br\"{u}ll$^{2e)}$,
W.J.~Burger$^{11f)}$,
J.~Ciborowski$^{15}$,
R.~van~Dantzig$^8$,
A.~Dyring$^{13}$,
H.~Engelien$^2$,
M.I.~Ferrero$^{12}$,
L.~Fluri$^7$,
U.~Gaul$^3$,
T.~Granier$^{9g)}$,
M.~Grosse-Perdekamp$^{2h)}$,
D.~von~Harrach$^{3i)}$,
M.~van~der~Heijden$^8$,
C.~Heusch$^{10}$,
Q.~Ingram$^{11}$,
M.~de~Jong$^{8b)}$,
E.M.~Kabu\ss$^{3i)}$,
R.~Kaiser$^2$,
T.J.~Ketel$^8$,
F.~Klein$^{5j)}$,
S.~Kullander$^{13}$,
U.~Landgraf$^2$,
T.~Lindqvist$^{13}$,
G.K.~Mallot$^{5}$,
C.~Mariotti$^{12k)}$,
G.~van~Middelkoop$^{8}$,
A.~Milsztajn$^9$,
Y.~Mizuno$^{3l)}$,
A.~Most$^{3m)}$,
A.~M\"{u}cklich$^3$,
J.~Nassalski$^{14}$,
D.~Nowotny$^{3}$,
J.~Oberski$^8$,
A.~Pai\'{c}$^7$,
C.~Peroni$^{12}$,
B.~Povh$^{3,4}$,
K.~Prytz$^{13n)}$,
R.~Rieger$^{5}$,
K.~Rith$^{3o)}$,
K.~R\"{o}hrich$^{5p)}$,
E.~Rondio$^{14b)}$,
L.~Ropelewski$^{15b)}$,
A.~Sandacz$^{14}$,
D.~Sanders$^{q)}$,
C.~Scholz$^{3}$,
R.~Seitz$^{5r)}$,
F.~Sever$^{1,8s)}$,
T.-A.~Shibata$^{4t)}$,
M.~Siebler$^1$,
A.~Simon$^{3u)}$,
A.~Staiano$^{12}$,
M.~Szleper$^{14}$,
W.~T{\l }acza{\l }a$^{14v)}$
Y.~Tzamouranis$^{3q)}$,
M.~Virchaux$^9$,
J.L.~Vuilleumier$^7$,
T.~Walcher$^5$,
R.~Windmolders$^6$,
A.~Witzmann$^2$,
K.~Zaremba$^{14v)}$,
F.~Zetsche$^{3w)}$}} \\
\vspace{0.6cm}
\end{center}

\begin{center}
{\footnotesize\it (submitted to Nuclear Physics)}
\end{center}
\noindent
\begin{abstract}
The muon-proton and muon-deuteron
inclusive deep inelastic scattering cross sections
were measured
in the kinematic range $ \, 0.002 < x < 0.60 \, $ and
$ \, 0.5 < Q^2 < 75 \, $ GeV$^2$
at incident muon energies of 90, 120, 200 and 280~GeV.
These results are based on the full data set collected by 
the New Muon Collaboration, including the data taken with 
a small angle trigger.
The extracted values of the
structure functions $F_2^p$ and $F_2^d$
are in good agreement with those from other experiments.
The data cover a sufficient range of $y$ to allow the determination
of the ratio of the longitudinally to transversely polarised virtual
photon absorption cross sections, $R=\sigma_L/\sigma_T$, 
for $0.002 < x < 0.12 \,$.
The values of $R$ are compatible with a perturbative
QCD prediction;
they agree with earlier measurements and extend to smaller $x$.
\end{abstract}
 
{\footnotesize {-----------------------------------\\
 
For footnotes see next page.}}
\newpage

\begin{tabbing}
~~~~\=+~~~\=Supported by Bundesministerium f\"{u}r Bildung und Forschung.\\
\>      ++\>    Supported in part by FOM, Vrije Universiteit Amsterdam and NWO.\\
\>       *\>    Supported by KBN SPUB Nr 621/E - 78/SPUB/P3/209/94. \\
\>      **\>    Laboratory of CEA, Direction des Sciences de la Mati\a`ere.\\
\> \> \\
\>      a)\>    Alexander von Humboldt fellow. \\
\>      b)\>    Now at CERN, 1211 Gen\a`eve 23, Switzerland. \\
\>      c)\>    Now at NIKHEF, 1009 DB Amsterdam, The Netherlands. \\
\>      d)\>    Now at University of Padova, 35131 Padova, Italy.\\
\>      e)\>    Now at MPI f\"{u}r Kernphysik, 69029 Heidelberg, Germany. \\
\>      f)\>    Now at Universit\a'e de Gen\a`eve, 1211 Gen\a`eve 4, Switzerland.\\
\>      g)\>    Now at DPTA, CEA, Bruy\a`eres-le-Chatel, France. \\
\>      h)\>    Now at Yale University, New Haven, 06511 CT, U.S.A.\\
\>      i)\>    Now at University of Mainz, 55099 Mainz, Germany. \\
\>      j)\>    Now at University of Bonn, 53115 Bonn, Germany. \\
\>      k)\>    Now at INFN-Instituto Superiore di Sanit\a`a, 00161 Roma, Italy.\\
\>      l)\>    Now at Osaka University, 567 Osaka, Japan.\\
\>      m)\>    Now at University of Michigan, Michigan, U.S.A.\\
\>      n)\>    Now at Stockholm University, 113 85 Stockholm, Sweden.\\
\>      o)\>    Now at University of Erlangen-N\"{u}rnberg, 91058 Erlangen, Germany.\\
\>      p)\>    Now at IKP2-KFA, 52428 J\"{u}lich, Germany.\\
\>      q)\>    Now at University of Houston, 77204 TX, U.S.A.\\
\>      r)\>    Now at Dresden University, 01062 Dresden, Germany.\\
\>      s)\>    Now at ESRF, 38043 Grenoble, France.\\
\>      t)\>    Now at Tokyo Institute of Technology, Tokyo, Japan. \\
\>      u)\>    Now at University of Freiburg, 79104 Freiburg, Germany. \\
\>      v)\>    Now at Warsaw University of Technology, Warsaw, Poland. \\
\>      w)\>    Now at Hamburg University, 22761 Hamburg, Germany.\\

\end{tabbing}
\vskip 0.5 cm
\vspace{0.8cm}

\vskip 1.5 cm
\section{Introduction}

The structure function $F_2(x,Q^2)$ reflects the momentum
distributions of the quarks in the nucleon.  Because the proton
and neutron have different contents of up and down
quarks, a measurement of
$F_2^p$ and $F_2^d$, together with isospin symmetry, provides
a constraint on the individual quark distributions;
this is
an important phenomenological input to the calculation of many
strong interaction processes.
In addition, the \q2 dependence of $F_2$
can be used to test
perturbative Quantum Chromodynamics (QCD),
and to determine the strong
coupling constant, $\alpha_s$, and the momentum distribution
of the gluons in the nucleon, $x G(x, Q^2)$.
Measurements of $R(x,Q^2)=\sigma_L/\sigma_T$, the ratio of the 
longitudinally to transversely polarised virtual
photon absorption cross sections,  
can also be compared to QCD predictions
computed from the measured
values of $F_2$ and $x G$, thus
providing an additional
consistency check of the theory.

In the deep inelastic scattering of charged leptons off nucleons,
the differential cross section for one photon exchange
can be written in
terms of  $F_2(x,Q^2)$ and $R(x,Q^2)$ as~:
{\it
\begin{eqnarray}
\frac{{\rm d}^2\sigma(x,Q^2,E)}{{\rm d}x {\rm d} Q^2} &=&
\frac{4 \pi \alpha^2 }{Q^4} \cdot
\frac{F_2(x,Q^2)}{x} \cdot \nonumber \\
 &&  \left\{ 1 -y -\frac{Q^2}{4E^2}
+ (1 - \frac{2m^2}{Q^2}) \cdot \frac{y^2 +Q^2/E^2}
{2\left(1+R(x,Q^2)\right)} \right\},
\label{eq:sigma}
\end{eqnarray} }
where $\alpha$ is the fine structure constant,
$-Q^2$ the four momentum transfer
squared, $E$ the energy of the incident muon
and $m$ the muon mass.
The Bjorken scaling variable, $x$, and $y$ are
defined as $x = Q^2/2M\nu$ and $y = \nu/E$,
where $\nu$
is the virtual photon energy in the laboratory frame and
$M$ is the proton mass\footnote{This expression
for the cross section,
appropriate for fixed target experiments,
can be expressed in relativistically invariant form by
replacing $Q^2 / E^2$ with $4 M^2 x^2 y^2 / Q^2$.}.
The one photon exchange cross section given by
eq.~(\ref{eq:sigma})
is obtained from the measured differential cross section by
correcting for higher order electroweak effects.
 
When $y$ is small enough, which is
the case for most of the data presented
here, the dependence of
the one photon cross section on $R$ is weak,
as can be seen from
eq.~(\ref{eq:sigma}),
and a measurement of the cross section is thus
mainly a measurement of $F_2$.
In contrast, $R$ can be obtained
from the small differences in the cross section
at given values of $x$ and \q2
but at different values of $y$,
i.e. at different incident muon energies $E$.
 
In a recent article \cite{f2nmc95}, we
presented the structure functions
of the proton and the deuteron, $F_2^p$ and $F_2^d$,
obtained from deep inelastic muon scattering
at incident energies of 90, 120, 200 and 280 GeV.
The data had been obtained between 1986 and 1989
using the large angle trigger of the experiment.
This trigger recorded only events with scattering
angles above 10~mrad and the accessible kinematic range
did not allow a precise measurement
of $R$. Hence the
$F_2^p$ and $F_2^d$
measurements of ref.~\cite{f2nmc95} were obtained using a
parametrisation \cite{rslac} of earlier results on $R$.
In the results presented here, the ranges of $x$ and $y$
covered have been extended downward by using data taken with
a small angle trigger at 200 and 280~GeV in 1989.
This has permitted the determination
of $R$ as well as $F_2$.

\section{The experiment and data selection }
The experiment (NMC--NA37) was performed at the M2 muon beam line
of the CERN SPS.
The proton and deuteron differential
cross sections were measured simultaneously with two pairs of 3 m long
targets placed alternately in the muon beam. In one pair the upstream
target was liquid hydrogen and the downstream one
liquid deuterium, while in the other pair the order was reversed.
The spectrometer acceptance was substantially
different for the upstream and downstream targets,
thereby giving two separate
measurements of the cross section for each material.
 
The integrated incident muon flux
was continuously measured both by
sampling the beam with a random trigger \cite{t10} and by
sampling the counts recorded in two scintillator
hodoscope planes used to determine incident beam tracks
\cite{f2nmc92}. In both cases the beam tracks present
in these triggers were reconstructed off-line,
in the same way as for scattered muon triggers,
to determine the integrated usable flux.
 
Uncertainties on the incident and scattered muon momenta
are important sources of systematic error. The forward
spectrometer which measured the scattered muon momentum
was calibrated to an accuracy of 0.2\% using the
reconstructed masses of J/$\psi$ and K$^0$ mesons.
The beam momentum spectrometer (BMS) was calibrated in dedicated
runs by measuring the average incident muon momentum in a purpose
built spectrometer \cite{bcs}.
The BMS was also calibrated relative to the forward spectrometer
in a series of runs using precision silicon microstrip
detectors.  The results of the two BMS calibrations were
averaged, leading to an accuracy of 0.2\%.
These calibrations are described in more detail
in ref.~\cite{annaphd}.
 
The reader is referred to
refs.~\cite{f2nmc95,longratio,apparatus,peterphd}
for more details on the spectrometer,
the large angle trigger and the analysis of the data obtained
from this trigger.  Properties specific to the
small angle trigger are described below and in ref.~\cite{annaphd}.
 
The small angle trigger (see also ref.~\cite{longratio})
used three hodoscopes of 14~mm high, slightly overlapping,
horizontal scintillator strips.
Each hodoscope was vertically separated into
two halves, above and below the muon beam.
Coincidences between elements of each of the hodoscopes,
which were about 5~m apart, selected particles whose trajectories
were consistent with muons scattered from the targets.
The narrow strips provided the necessary angular resolution and
reduced the count rate per strip.
 
The small angle trigger collected events with scattering
angles in the range $5 < \theta < 25$~mrad.
Thus, this trigger detected events with smaller $x$ and $Q^2$
than the large angle trigger ($\theta > 10$~mrad).
The small angle data cover the kinematic
range $0.002 < x < 0.14$ and $0.5 < Q^2 < 25$~GeV$^2$.
 
The following selections were applied in the analysis of
the small angle trigger data
(see table \ref{tab:cutpar}).
The scattered muon momentum, $p'$, was required to be above a
minimum value to suppress muons
from hadron  decays.
Events with small $\nu$ (or $y$),
where the spectrometer resolution was poor, were rejected.
By requiring minimum scattering angles, $\theta_{min}$,
regions with rapidly varying acceptance were excluded.
We imposed a maximum value of $y$ in order to limit the
contribution from higher order electroweak processes.
In addition, the position of the reconstructed
vertex was constrained to be within one of the targets.
At each value of $x$, data in regions of $Q^2$ where the
acceptance was less than 1\%, or less than
25\% of the maximum acceptance at that $x$, were removed.
Regions close to the muon beam, whose efficiencies were
difficult to describe, have been excluded from the analysis.
 
The number of events from the small angle trigger surviving
these cuts was 0.54 million, while that from the large angle
trigger was 1.82 million~\cite{f2nmc95}.
This is the full NMC data set covering the kinematic range
$0.002 < x < 0.60$ and
$0.5 < Q^2 < 75$ GeV$^2$.

\section{Data analysis }
In order to measure the deep inelastic cross sections, one needs
to take into account the effect
of the spectrometer's resolution, which
requires knowledge of the event distributions
in the $(x,Q^2)$ plane
and thus knowledge of the cross sections.
In addition, the extraction of the structure functions
from the measured cross sections
requires the evaluation of the
higher order electroweak contributions and thus
a priori knowledge of the structure functions themselves.
For these reasons,
the structure functions were determined iteratively, using a
Monte Carlo simulation of the experiment.
Separate simulations were performed for each period of
data taking, which enabled changes in the beam
and the detector to be taken into account.
All Monte Carlo events were processed through the same chain
of reconstruction programs as the data,
and weighted with the best
knowledge of the total cross section.
The accuracy of the Monte Carlo simulation was checked
after the iterative process had converged
by comparing, for each data period,
the distributions of data and Monte Carlo
events in variables not or only weakly related to $x$ and $Q^2$.
 
We discuss
two iterative methods to compute $F_2$ and $R$.  Method~A had
been used in our previous publications \cite{f2nmc92,f2nmc95},
whereas method~B was used in the present analysis.
 
\subsection{Method A }
In this method $F_2$ was calculated from all the data simultaneously.
Parametrisations of $F_2$ and $R$ were used, together with calculated higher order
electroweak contributions, to determine cross sections which were used as the
weights for the Monte Carlo events.
The differences between the distributions of the accepted Monte Carlo events and the data
led to a new parametrisation of $F_2$, for input into the next iteration step.
The parametrisation of $R$ was taken from ref.~\cite{rslac} and kept fixed in the iteration.
Convergence was considered reached when the values of $F_2$ changed by less than 0.2\%, in
practice after two iterations.

In this determination of $F_2$ it was assumed that the parametrisation of $R$,
 valid for $ x \ge 0.015 $ \cite{rslac}, could be extrapolated to small values of $x$;
the validity of this assumption is discussed in section~6.  The main advantage of method~A
is that it uses the information on $F_2$ over the full kinematic range
at each step of the iterative process.
 
\subsection{Method B }
In this method, used in the present publication, the iterative procedure was
split in two~\cite{peterphd}, which facilitates the extraction
of $R$ as well as $F_2$ from the data.
\begin{itemize}
\item
First, the scattering cross sections were extracted separately for each
data set taken at a given energy and target position with a given trigger.
For each such data set, the cross sections were parametrised
using the product of a 10-parameter function~\cite{peterphd}
and the Mott cross section and used as
the weights for the Monte Carlo events.
In this iterative procedure the cross sections were  determined 
but were not corrected for higher order electroweak processes. 
Convergence to better than 0.2\% was reached after typically three iterations.
\item
In the second step all the extracted cross sections were
corrected for higher order electroweak processes and
used simultaneously to determine $F_2$ and $R$.
Again this was done
iteratively, and typically convergence was reached after two
iterations.
In this procedure, $F_2^p$ and $F_2^d$ were parametrised
using the 15-parameter function
of ref.~\cite{f2nmc95};
the parametrisation of $R$ is discussed in section~4.
\end{itemize}
 
The individual extraction of the cross sections for each
data set in the first step of method~B leads to a different
sensitivity to uncertainties in the acceptance
to that of method~A.
Thus a comparison of the results from the two methods provides
information on the overall acceptance uncertainty as well
as on the reliability of the methods used to extract $F_2$
and $R$. 
The results of the two methods were found to agree
to a fraction of a percent, except at the largest $x$,
if the values of $R$ in method~B were taken to be the same as
in method~A.

\subsection{Relative normalisations }
The cross section measurements are
affected by normalisation
uncertainties which are of the order of
2\%~\cite{f2nmc95}.
To estimate possible normalisation shifts, we have
fitted simultaneously the proton and deuteron structure
functions $F_2^p$ and $F_2^d$ from the present analysis together
with those from BCDMS \cite{BCDMS} and
SLAC \cite{slac92}, again using the
15-parameter function of ref.~\cite{f2nmc95}.
For the NMC data there were six normalisation parameters
(one per energy and trigger), and the same normalisation
parameters were used for the proton and deuteron data owing to the
simultaneous data taking; the other four data sets each had
one normalisation parameter.  The $\chi^2$ of the fit
included the shifts,
weighted according to the quoted normalisation uncertainties
of the experiments.  In order to reduce the sensitivity
of the fitted parameters to systematic uncertainties
and to the value of $R$, we excluded
data at small and large $y$ from the fit.
The resulting shifts are given in table~\ref{tab:normpar};
they were determined using data in the range $0.2 < y < 0.5 \,$.
We checked that they do not significantly depend on
the particular $y$ range used.
The shifts are all compatible with the
normalisation uncertainties quoted by the experiments;
they are close to those obtained in ref.~\cite{f2nmc95}.
For the results presented in the rest of this paper
the normalisation shifts of table~\ref{tab:normpar}
have been applied.

\section{The extraction of $F_2$ and $R$ }
 
Where at least two cross section measurements
are available at a given $x$ and \q2 
for different incident muon
energies, it is possible to determine the function
\rx from the weak dependence of the cross section on $y$,
as can be seen from eq.~(1).  Data points
at small $y$ are mainly sensitive to $F_2$, while at large $y$
they are also sensitive to $R$.
The inclusion in the present analysis
of the small angle trigger data, which have smaller average values of $y$
and are thus {\it less} sensitive to $R$ than the large angle
trigger data, allows in principle a simultaneous
determination of $F_2$ and $R$.
In practice, the extraction of $R$ was limited as follows:
\begin{itemize}
\item
The sensitivity of our data to $R$ is substantial only at small $x$
(see figure~1).  
For $ x > 0.12 \,$ almost all our data are at $y < 0.40$
and there is little sensitivity to $R$.
Thus the simultaneous determination of $F_2$ and $R$
was not possible over the full kinematic range
of our measurement.
\item
The study of systematic uncertainties on $R$ showed
that the present data do not allow a measurement of
both its $x$ and \q2 dependence.
\item
We have extracted the average $R$ for
the proton and the deuteron.  This is justified by the
NMC results on the cross section ratio
$\sigma^d / \sigma^p$ \cite{nmcnp96}, with
more events and smaller systematic errors than
the data described here, which show that in the kinematic
range $ 0.002 < x < 0.40 \,$, $R^d - R^p$ is compatible
with zero to within 0.02 
(see also refs.\cite{rslac,nmcdr92}).
\end{itemize}
 
As a consequence, we have chosen the following procedure for the
extraction of $F_2$ and $R$.
For $ x < 0.12 $, we determined in each bin of $x$ one $Q^2$ averaged
value of $R$, together with the values of
$F_2^p(x,Q^2)$ and $F_2^d(x,Q^2)$.
In the lowest $x$ bin, $ x = 0.0045 $,
where our data span a very small $y$ interval,
the determination of $R$ relies on the
extrapolation of $F_2^p$ and $F_2^d$ from neighbouring
bins at larger $x$.
At $x > 0.12 \,$, where $R$ is known from other measurements,
we used the $R$ parametrisation given in ref.~\cite{rslac}
in the determination of the structure functions,
$F_2$.\footnote{We remark that most of the 139
measurements of $R$ used to obtain the parametrisation
in ref.~\cite{rslac} are for $ x \ge 0.10 \,$.
In contrast, for $ x < 0.10 $
where our data are most sensitive to $R$, the
remaining 5 measurements of $R$
have total errors of typically 40\%.}
 
To summarize, the final results of this analysis,
given below in sections 5 and 6, were obtained using
the method described in this section and in
sections 3.2 and 3.3\,.

\subsection{Systematic uncertainties on $F_2$}
The systematic uncertainties on the large angle trigger data
have been discussed in ref.~\cite{f2nmc95}.
Here, we discuss only those for the small angle trigger data.
 
As in ref.~\cite{f2nmc95} the uncertainty in the determination
of the spectrometer acceptance was studied by comparing the
structure functions determined separately from the upstream and
the downstream targets, for which the spectrometer has largely
different acceptances.
This uncertainty was also studied by comparing the structure
function results obtained from methods~A and B described above.
These studies led to an estimated contribution to the
systematic error on $F_2$ usually in the range
0.5-2\%, reaching 4\% at the edge of the kinematic domain in $x$ and $Q^2$.
The spectrometers' calibration uncertainties of 0.2\% discussed
in section~2 contributed at most
1.5\% to the systematic uncertainty on $F_2$.

The effect of hadronic and electromagnetic showers on the reconstruction
efficiency was determined from a simulation of the full final state, 
following the method described in refs.\cite{f2nmc92,annaphd}.
For the small angle trigger the loss of efficiency due to these showers
was much larger than for the large angle trigger.
The resulting inefficiency was found to be independent of $Q^2$ and
thus was parametrised as a function of $x$ only. At the smallest $x$ ($x=0.0045$)
it amounts to about 20\% decreasing linearly with ln$x$ to zero at $x\sim0.1$.
The consequent systematic error on $F_2$ was estimated to be 1.5 to 4\%.
 
The higher order electroweak contributions to the cross
sections and their uncertainties were calculated as in
ref.~\cite{f2nmc95} using the method of ref. \cite{radcor}.
These contributions were at most 20\% (at $x = 0.005$
and $y = 0.75$) and generally much smaller.
The consequent systematic errors on $F_2$
are typically less than 1\%, but can reach 2\% at large $y$.
They arise predominantly from the uncertainties
in $R$, in the proton form factor and in the suppression
of the quasi-elastic scattering on the deuteron.
The uncertainty in $R$ contributes both here and through the
calculation of the one photon exchange cross section.
 
The normalisation uncertainty of the data at each incident
energy, relative to the fitted function describing
$F_2$ used in the iteration, is estimated to be 2\%
(see also section~3.3 and table~\ref{tab:normpar}).
This is included in the 2.5\% total normalisation uncertainty of
the combined data.
Because the hydrogen and deuterium targets were simultaneously
exposed to the beam, the uncertainty on the relative
normalisation of $F_2^p$ and $F_2^d$ is negligible.
 
The total systematic uncertainties on $F_2$, obtained by adding
in quadrature all the above contributions apart from the
normalisation uncertainty, range from 1\% to 5\%, with
a median value of 2\%.
 
\subsection{Systematic uncertainties on $R$}
The largest contribution to the systematic error on $R$
stems from the 2\% uncertainties in the relative
normalisations of the six data sets.
It was estimated by varying the normalisation of
each data set in turn and recalculating $R$ and $F_2$.
The quadratic sum of the resulting variations in $R$
was taken as one contribution to the systematic uncertainty.
 
The other sources of uncertainty are the same as
those for $F_2$, discussed above.
These are the uncertainties in the incident and scattered
muon momenta, the higher order electroweak contributions,
the acceptance and the reconstruction losses;  the last three
were assumed to be fully correlated between data sets.
The contributions of these uncertainties to the systematic error
on $R$ were determined by recalculating $R$ with the cross
sections shifted according to each uncertainty in turn.
 
The individual contributions to the systematic error
on $R$ were added in quadrature giving total
uncertainties ranging from 0.11 at $x = 0.0045$ down
to 0.07 at $x = 0.11 \,$; the uncertainties are largely correlated.
 
\subsection{Alternative extractions of $R$}
The extraction of $R$ described above used parametrisations
of $F_2^p$ and $F_2^d$ over the full kinematic domain of
the measurement.  This has the advantage of being less
sensitive to \q2 dependent  and data set dependent systematic uncertainties
of the cross sections, but introduces correlations between the
statistical errors on $R$ in neighbouring $x$ bins.
 
In order to cross check our standard method of calculating $R$,
we performed two other extractions whose results are shown in figure~2.
Both checks used  the one photon exchange cross sections
obtained in method B (section 3.2)
adjusted by the normalisation shifts of table~\ref{tab:normpar}.
The corrections for higher order electroweak effects were calculated 
with the SLAC parametrisation for $R$, but without any iteration.
 
The first check, labelled check~1 in figure~2, evaluated $R$ and $F_2$ at each $x$ bin
independently, using all the measured cross sections in that bin.
We fitted an average value of $R$ and a parametrisation
of $F_2^d$ to the one photon exchange cross sections
for each $x$ bin in the range $ 0.006 < x < 0.12 $ in turn.
The parametrisation for  $F_2^d$ which we used, and which we checked
to be sufficiently flexible to describe our data, was
\begin{center}
$ \ln F_2^d(x_i,Q^2)  =  ( a_i + b_i \ln(Q^2)) +
   \ln (1 + c_i / Q^2) \; , $
\end{center}
while $F_2^p$ was obtained by dividing
the $F_2^d$ parametrisation by our measured ratio
$F_2^d / F_2^p$~\cite{nmcnp96}.
 
The second check (check~2) followed the traditional
$\varepsilon$-method, as e.g. in the SLAC analysis~\cite{rslac},
which uses the individual $(x, Q^2 )$ bins one at a time.
The ratio $R \, ( \, = \sigma_L/\sigma_T$) was obtained from the
variation of the cross section as a function of $y$ within each bin,
using the cross sections expressed in a form
proportional to $ \sigma_T + \varepsilon \sigma_L $, where
$ \varepsilon^{-1}  \simeq  1 \, + \, (y^2 +Q^2 / E^2) /
2 (1 - y - Q^2 / 4 E^2 ) \,$.
A linear fit to the cross section as a function of $\varepsilon$
yielded a determination of $R$ in each $(x,Q^2)$ bin in which measurements
from at least three energies are available.
 
 As can be seen in figure~2, the results of the two checks agree well with
those of the present analysis, given their different sensitivities 
to systematic errors.
 
\section{Results for $F_2$ }
The results obtained for the structure functions $F_2^p$
and $F_2^d$ for all energies and triggers are shown
in figures~3 and 4
as a function of \q2 for fixed values of $x < 0.12 \,$.
They were obtained for all six data sets using method~B
(the data of ref.~\cite{f2nmc95} were re-evaluated).
In these figures, the error bars represent the
quadratic sum of the statistical and systematic errors.
The small angle data extend the measured $x$ domain downwards
to $x = 0.0045 \,$;  they also fill in gaps in the \q2
range covered by the large angle trigger data.
Within the quoted uncertainties, there is good agreement
between all data sets.
It may be noted that if $R = 0$ was used in the analysis,
the \q2 dependence of the $F_2$ data in the small $x$
bins became noticeably less smooth.
 
The values of $F_2^p(x,Q^2)$ and $F_2^d(x,Q^2)$
and the various contributions to the systematic errors are
given in tables 3~(a--d) and 4~(a--d);
the scattering cross sections, $d^2\sigma^{meas}/dxdQ^2$, 
not corrected for higher order electroweak effects, are also given.
In these tables, the results from the two triggers
in the data taken at 200 and 280~GeV were averaged.
 
In figures~5 and~6 we show the measured $F_2^p$ and $F_2^d$
averaged over the four energies.
The corresponding values are given in tables~5 and~6.
We also give in these tables values of $\tilde{F}_2^{\rm p}$ and $\tilde{F}_2^{\rm d}$ 
determined if we use the
SLAC parametrisation of $R$ ($R_{1990}$); the difference becomes sizeable at small $x$.

Our results are in good agreement with those of SLAC
\cite{slac92} and BCDMS \cite{BCDMS}
as is shown in figure~7 for the deuteron.
Recently, final results extending down to $x = 0.0008 \, $
have become available from the
fixed target muon scattering experiment E665 at Fermilab
\cite{ref:E665}. 
In figure~8 our data on $F_2^p$ and $F_2^d$ are compared
with the E665 results for $ 0.004 < x < 0.04 \,$.
The agreement is generally good, except perhaps
around $ x = 0.01 $ for $F_2^d$.
Note that E665 uses $R_{1990}$ while 
we use in this small $x$ region our own measurement for $R$.
However, this does not influence the comparison since the 
E665 data are not very sensitive
to $R$ (because the cover the small $y$ region).
 
In figure~9 we present a comparison of the NMC data
with the results obtained by the H1 \cite{ref:h1} and
ZEUS \cite{ref:zeus} collaborations at HERA
from their 1994 data.
These new HERA results agree much better with the NMC data than the earlier
ones and the gap in \q2
between the HERA and NMC data has become smaller (see e.g. ref.\cite{f2nmc95}).
 
\section{Results for $R$}
In this section, we present the results on $R$,
obtained from the combined proton and deuteron data,
in the range $ 0.002 < x < 0.12 \,$.
Figure~10(a) shows these results for $R$ as a function
of $x$ at an average \q2 ranging from
$\langle Q^2 \rangle \, = \, $ 1.4 to 20.6~GeV$^2$ for the different $x$-bins.
The corresponding values are given in table~7.
In the figure the error bars indicate
the quadratic sum of statistical and systematic uncertainties.
The systematic errors are 1.5 to 3 times larger than the
statistical ones; they are dominated by the normalisation
uncertainty and are thus largely correlated.
 
In figure~10(a), we also compare the present results for $R$
with the parametrisation of ref.~\cite{rslac} used in
previous measurements of $F_2$, and with a QCD prediction.
Within the largely correlated systematic errors, the agreement
with the parametrisation is good, even at small $x$ where the
parametrisation is an extrapolation of previous measurements.
 
The QCD prediction for $R$ was calculated as
\begin{equation}
  R(x,Q^2) \, =  \, {F_L(x,Q^2) + {4M^2x^2 \over Q^2} F_2(x,Q^2)
 \over F_2(x,Q^2) - F_L(x,Q^2)} \, ,
\end{equation}
using the formula of ref.~\cite{altamar} for the
longitudinal structure function $F_L$ in
next to leading order QCD, 
\begin{equation}
    F_L(x,Q^2) \, =  \, {\alpha_S(Q^2) \over 2 \pi} \,
 x^2 \int_x^1 \, {{\rm d}w \over w^3}  \,
 \left[ {8 \over 3} \, F_2^{SI}(w,Q^2) + {40 \over 9}    \,
(1-{x \over w}) \, w \, G(w,Q^2) \right] \, ,
\label{eq:flqcd}
\end{equation}
where $F_2^{SI}$ is the singlet quark and $xG$ the gluon
momentum distribution.    In evaluating this expression,
we have taken the $F_2^d$ parametrisation
from ref.~\cite{f2nmc95} for $F_2^{SI}$, and used
the gluon distribution from a QCD analysis of BCDMS,
NMC and H1 data~\cite{ref:h1}; as in this analysis, we set
$\alpha_S ( 50 $GeV$^2 ) = 0.180 $~\cite{virmil}.
The agreement is good, even at small \q2 and $x$ where the
validity of the Altarelli--Martinelli relation might be
questioned, and the uncertainty on the gluon distribution
is large.
 
In figure~10(b) we compare the present $R$ measurement to
other results obtained at comparable $x$ and \q2 
by BCDMS \cite{BCDMS,BCDMSC} and CDHSW \cite{CDHSW}.
Good agreement is observed; the present results are lower
but the difference is within the largely correlated
uncertainties of the experiments.
The NMC results improve the knowledge of $R$
for $ x < 0.1 $ considerably.

In figures~10(b) and 11, the $R$ measurements shown have been
obtained from very different target materials.  There is no
evidence for a significant $A$ dependence of $R$, consistent with
perturbative QCD expectations. Similar conclusions were reached from 
measurements of differences between $R$ for various 
nuclei \cite{nmcdr92,E140,E140X,snc}.
 
In figure~11, we compare in \q2 bins
the present results with others, mostly
obtained at SLAC
\cite{rslac,E140,E140X} at larger $x$.
In these plots, no significant $x$-dependence is
observed for $ Q^2$ between 1.5 and 20~GeV$^2$.
Thus one may conclude that the $x$ dependence observed
in figure~10(b) is a reflection of the variation of the
average \q2 with $x$.

\section{Summary}
 
We have extracted $F_2^p$, $F_2^d$ and $R$ from measurements
of inclusive deep inelastic muon scattering on the proton and
the deuteron at 90, 120, 200 and 280~GeV.
The results for $F_2^p$ and $F_2^d$ cover
the kinematic range $ 0.002 < x < 0.60 $ and
$ 0.5 < Q^2 < 75$ GeV$^2 $, with high statistical accuracy
and with systematic uncertainties between 1\% and 5\%.
The data are in good agreement with the earlier high
statistics results from SLAC and BCDMS, as well as with more
recent data at small $x$
from the Fermilab E665 experiment and from H1 and ZEUS at HERA.
Results for $R$ cover
the kinematic range $0.002 < x < 0.12$ and
$1.0 < Q^2 < 25$~GeV$^2$,
extending its knowledge to small values of $x$.
They are in agreement with earlier measurements as well as 
with expectations from perturbative QCD.

\vspace{2.2cm}

\newpage
\begin{table}[p]
\centering
\begin{tabular}{|c|cccccc|cc|} \hline
 Incident energy & $p'_{min}$ & $\nu_{min}$ & $\theta^{up}_{min}$ &
 $\theta^{down}_{min}$ & $y_{min}$ & $y_{max}$ & $N_p$ & $N_d$ \\
 ~(GeV)   & (GeV)      & (GeV) & (mrad)              &
(mrad)                &     &           & ($10^3$)  &  ($10^3$) \\
 \hline
200 & 35 & 15 & 6-6.5 & 6.5-7 &  -- & 0.8 &  78 & 162 \\
280 & 40 & 30 & 6-7   & 6-7.5 & 0.2 & 0.8 &  97 & 207 \\
\hline
\end{tabular}
\caption{Cuts applied to the small angle trigger data,
as explained in the text.
Different values of $\theta_{min}$ were used
for the upstream and downstream targets
and for different data taking periods.
$N_p$ and $N_d$ are the total number of events for
protons and deuterons, respectively, after applying all cuts.
}
\label{tab:cutpar}
\end{table}
 
\begin{table}[p]
\centering
\begin{tabular}{|l|rrr|}
\hline
Data set & Proton   & ~ & Deuteron         \\
 ~       & ~ & Both & ~                \\
\hline
SLAC            &  --0.4\%  & ~  &  +0.9\%      \\
BCDMS           &  --1.8\%  & ~  & --0.7\%      \\
NMC  90 GeV     &   ~   & --2.7\%    & ~  \\
NMC 120 GeV     &   ~   &  +1.1\%    & ~  \\
NMC 200 GeV T1  &   ~   &  +1.1\%    & ~  \\
NMC 280 GeV T1  &   ~   &  +1.7\%    & ~  \\
NMC 200 GeV T2  &   ~   & --2.9\%    & ~  \\
NMC 280 GeV T2  &   ~   &  +2.0\%    & ~  \\
\hline
\end{tabular}
\caption{Normalisation changes for the different data sets.
All numbers were obtained from the fits described in section~3.3\,.
The normalisation uncertainties quoted by the experiments are
respectively 2\%, 3\% and 2\% for SLAC, BCDMS and NMC.
}
\label{tab:normpar}
\end{table}
 
\begin{table}[p]
\label{tab:f2p4nrg}
\end{table}
 
\begin{table}[p]
\label{tab:f2d4nrg}
\end{table}
 
\begin{table}[p]
\label{tab:f2pavg}
\end{table}
 
\begin{table}[p]
\label{tab:f2davg}
\end{table}
 
\setcounter{table}{6}
\begin{table}[p]
\centering
\begin{tabular}{|cccccc|}
\hline
 $x$ & $\langle Q^2 \rangle$& range & $R$ & statistical & systematic   \\
   ~ &       (GeV$^2$)     & of $y$  &   ~ &    error    &   error      \\
\hline
 0.0045 &   1.38 & 0.63-0.74 & 0.537 & 0.067 & 0.110       \\
 0.0080 &   1.31 & 0.49-0.77 & 0.337 & 0.044 & 0.112       \\
 0.0125 &   2.20 & 0.40-0.75 & 0.246 & 0.037 & 0.110       \\
 0.0175 &   3.12 & 0.28-0.73 & 0.190 & 0.036 & 0.109       \\
 0.025  &   4.5  & 0.20-0.74 & 0.099 & 0.032 & 0.090       \\
 0.035  &   7.0  & 0.20-0.73 & 0.108 & 0.030 & 0.080       \\
 0.050  &  10.2  & 0.17-0.69 & 0.117 & 0.030 & 0.079       \\
 0.070  &  14.1  & 0.12-0.69 & 0.113 & 0.049 & 0.076       \\
 0.090  &  16.9  & 0.10-0.57 & 0.096 & 0.036 & 0.079       \\
 0.110  &  20.6  & 0.10-0.60 & 0.043 & 0.041 & 0.067       \\
\hline
\end{tabular}
\caption{Results for $R$ as a function of $x$; 
$\langle Q^2 \rangle$ is the mean value of
\q2 at which $R$ was determined. The range of $y$ covered by the
data is also given.
}
\label{tab:rnmc}
\end{table}

 
\begin{figure}[htb]
\vspace{1cm}
\begin{center}
\epsfig{figure=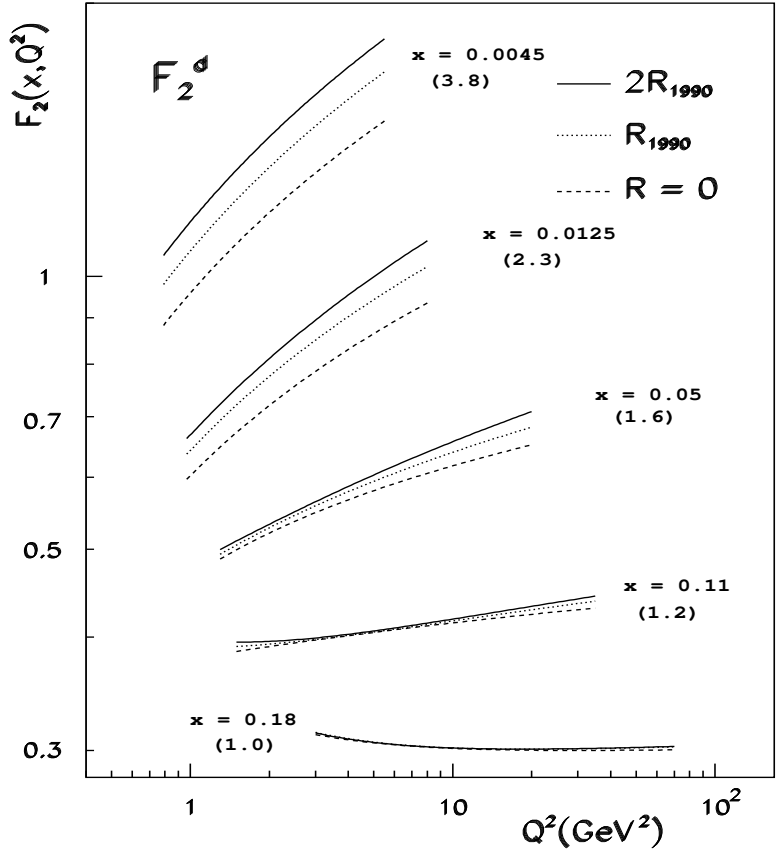,height=12cm,width=11cm}
\end{center}
\caption{The sensitivity of the measured structure
function $F_2^d$ to the assumed value of $R$.
This is illustrated for five $x$ bins.
The lines indicate the results of fits to $F_2^d$
computed assuming $R$ to be given by the $R_{1990}$
parametrisation of ref.~{\protect\cite{rslac}} (dotted),
twice that parametrisation (solid), or zero (dashed)
for five $x$ bins, with $F_2$ scaled by the numbers in brackets.}
\label{fig:sensit}
\end{figure}

\begin{figure}[h]
\vspace{1cm}
\begin{center}
\epsfig{figure=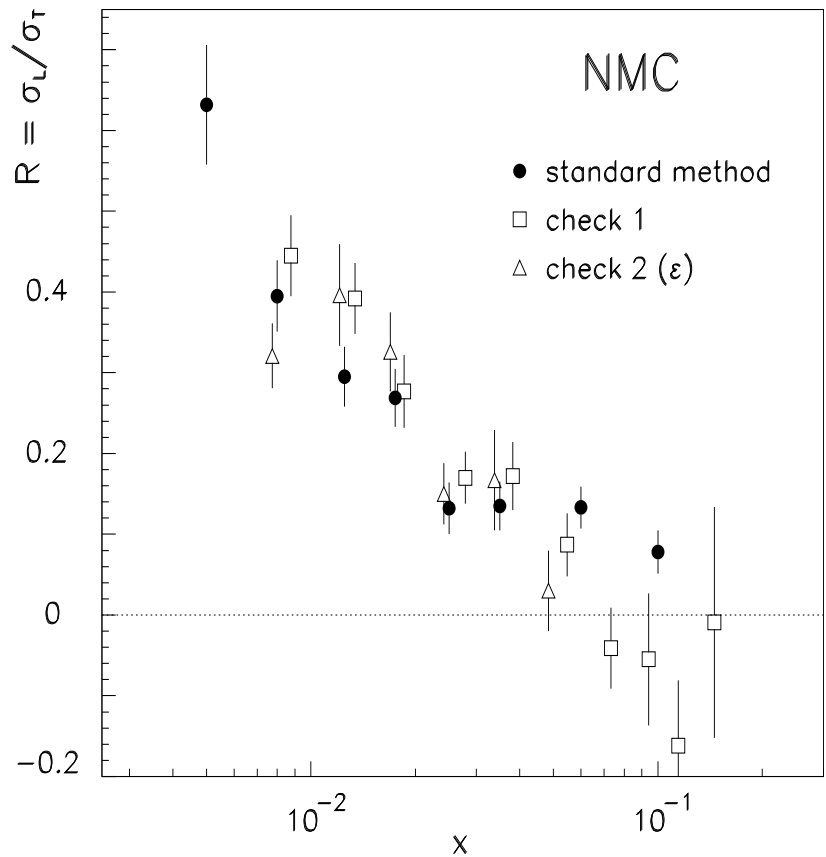,height=12cm,width=11cm}
\end{center}
\caption{
The standard NMC method for calculating $R$ (filled circles)
compared to two alternative extractions described in the text,
per $x$ bin (open squares) or per $(x,Q^2)$ bin (open triangles).
Only statistical errors are shown.
The three extractions are compared before any iteration is performed;
in particular, the results shown for the standard method are not
the final ones (these are shown in figs.~10 and 11).
}
\label{fig:Rmethod}
\end{figure}

\begin{figure}[h]
\vspace{1cm}
\begin{sideways}
\begin{minipage}[b]{\textheight}
\begin{center}
\begin{tabular}{c c}
\epsfig{figure=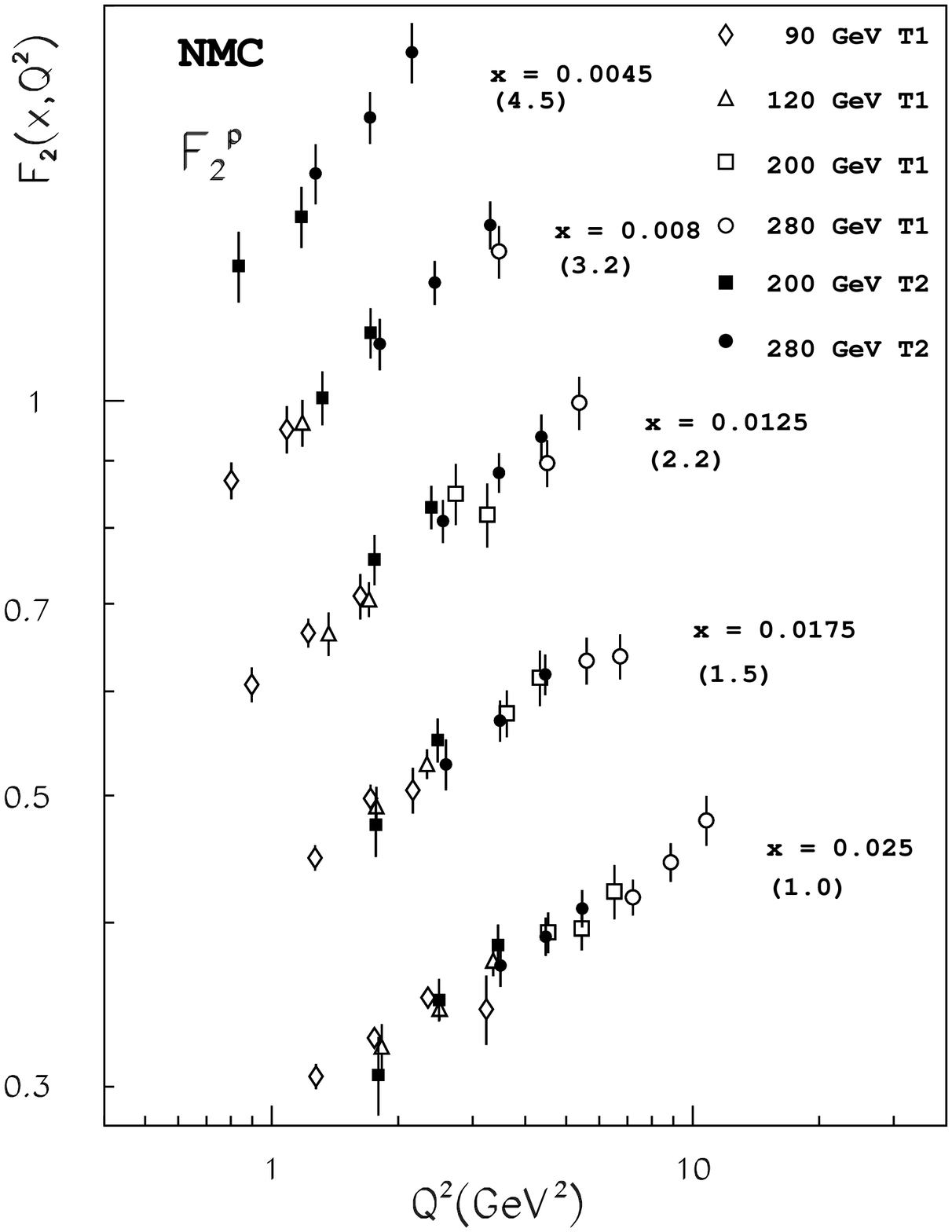,height=12cm,width=11cm} &
\epsfig{figure=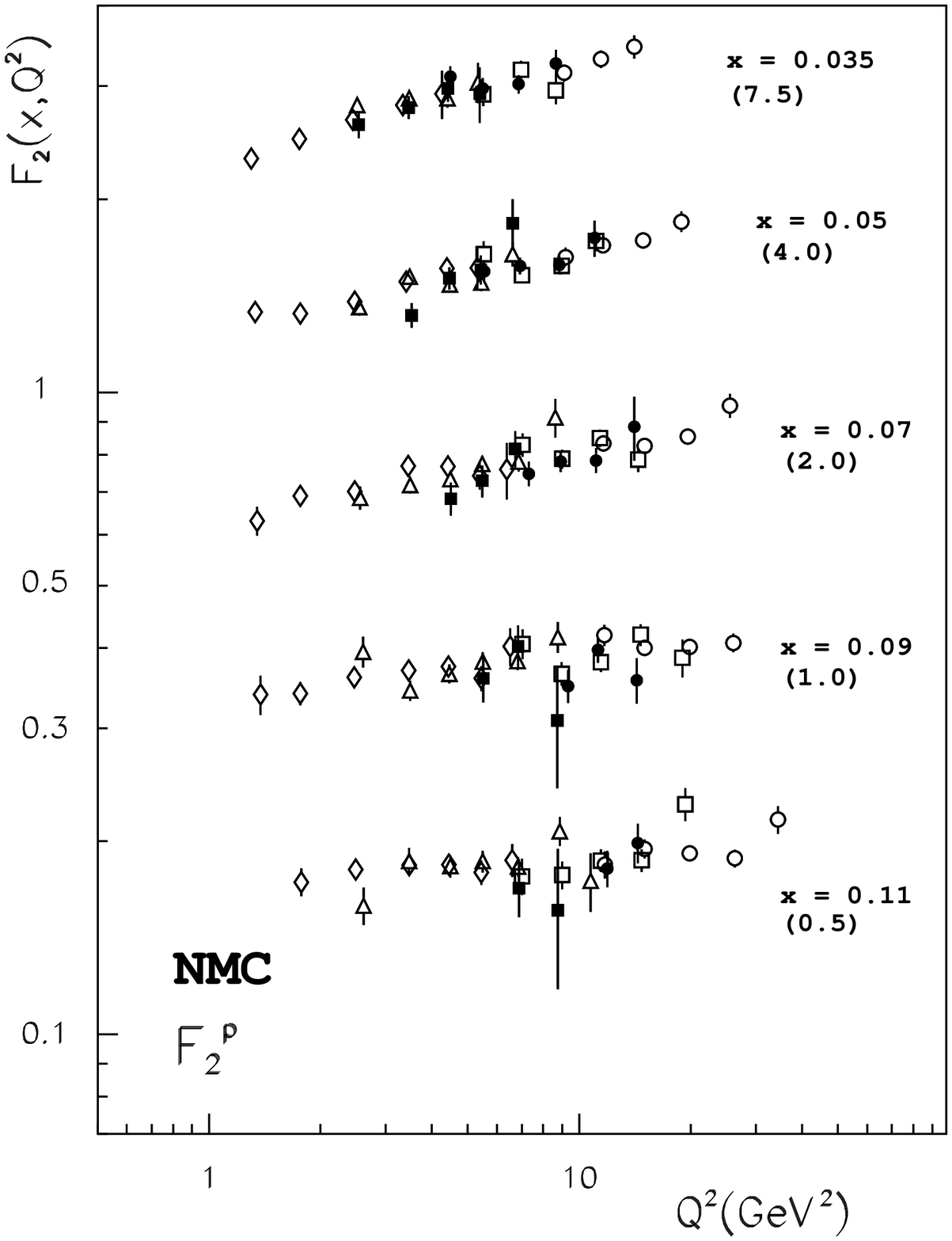,height=12cm,width=11cm}
\end{tabular}
\end{center}
\caption{
The structure function $F_2^p(x, Q^2)$ versus $Q^2$ in bins of $x$.
The six data sets correspond to four different incident energies and
two triggers.
The new small angle trigger data are shown as
filled symbols;
bins of $x$ where no small angle data exist are not shown.
The data in each $x$ bin have been scaled by the
factors indicated in brackets.
The error bars represent the quadratic sum of systematic
and statistical errors. The normalisation uncertainties are
not included.
}
\label{fig:NMCprot}
\end{minipage}
\end{sideways}
\end{figure}
 
\newpage
 
\begin{figure}[htb]
\vspace{1cm}
\begin{sideways}
\begin{minipage}[b]{\textheight}
\begin{center}
\begin{tabular}{c c}
\epsfig{figure=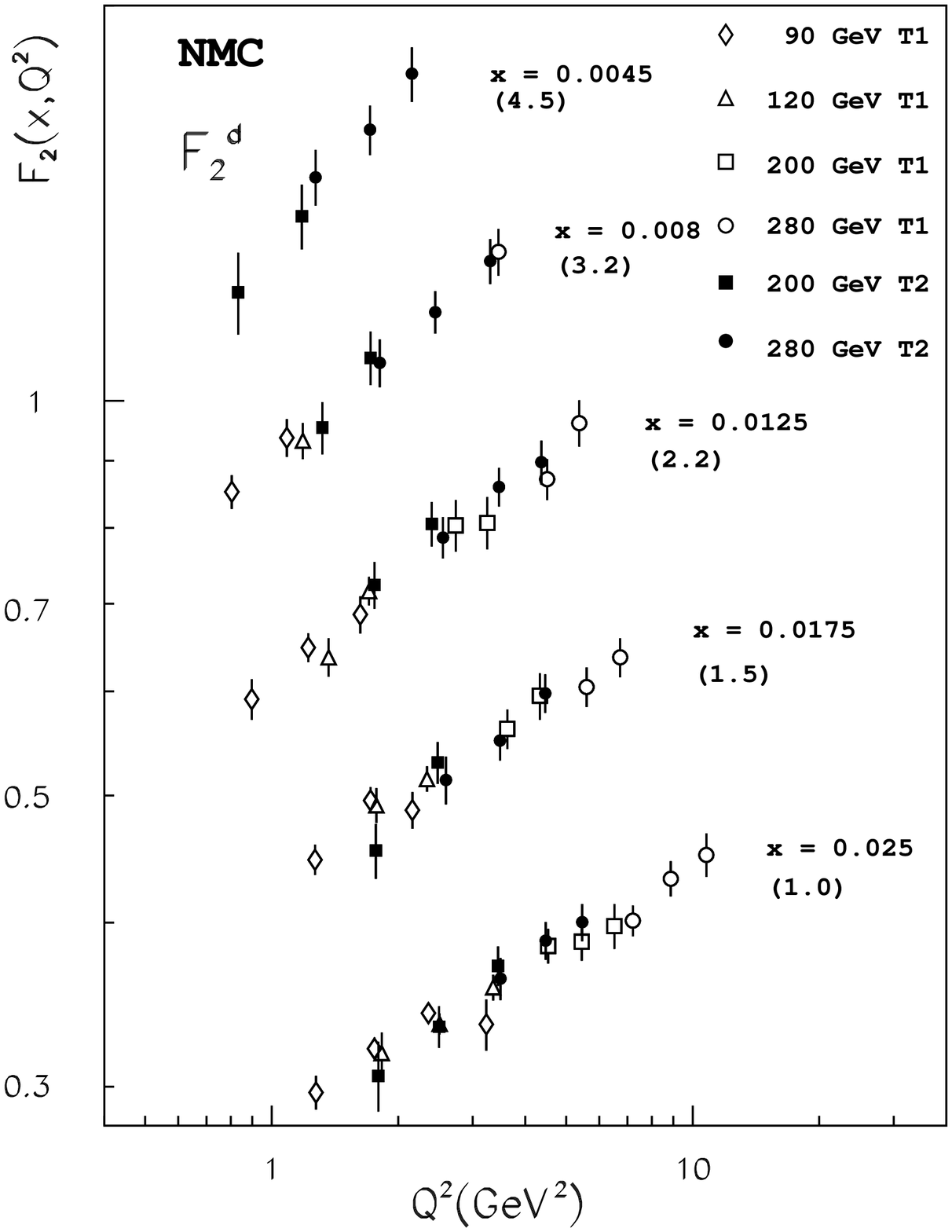,height=12cm,width=11cm} &
\epsfig{figure=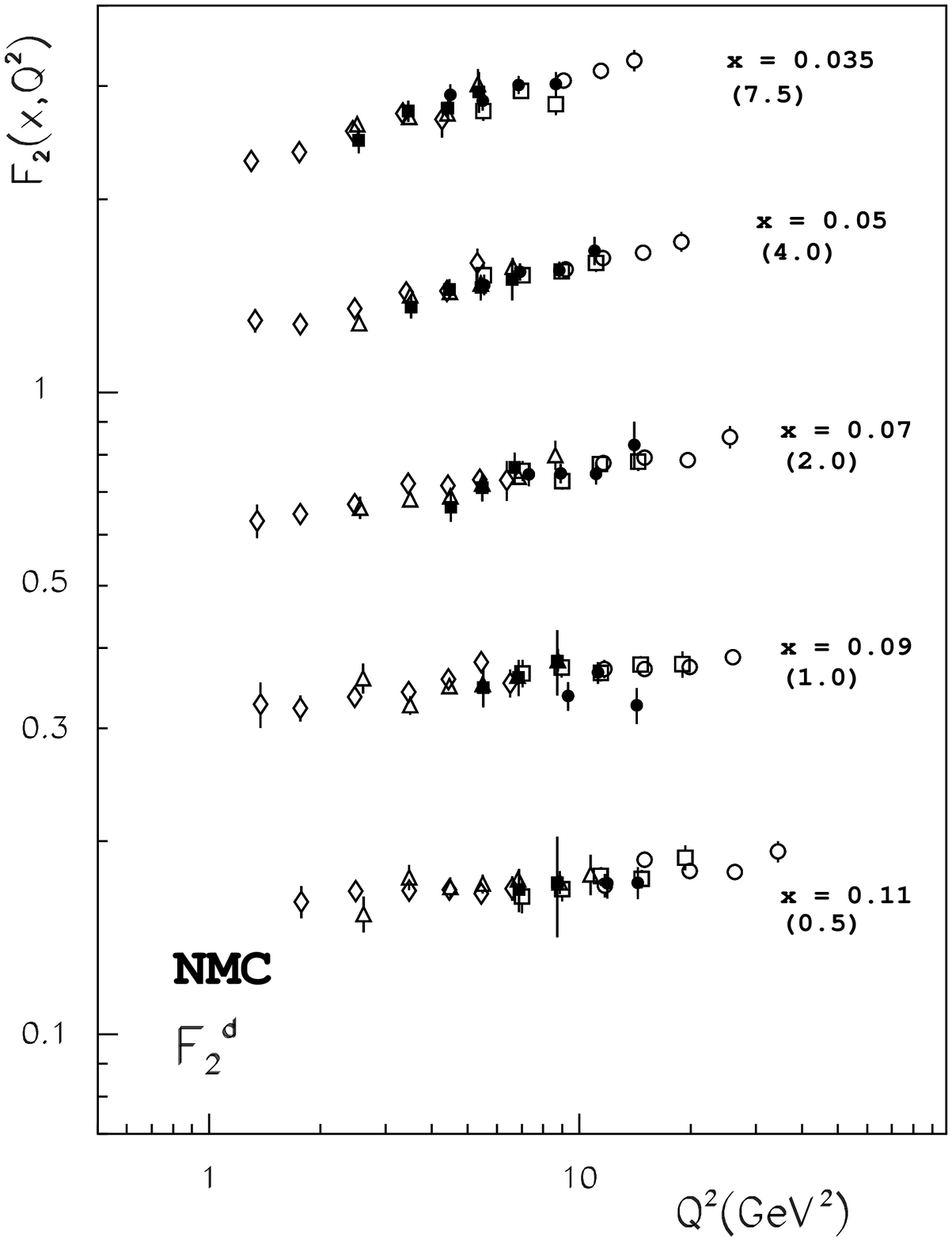,height=12cm,width=11cm}
\end{tabular}
\end{center}
\caption{
The structure function $F_2^d(x, Q^2)$ versus $Q^2$ in bins of $x$.
The six data sets correspond to four different incident energies and
two triggers.
The new small angle trigger data are shown as
filled symbols;
bins of $x$ where no small angle data exist are not shown.
The data in each $x$ bin have been scaled by the
factors indicated in brackets.
The error bars represent the quadratic sum of systematic
and statistical errors. The normalisation uncertainties are
not included.
}
\label{fig:NMCdeut}
\end{minipage}
\end{sideways}
\end{figure}

\begin{figure}[h]
\vspace{1cm}
\begin{sideways}
\begin{minipage}[b]{\textheight}
\begin{center}
\begin{tabular}{c c}
\epsfig{figure=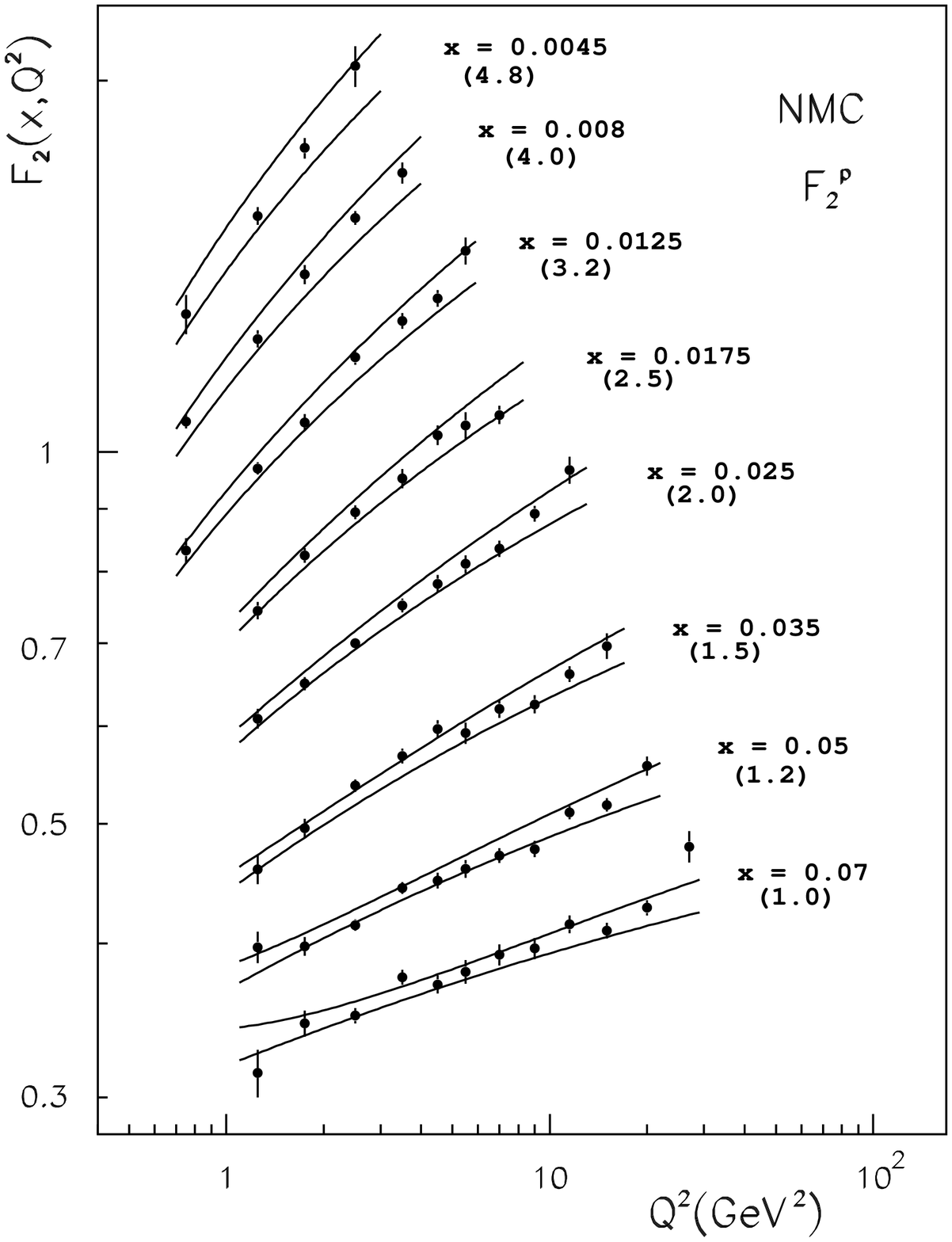,height=12cm,width=11cm} &
\epsfig{figure=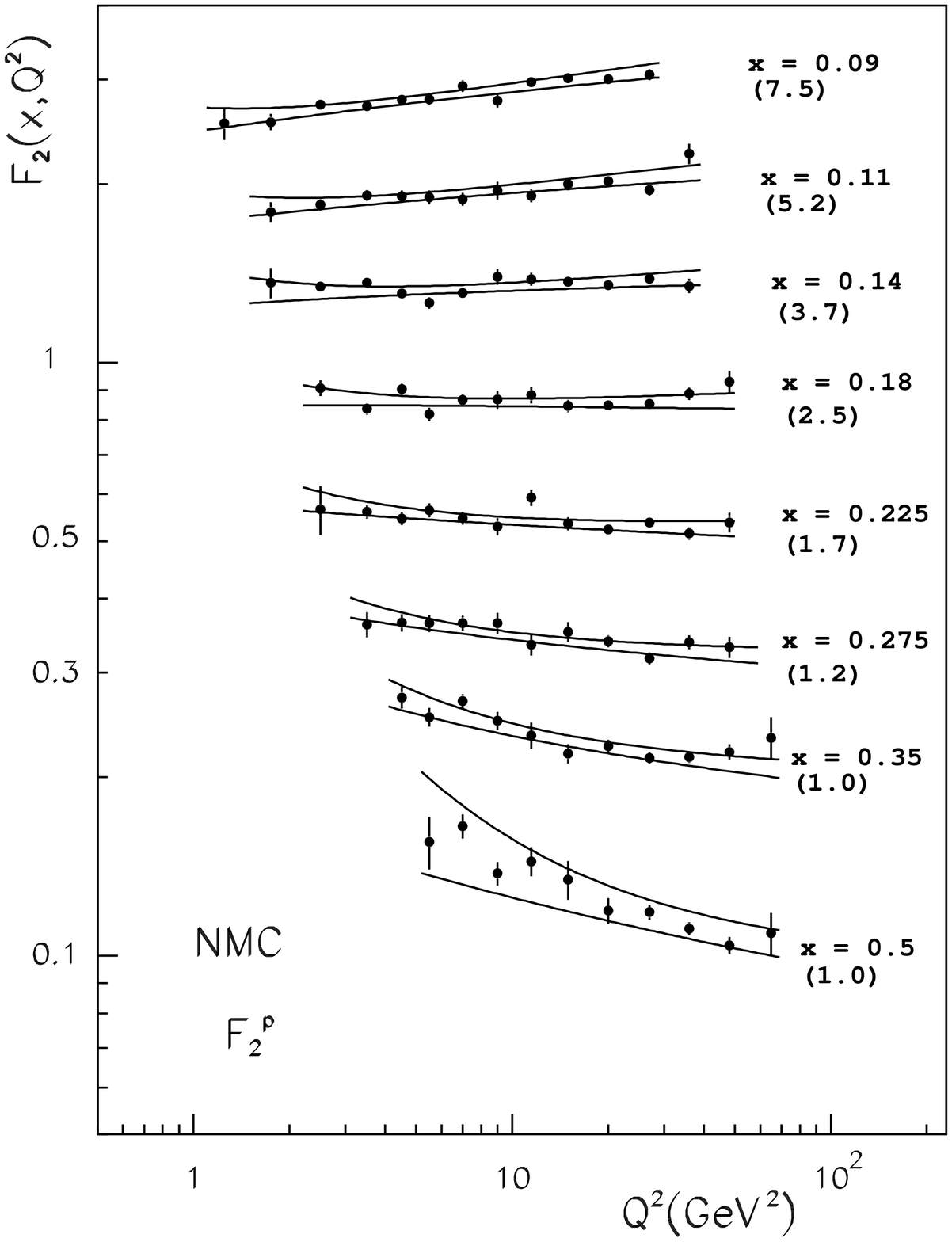,height=12cm,width=11cm}
\end{tabular}
\end{center}
\caption{
The structure function $F_2^p(x, Q^2)$ versus $Q^2$
in bins of $x$.
The corresponding values are given in table~5.
The six data sets of fig.~3 have been averaged.
The data in each $x$ bin have been scaled by the
factors indicated in brackets.
The error bars represent the statistical errors;
the total systematic uncertainties, apart from the
2.5\% normalisation uncertainty, are indicated by the bands.
}
\label{fig:NMCpavg}
\end{minipage}
\end{sideways}
\end{figure}

\begin{figure}[h]
\vspace{1cm}
\begin{sideways}
\begin{minipage}[b]{\textheight}
\begin{center}
\begin{tabular}{c c}
\epsfig{figure=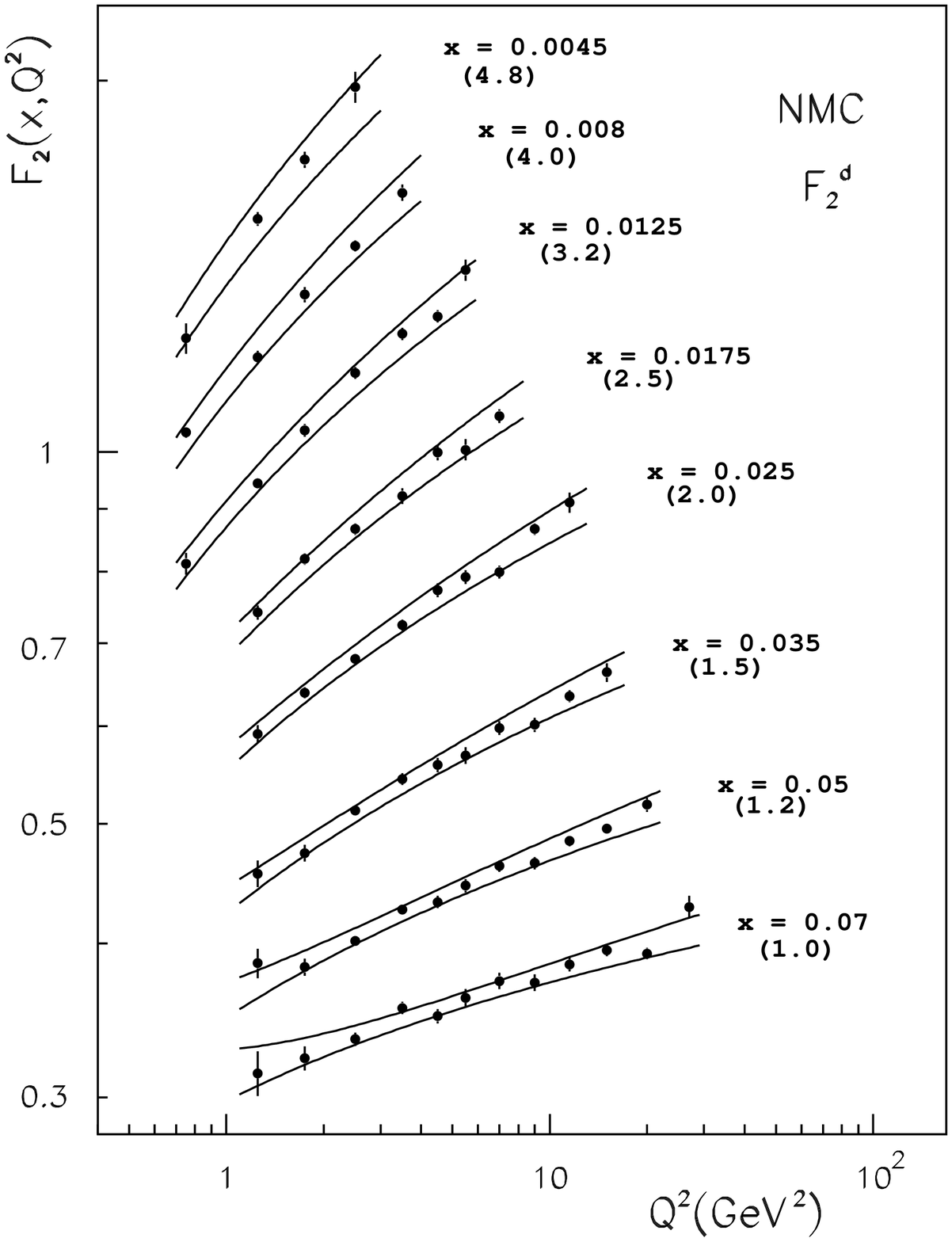,height=12cm,width=11cm} &
\epsfig{figure=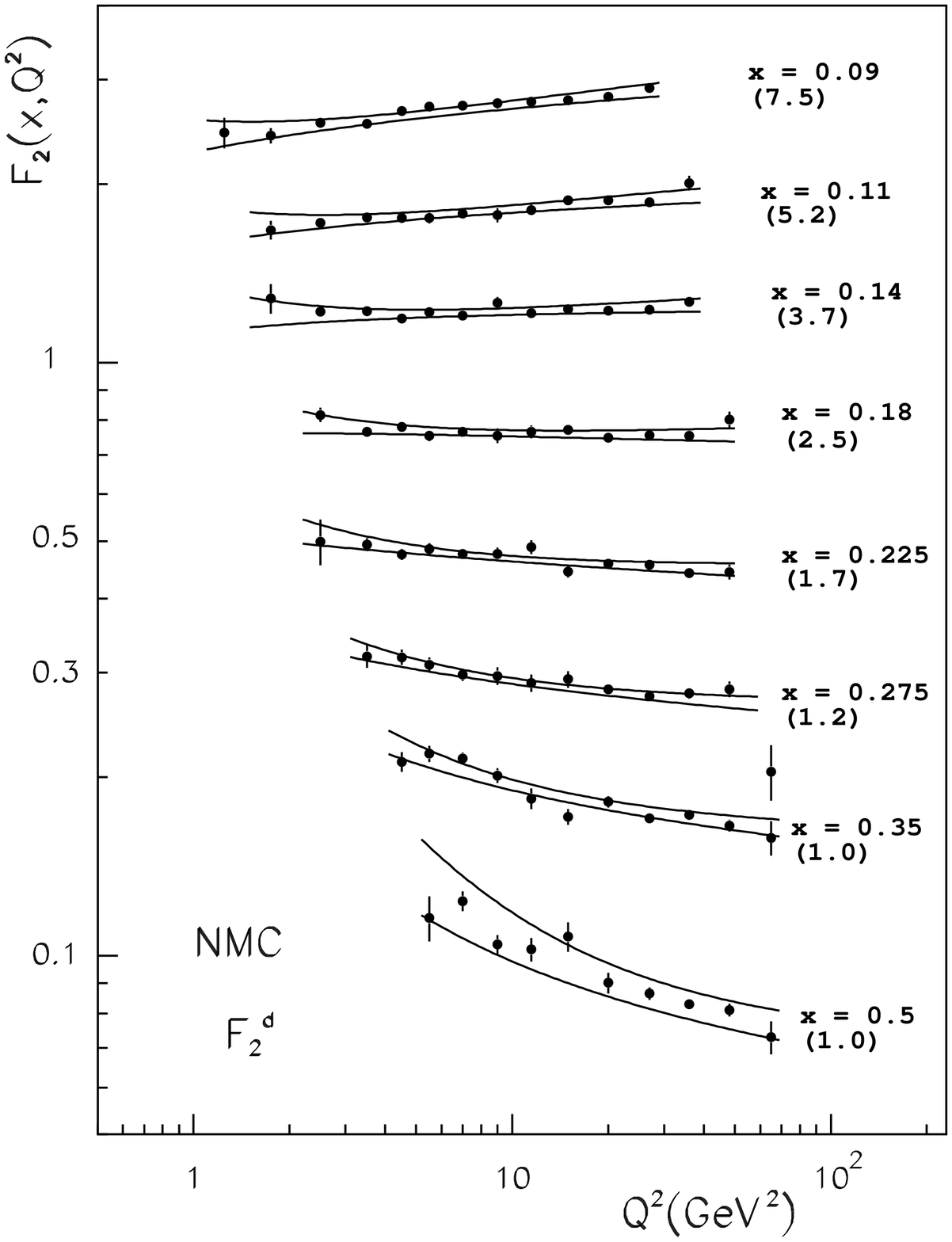,height=12cm,width=11cm}
\end{tabular}
\end{center}
\caption{
The structure function $F_2^d(x, Q^2)$ versus $Q^2$
in bins of $x$.
The corresponding values are given in table~6.
The six data sets of fig.~4 have been averaged.
The data in each $x$ bin have been scaled by the
factors indicated in brackets.
The error bars represent the statistical errors;
the total systematic uncertainties, apart from the
2.5\% normalisation uncertainty, are indicated by the bands.
}
\label{fig:NMCdavg}
\end{minipage}
\end{sideways}
\end{figure}
 
\newpage
 
\begin{figure}[h]
\vspace{1cm}
\begin{sideways}
\begin{minipage}[b]{\textheight}
\begin{center}
\begin{tabular}{c c}
\epsfig{figure=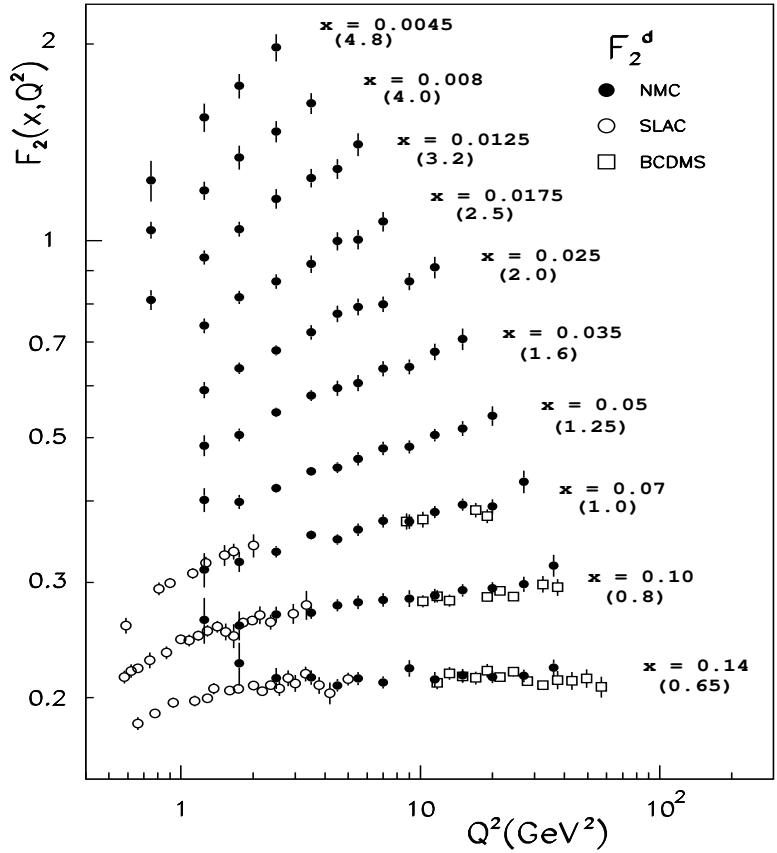,height=12cm,width=11cm} &
\epsfig{figure=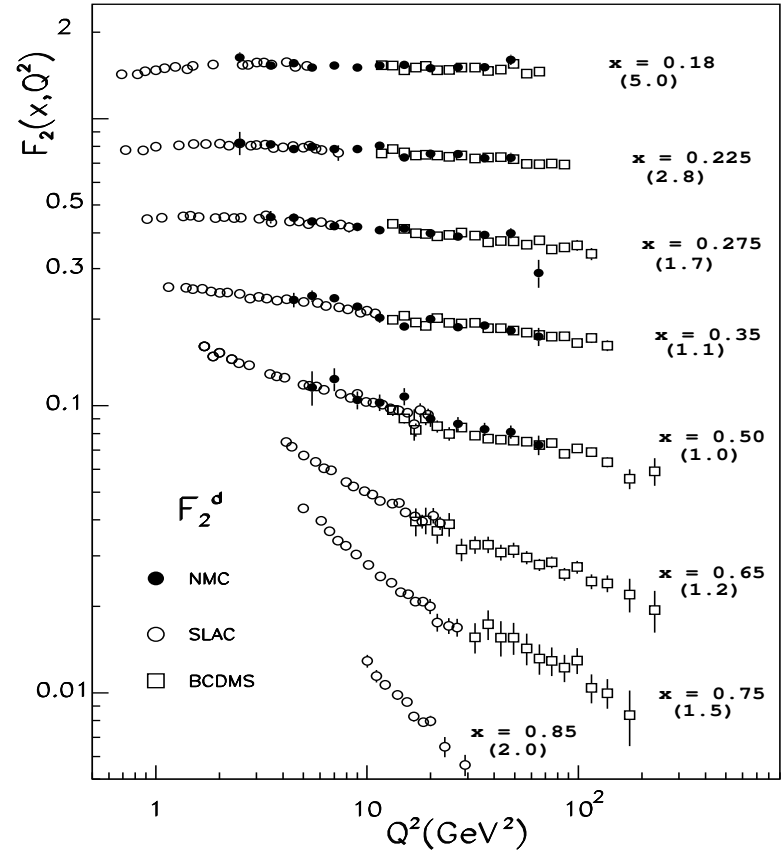,height=12cm,width=11cm}
\end{tabular}
\end{center}
\caption{
The present results for $F_2^d$ (filled circles) compared to
those of BCDMS~\protect\cite{BCDMS} (open squares)
and SLAC~\protect\cite{slac92} (open circles).
The data have been adjusted by the normalisation shifts
of table~2, and the BCDMS data have been corrected by their
calibration uncertainty, as in ref.~\protect\cite{f2nmc95}.
At $x = 0.50$ the BCDMS and SLAC data
have been interpolated.
The data in each $x$ bin have been scaled by the
factors indicated in brackets.
The error bars represent the total errors,
apart from normalisation uncertainties.
}
\label{fig:NMC+B+S}
\end{minipage}
\end{sideways}
\end{figure}

\begin{figure}[h]
\vspace{1cm}
\begin{center}
\epsfig{figure=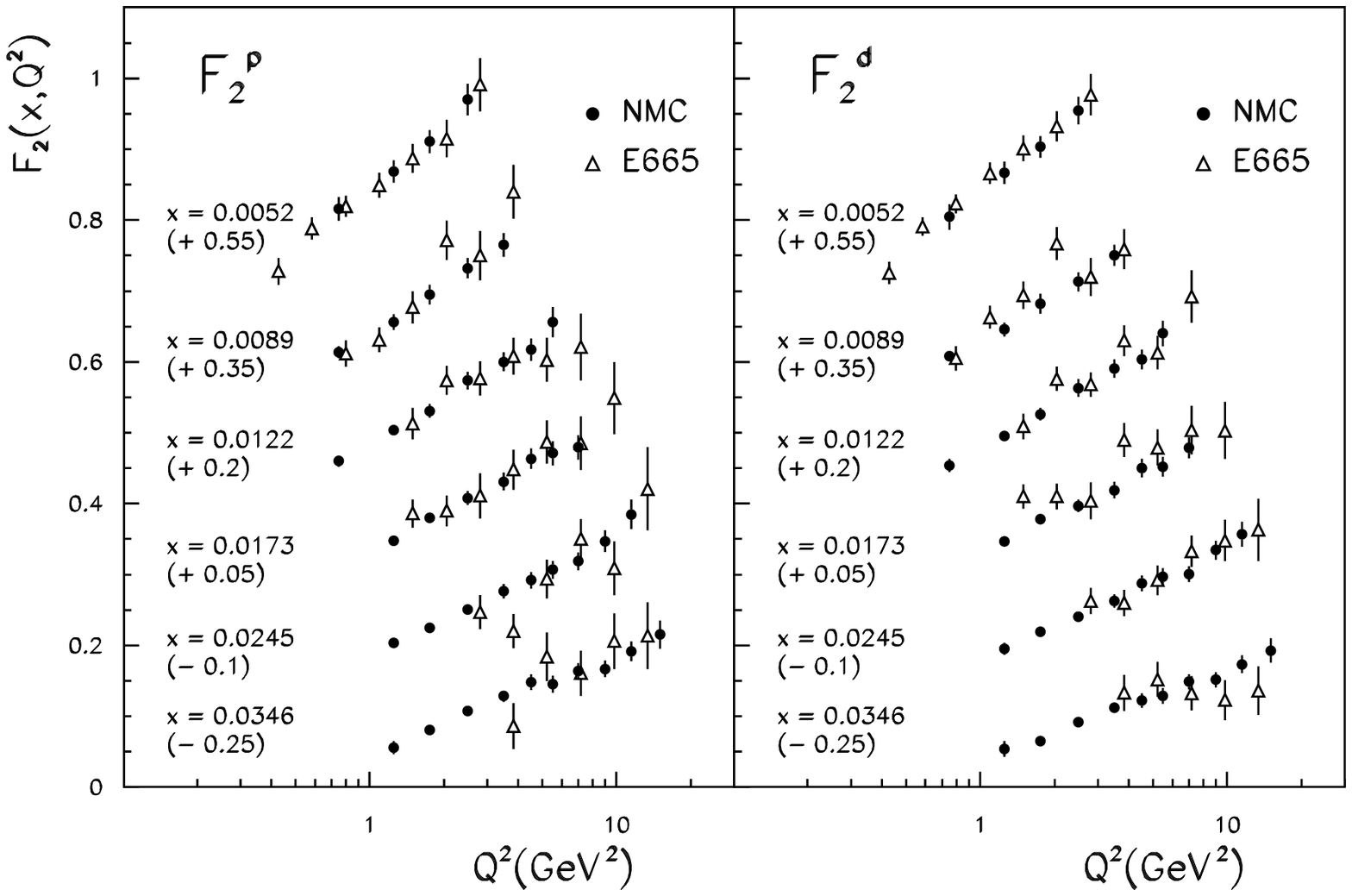,height=21cm,width=16cm}
\end{center}
\caption{
The structure function $F_2^p$ and $F_2^d$ from the present
measurement (filled circles) compared to those of the E665
experiment at Fermilab~\protect\cite{ref:E665}
(open triangles), in the range $ 0.004 < x < 0.04 \,$.
The NMC data have been interpolated to the E665 $x$ bins.
The data in each $x$ bin have been offset by the
amounts indicated in brackets.
The error bars represent the total errors,
apart from normalisation uncertainties.
}
\label{fig:NMC+E665}
\end{figure}

\begin{figure}[t]
\vspace{1cm}
\begin{center}
\epsfig{figure=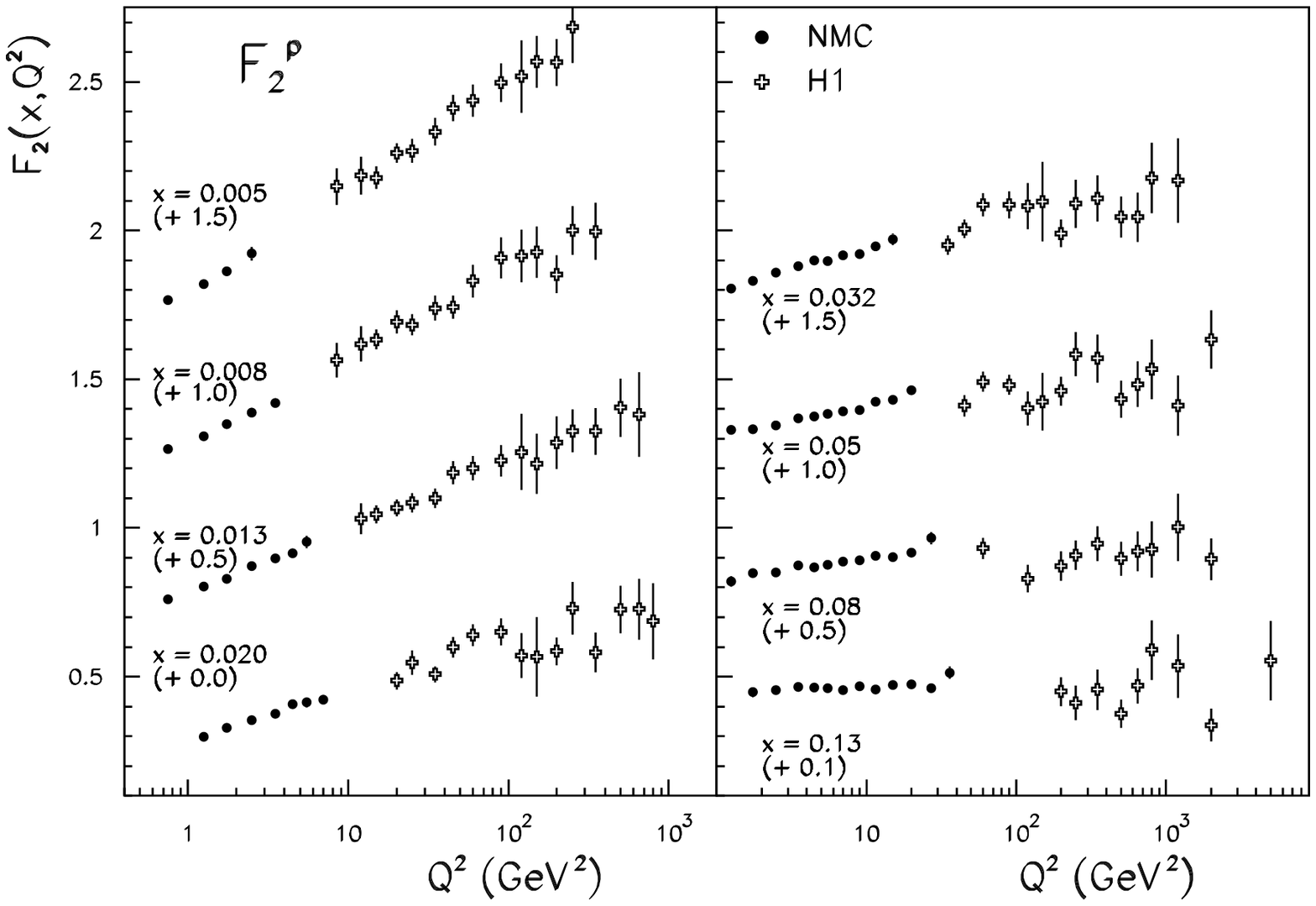,height=9.0cm,width=14cm} 
\epsfig{figure=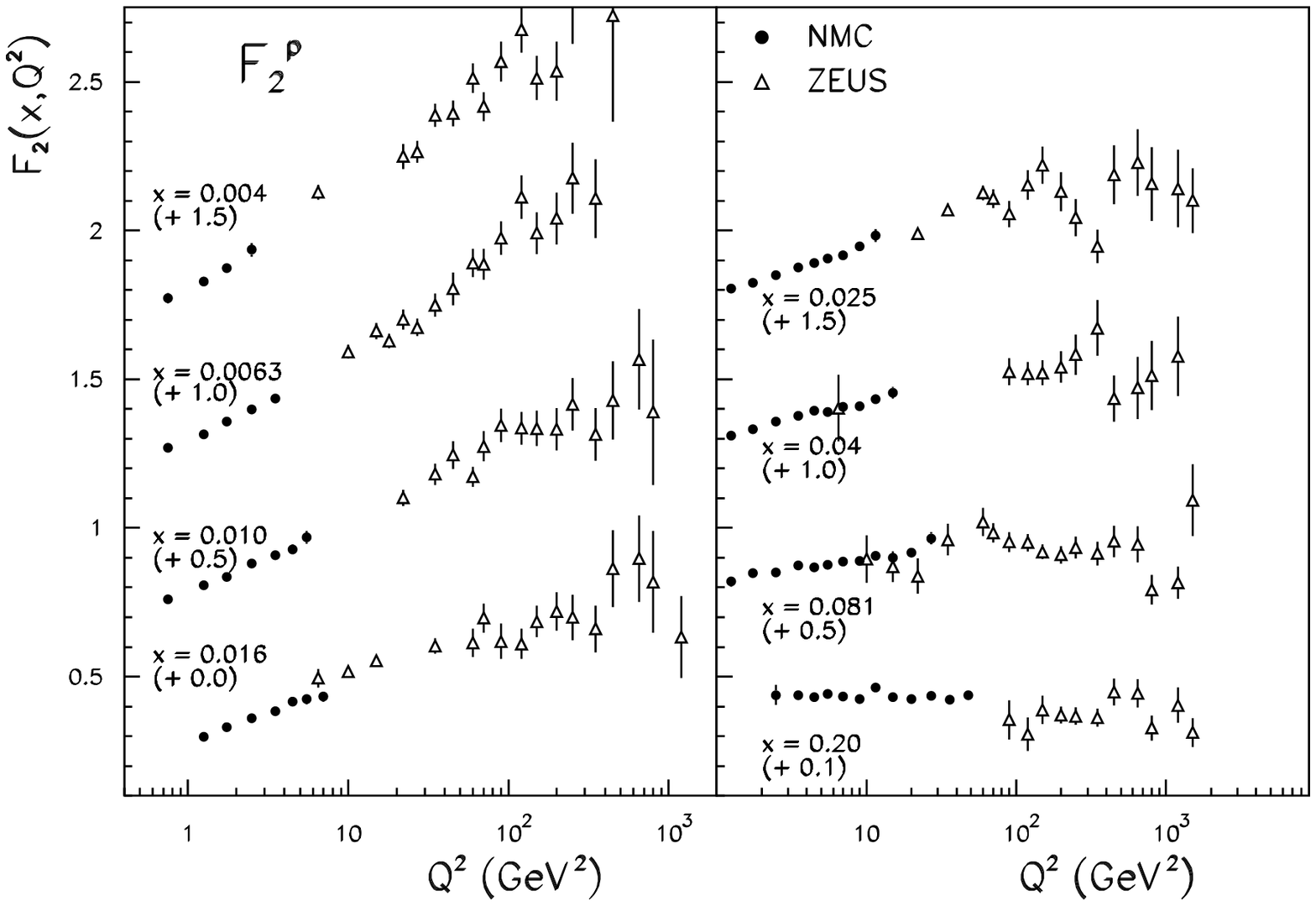,height=9.0cm,width=14cm}
\end{center}
\caption{
The structure function $F_2^p$ from the present measurement
(filled circles) compared to those of the DESY experiments
H1~\protect\cite{ref:h1} (top, open crosses) and
ZEUS~\protect\cite{ref:zeus} (bottom, open triangles),
in the range $ 0.004 < x < 0.20 \,$.
The NMC data have been interpolated to the H1 and ZEUS
$x$ bins.
The data in each $x$ bin have been offset by the
amounts indicated in brackets.
The error bars, wherever visible,
represent the total errors, apart from normalisation
uncertainties.
}
\label{fig:NMC+H1}
\end{figure}

\newpage
 
\begin{figure}[h]
\vspace{1cm}
\begin{sideways}
\begin{minipage}[b]{\textheight}
\begin{center}
\begin{tabular}{c c}
\epsfig{figure=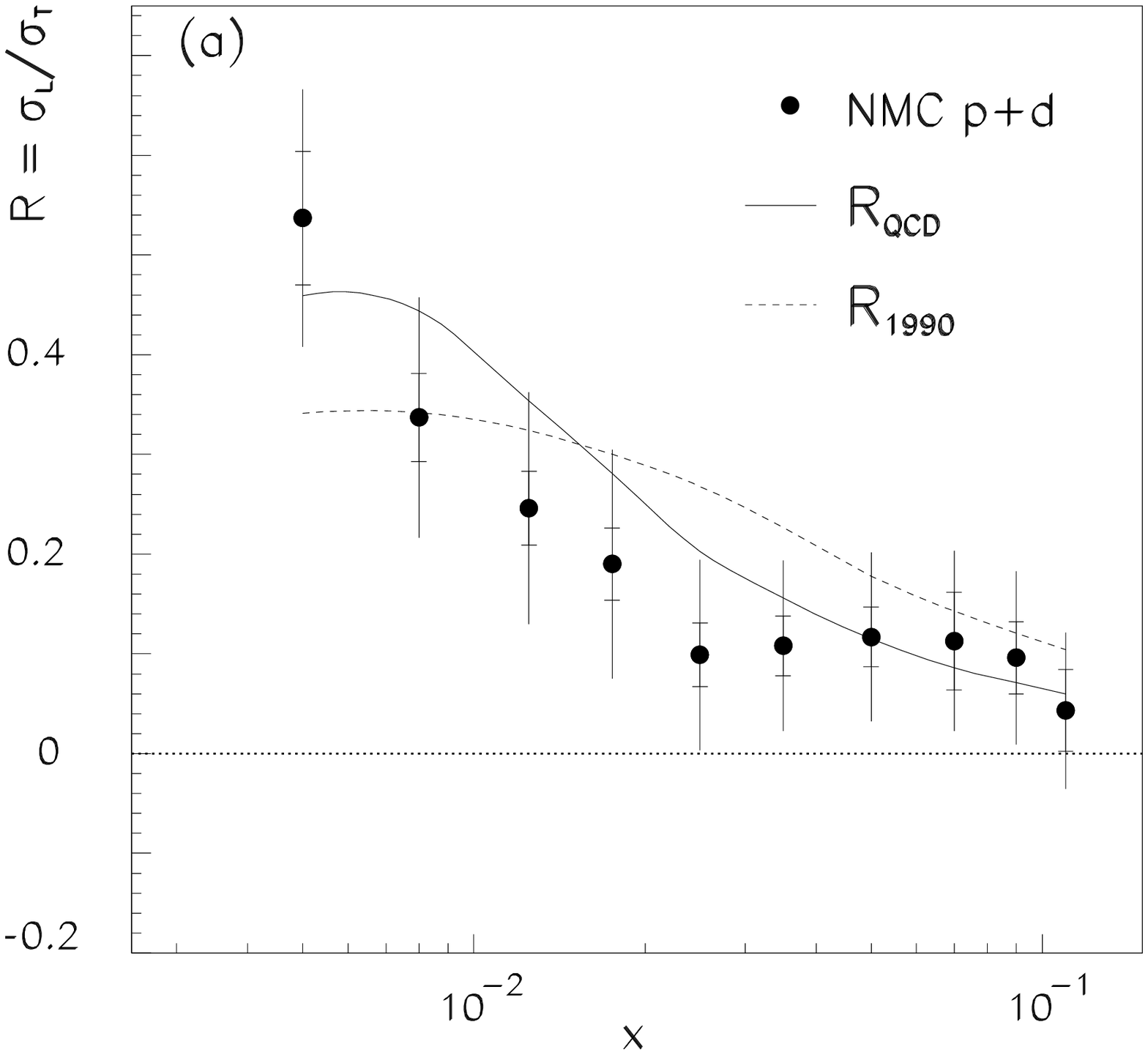,height=12cm,width=11cm}  &
\epsfig{figure=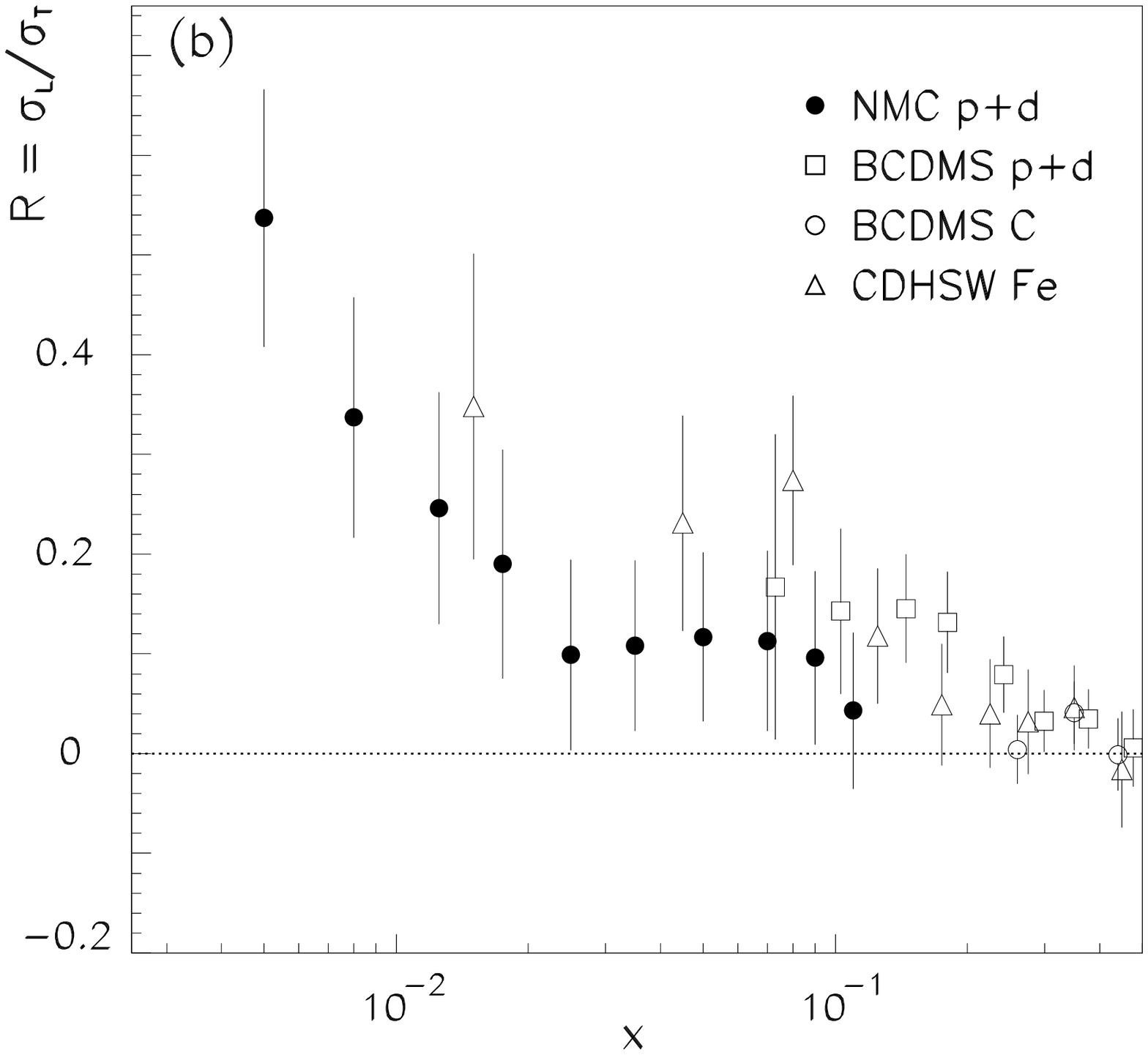,height=12cm,width=11cm}
\end{tabular}
\end{center}
\caption{~~(a) The ratio $R$ measured in this experiment,
averaged for the proton and the deuteron.
The outer error bars represent the quadratic sum
of the statistical and systematic uncertainties;
the inner bars indicate the statistical errors.
The dashed line represents the $R$ parametrisation
from ref.~{\protect\cite{rslac}}
at the average $Q^2$ of each $x$ bin.
The solid line indicates a QCD calculation described in the text.
~~~(b) The ratio $R$ measured in this experiment
(filled circles), compared
to previous results at comparable $x$ and $Q^2$ obtained by
BCDMS~{\protect\cite{BCDMS,BCDMSC}}
(open squares and circles)
and by CDHSW~{\protect\cite{CDHSW}}
(open triangles).
The error bars represent the total errors.}
\label{fig:NMCrx}
\end{minipage}
\end{sideways}
\end{figure}

\begin{figure}[htb]
\vspace{1cm}
\begin{center}
\epsfig{figure=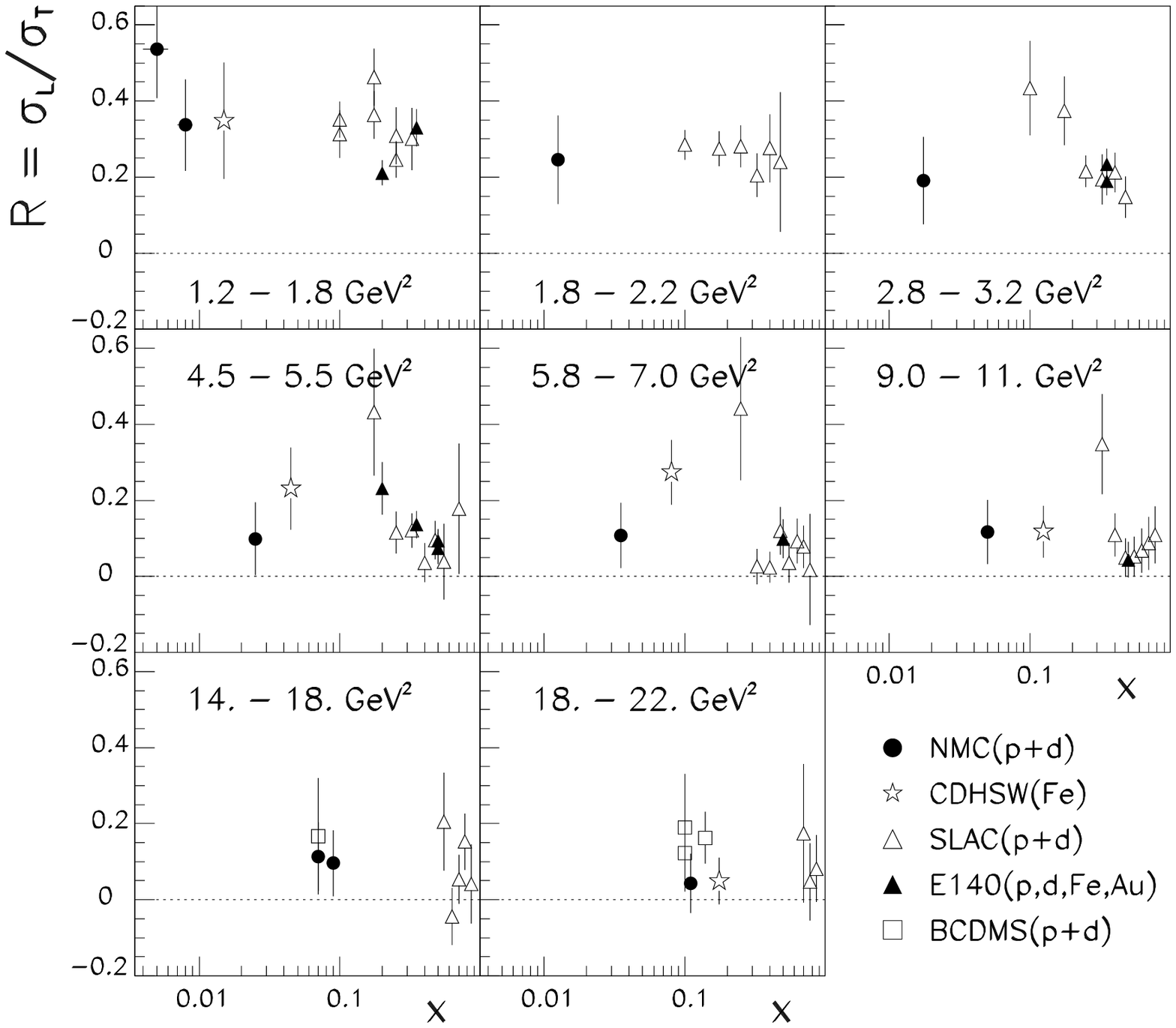,height=20cm,width=15cm}
\end{center}
\caption{The $x$ dependence of the ratio $R$ in bins of $Q^2$
as measured in this (NMC, filled circles) and previous experiments.
These are mainly the SLAC hydrogen and deuterium
scattering experiments~\protect\cite{rslac}
(open triangles) and the SLAC E140
experiments~\protect\cite{E140,E140X} (filled triangles);
also shown are results from
CDHSW~\protect\cite{CDHSW}
(open stars)
and BCDMS~\protect\cite{BCDMS}
(open squares).
The error bars represent the total errors.
}
\label{fig:NMCrq}
\end{figure}
 

\begin{thebibliography}{99}
\bibitem{f2nmc95}
NMC, M.~Arneodo et al., Phys. Lett. {\bf B 364} (1995) 107, \\
and preprint CERN-PPE/95-138 for the tables.
%
\bibitem{rslac}
L.W.~Whitlow et al., Phys. Lett. {\bf B 250} (1990) 193.
%
\bibitem{t10}
R.P.~Mount, Nucl. Instrum. Methods {\bf 187} (1981) 401.
%
\bibitem{f2nmc92}
NMC, P.~Amaudruz et al., Phys. Lett. {\bf B 295} (1992) 159.
%
\bibitem{bcs}
M.~Arneodo, Ph.D. Thesis, Princeton University (1992).
%
\bibitem{annaphd}
A. Dyring, Ph.D. Thesis, Uppsala University (1995).
%
\bibitem{longratio}
NMC, P.~Amaudruz et al., Nucl. Phys. {\bf B 371} (1992) 3.
%
\bibitem{apparatus}
M. van der Heijden, Ph.D. Thesis, University of Amsterdam
(1991);\\ I.G. Bird, Ph.D. Thesis, Free University, Amsterdam (1992);\\
A. Br\"{u}ll, Ph.D. Thesis, Freiburg University (1992)(in German);\\
T. Granier, Ph.D. Thesis, Universit\'{e} Pierre et Marie Curie,
Paris (1994)(in French).
%
\bibitem{peterphd}
P. Bj\"{o}rkholm, Ph.D. Thesis, Uppsala University (1995).
%
\bibitem{BCDMS}
BCDMS Collab., A.C. Benvenuti et al., Phys. Lett. {\bf B
233} (1989) 485;\\
BCDMS Collab., A.C. Benvenuti et al., Phys. Lett. {\bf B
237} (1990) 592.
%
\bibitem{slac92}
L.W. Whitlow et al., Phys.Lett {\bf B 282} (1992) 475.
%
\bibitem{nmcnp96}
NMC, M.~Arneodo et al.,
subm. to Nucl. Phys. B,
%
\bibitem{nmcdr92}
NMC, P.~Amaudruz et al., Phys. Lett. {\bf B 294} (1992) 120. \\
%
\bibitem{radcor}
A.A.~Akhundov et al., Sov. J. Nucl. Phys. {\bf 26} (1977) 660;
44 (1986) 988;\\
JINR-Dubna preprints E2-10147 (1976), E2-10205 (1976),
E2-86-104 (1986);\\
D.~Bardin and N.~Shumeiko, Sov. J .Nucl. Phys. {\bf 29} (1979) 499.
%
\bibitem{ref:E665}
E665 Collab., preprint Fermilab-Pub-95/396-E,
subm. to Phys. Rev. D; \\
A.V. Kotwal, Ph.D. Thesis, Harvard University (1995).
%
\bibitem{ref:h1}
H1 Collab., S. Aid et al., Nucl. Phys. {\bf B 470} (1996) 3.
%
\bibitem{ref:zeus}
ZEUS Collab., M. Derrick et al., preprint DESY 96-076, subm.
to Z. Phys. C.
%
\bibitem{altamar}
G. Altarelli and G. Martinelli, Phys. Lett. {\bf B 76} (1978) 89.
%
\bibitem{virmil}
M.~Virchaux and A.~Milsztajn, Phys. Lett. {\bf B 274} (1992) 221.
%
\bibitem{BCDMSC}
BCDMS Collab., A.C. Benvenuti et al., Phys. Lett. {\bf B
195} (1987) 91.
%
\bibitem{CDHSW}
CDHSW Collab., P. Berge et al., Z. Phys. {\bf C 49} (1991) 187.
%
\bibitem{E140}
E140 Collab., S. Dasu et al., Phys. Rev. Lett. {\bf 61} (1988) 1061;\\
E140 Collab., S. Dasu et al., Phys. Rev. {\bf D 49} (1994) 5641.
%
\bibitem{E140X}
E140X Collab., L.H. Tao et al., Z. Phys. {\bf C 70} (1996) 387.
%
\bibitem{snc}
NMC, M.~Arneodo et al., subm. to Nucl.Phys. B.
 
\end{thebibliography}
\end{document}


\setcounter{table}{2}
\begin{table}[t]
\centering

\tiny
\begin{tabular}{|c|c|c||c||c|r|c|c|c||c|r|}\hline
$x$ & $Q^2$ & $y$ &  $d^2\sigma_p^{meas}/dxdQ^2$
& $E$ & $E^{\prime}$ & $AC$ &  $RC$ & $RE$  & $R$ & $F_2^p\;\pm\Delta{F^{stat}_2}\;\pm\Delta{F^{syst}_2}$ \\
  & (${\rm GeV^2}$) & & $({\rm b\cdot GeV^{-2}})$ &  [\%]  &  [\%]  &  [\%]  &  [\%] &  [\%]  & & \\
 \hline
\hline
0.0078 &   0.80 & 0.635 & 8.600e-06 &   0.5 & -0.1 &  2.3 &  1.7 &  0.3 & 0.337 & 0.2720 $\pm$ 0.0034 $\pm$ 0.0080\\
0.0092 &   1.09 & 0.709 & 4.125e-06 &   0.4 &  0.1 &  2.1 &  2.1 &  0.5 & 0.337 & 0.2943 $\pm$ 0.0082 $\pm$ 0.0091\\
\hline
0.0120 &   0.90 & 0.462 & 5.290e-06 &   0.7 & -0.3 &  1.6 &  0.8 &  0.2 & 0.246 & 0.2763 $\pm$ 0.0064 $\pm$ 0.0054\\
0.0125 &   1.22 & 0.596 & 2.707e-06 &   0.6 & -0.1 &  1.6 &  1.4 &  0.3 & 0.246 & 0.3022 $\pm$ 0.0037 $\pm$ 0.0067\\
0.0139 &   1.62 & 0.703 & 1.387e-06 &   0.4 &  0.0 &  1.5 &  1.8 &  0.5 & 0.246 & 0.3204 $\pm$ 0.0101 $\pm$ 0.0078\\
\hline
0.0173 &   1.27 & 0.448 & 2.013e-06 &   0.7 & -0.2 &  1.1 &  0.7 &  0.2 & 0.190 & 0.2988 $\pm$ 0.0048 $\pm$ 0.0045\\
0.0176 &   1.72 & 0.593 & 1.078e-06 &   0.6 & -0.1 &  1.2 &  1.2 &  0.3 & 0.190 & 0.3316 $\pm$ 0.0054 $\pm$ 0.0059\\
0.0185 &   2.16 & 0.703 & 6.251e-07 &   0.5 &  0.0 &  1.1 &  1.5 &  0.4 & 0.190 & 0.3354 $\pm$ 0.0118 $\pm$ 0.0065\\
\hline
0.0246 &   1.28 & 0.320 & 1.601e-06 &   1.0 & -0.4 &  0.4 &  0.6 &  0.1 & 0.099 & 0.3053 $\pm$ 0.0056 $\pm$ 0.0039\\
0.0247 &   1.75 & 0.437 & 8.198e-07 &   0.7 & -0.2 &  0.7 &  0.6 &  0.2 & 0.099 & 0.3268 $\pm$ 0.0044 $\pm$ 0.0038\\
0.0254 &   2.35 & 0.565 & 4.349e-07 &   0.6 & -0.1 &  0.8 &  0.8 &  0.3 & 0.099 & 0.3504 $\pm$ 0.0046 $\pm$ 0.0046\\
0.0276 &   3.24 & 0.707 & 1.950e-07 &   0.5 &  0.1 &  1.0 &  1.7 &  0.5 & 0.099 & 0.3415 $\pm$ 0.0195 $\pm$ 0.0070\\
\hline
0.0348 &   1.30 & 0.231 & 1.182e-06 &   1.2 & -0.7 &  0.3 &  0.2 &  0.1 & 0.108 & 0.3082 $\pm$ 0.0085 $\pm$ 0.0045\\
0.0348 &   1.76 & 0.309 & 6.482e-07 &   1.0 & -0.4 &  0.2 &  0.2 &  0.1 & 0.108 & 0.3306 $\pm$ 0.0059 $\pm$ 0.0037\\
0.0349 &   2.44 & 0.427 & 3.239e-07 &   0.8 & -0.2 &  0.4 &  0.4 &  0.2 & 0.108 & 0.3541 $\pm$ 0.0046 $\pm$ 0.0035\\
0.0355 &   3.34 & 0.570 & 1.622e-07 &   0.6 &  0.0 &  0.6 &  1.1 &  0.3 & 0.108 & 0.3732 $\pm$ 0.0092 $\pm$ 0.0053\\
0.0364 &   4.26 & 0.705 & 9.547e-08 &   0.5 &  0.1 &  0.8 &  2.4 &  0.4 & 0.108 & 0.3878 $\pm$ 0.0317 $\pm$ 0.0101\\
\hline
0.0495 &   1.33 & 0.168 & 8.974e-07 &   1.6 & -1.0 &  0.9 &  0.1 &  0.1 & 0.117 & 0.3331 $\pm$ 0.0098 $\pm$ 0.0071\\
0.0493 &   1.76 & 0.221 & 4.879e-07 &   1.3 & -0.7 &  0.4 &  0.1 &  0.1 & 0.117 & 0.3314 $\pm$ 0.0058 $\pm$ 0.0051\\
0.0494 &   2.48 & 0.309 & 2.378e-07 &   1.0 & -0.4 &  0.2 &  0.2 &  0.1 & 0.117 & 0.3456 $\pm$ 0.0040 $\pm$ 0.0038\\
0.0501 &   3.41 & 0.415 & 1.220e-07 &   0.8 & -0.2 &  0.2 &  0.6 &  0.2 & 0.117 & 0.3721 $\pm$ 0.0057 $\pm$ 0.0038\\
0.0509 &   4.39 & 0.526 & 6.941e-08 &   0.6 & -0.1 &  0.3 &  1.1 &  0.3 & 0.117 & 0.3893 $\pm$ 0.0122 $\pm$ 0.0053\\
0.0526 &   5.31 & 0.608 & 4.373e-08 &   0.5 &  0.0 &  0.5 &  1.7 &  0.3 & 0.117 & 0.3895 $\pm$ 0.0286 $\pm$ 0.0072\\
\hline
0.0701 &   1.35 & 0.120 & 6.060e-07 &   2.1 & -1.5 &  1.1 &  0.1 &  0.1 & 0.113 & 0.3156 $\pm$ 0.0139 $\pm$ 0.0089\\
0.0699 &   1.77 & 0.156 & 3.741e-07 &   1.7 & -1.1 &  0.6 &  0.1 &  0.1 & 0.113 & 0.3446 $\pm$ 0.0084 $\pm$ 0.0070\\
0.0694 &   2.48 & 0.219 & 1.839e-07 &   1.3 & -0.7 &  0.3 &  0.1 &  0.1 & 0.113 & 0.3503 $\pm$ 0.0053 $\pm$ 0.0052\\
0.0696 &   3.45 & 0.303 & 9.630e-08 &   1.0 & -0.4 &  0.3 &  0.3 &  0.1 & 0.113 & 0.3841 $\pm$ 0.0065 $\pm$ 0.0044\\
0.0703 &   4.42 & 0.382 & 5.433e-08 &   0.8 & -0.2 &  0.3 &  0.6 &  0.1 & 0.113 & 0.3832 $\pm$ 0.0098 $\pm$ 0.0042\\
0.0703 &   5.38 & 0.463 & 3.306e-08 &   0.7 & -0.1 &  0.3 &  0.9 &  0.2 & 0.113 & 0.3705 $\pm$ 0.0174 $\pm$ 0.0043\\
0.0712 &   6.38 & 0.536 & 2.254e-08 &   0.6 &  0.0 &  0.4 &  1.0 &  0.3 & 0.113 & 0.3785 $\pm$ 0.0379 $\pm$ 0.0050\\
\hline
0.0903 &   1.38 & 0.095 & 4.879e-07 &   2.6 & -1.9 &  0.8 &  0.0 &  0.1 & 0.096 & 0.3384 $\pm$ 0.0211 $\pm$ 0.0110\\
0.0900 &   1.76 & 0.121 & 2.938e-07 &   2.1 & -1.5 &  0.4 &  0.0 &  0.1 & 0.096 & 0.3395 $\pm$ 0.0107 $\pm$ 0.0088\\
0.0898 &   2.47 & 0.169 & 1.524e-07 &   1.5 & -0.9 &  0.2 &  0.1 &  0.1 & 0.096 & 0.3599 $\pm$ 0.0068 $\pm$ 0.0066\\
0.0898 &   3.46 & 0.236 & 7.497e-08 &   1.2 & -0.6 &  0.3 &  0.2 &  0.1 & 0.096 & 0.3686 $\pm$ 0.0072 $\pm$ 0.0051\\
0.0900 &   4.43 & 0.299 & 4.376e-08 &   1.0 & -0.4 &  0.4 &  0.3 &  0.1 & 0.096 & 0.3739 $\pm$ 0.0100 $\pm$ 0.0043\\
0.0903 &   5.44 & 0.366 & 2.626e-08 &   0.8 & -0.2 &  0.4 &  0.5 &  0.1 & 0.096 & 0.3584 $\pm$ 0.0155 $\pm$ 0.0040\\
0.0897 &   6.50 & 0.436 & 1.962e-08 &   0.7 & -0.1 &  0.5 &  0.7 &  0.2 & 0.096 & 0.4020 $\pm$ 0.0269 $\pm$ 0.0047\\
\hline
0.1091 &   1.78 & 0.101 & 2.456e-07 &   2.4 & -1.8 &  0.6 &  0.0 &  0.1 & 0.043 & 0.3453 $\pm$ 0.0133 $\pm$ 0.0105\\
0.1097 &   2.49 & 0.140 & 1.261e-07 &   1.8 & -1.2 &  0.2 &  0.0 &  0.1 & 0.043 & 0.3615 $\pm$ 0.0082 $\pm$ 0.0079\\
0.1099 &   3.48 & 0.193 & 6.254e-08 &   1.4 & -0.8 &  0.3 &  0.1 &  0.1 & 0.043 & 0.3674 $\pm$ 0.0083 $\pm$ 0.0058\\
0.1100 &   4.45 & 0.246 & 3.642e-08 &   1.1 & -0.5 &  0.4 &  0.2 &  0.1 & 0.043 & 0.3676 $\pm$ 0.0104 $\pm$ 0.0049\\
0.1105 &   5.44 & 0.298 & 2.263e-08 &   1.0 & -0.4 &  0.4 &  0.2 &  0.1 & 0.043 & 0.3575 $\pm$ 0.0148 $\pm$ 0.0039\\
0.1096 &   6.60 & 0.363 & 1.536e-08 &   0.8 & -0.2 &  0.6 &  0.3 &  0.1 & 0.043 & 0.3740 $\pm$ 0.0217 $\pm$ 0.0042\\
\hline
0.1284 &   1.86 & 0.090 & 2.038e-07 &   2.9 & -2.2 &  1.1 &  0.0 &  0.1 & 0.296 & 0.3673 $\pm$ 0.0219 $\pm$ 0.0137\\
0.1385 &   2.49 & 0.111 & 1.020e-07 &   2.2 & -1.5 &  0.7 &  0.0 &  0.1 & 0.271 & 0.3631 $\pm$ 0.0071 $\pm$ 0.0101\\
0.1398 &   3.48 & 0.154 & 5.055e-08 &   1.6 & -1.0 &  0.3 &  0.1 &  0.1 & 0.233 & 0.3705 $\pm$ 0.0071 $\pm$ 0.0068\\
0.1396 &   4.45 & 0.195 & 2.952e-08 &   1.3 & -0.7 &  0.2 &  0.1 &  0.1 & 0.201 & 0.3653 $\pm$ 0.0083 $\pm$ 0.0054\\
0.1399 &   5.46 & 0.238 & 1.856e-08 &   1.1 & -0.5 &  0.3 &  0.1 &  0.1 & 0.176 & 0.3601 $\pm$ 0.0115 $\pm$ 0.0048\\
0.1409 &   6.69 & 0.289 & 1.228e-08 &   1.0 & -0.3 &  0.5 &  0.2 &  0.1 & 0.154 & 0.3768 $\pm$ 0.0144 $\pm$ 0.0044\\
\hline
0.1741 &   2.67 & 0.095 & 7.069e-08 &   2.4 & -1.7 &  1.2 &  0.0 &  0.1 & 0.246 & 0.3630 $\pm$ 0.0109 $\pm$ 0.0114\\
0.1807 &   3.47 & 0.119 & 3.690e-08 &   1.9 & -1.2 &  0.7 &  0.0 &  0.1 & 0.210 & 0.3386 $\pm$ 0.0081 $\pm$ 0.0079\\
0.1802 &   4.46 & 0.152 & 2.393e-08 &   1.5 & -0.9 &  0.3 &  0.0 &  0.1 & 0.179 & 0.3727 $\pm$ 0.0102 $\pm$ 0.0066\\
0.1808 &   5.46 & 0.185 & 1.385e-08 &   1.3 & -0.6 &  0.1 &  0.1 &  0.1 & 0.155 & 0.3337 $\pm$ 0.0119 $\pm$ 0.0049\\
0.1804 &   6.77 & 0.229 & 9.398e-09 &   1.1 & -0.4 &  0.4 &  0.1 &  0.1 & 0.135 & 0.3616 $\pm$ 0.0141 $\pm$ 0.0046\\
0.1771 &   8.50 & 0.289 & 6.257e-09 &   0.9 & -0.3 &  0.8 &  0.2 &  0.1 & 0.118 & 0.3921 $\pm$ 0.0380 $\pm$ 0.0050\\
\hline
0.2051 &   2.90 & 0.087 & 4.686e-08 &   2.3 & -1.7 &  1.0 &  0.0 &  0.1 & 0.223 & 0.3346 $\pm$ 0.0316 $\pm$ 0.0102\\
0.2210 &   3.51 & 0.099 & 2.905e-08 &   2.0 & -1.3 &  0.7 &  0.0 &  0.1 & 0.192 & 0.3310 $\pm$ 0.0088 $\pm$ 0.0082\\
0.2262 &   4.47 & 0.122 & 1.675e-08 &   1.6 & -0.9 &  0.3 &  0.0 &  0.1 & 0.161 & 0.3240 $\pm$ 0.0092 $\pm$ 0.0062\\
0.2263 &   5.47 & 0.149 & 1.182e-08 &   1.4 & -0.7 &  0.2 &  0.0 &  0.1 & 0.140 & 0.3498 $\pm$ 0.0126 $\pm$ 0.0055\\
0.2280 &   6.81 & 0.184 & 6.709e-09 &   1.1 & -0.5 &  0.5 &  0.1 &  0.1 & 0.120 & 0.3205 $\pm$ 0.0125 $\pm$ 0.0042\\
0.2215 &   8.64 & 0.236 & 3.930e-09 &   1.0 & -0.3 &  0.8 &  0.1 &  0.1 & 0.104 & 0.3062 $\pm$ 0.0265 $\pm$ 0.0039\\
\hline
0.2657 &   3.75 & 0.088 & 1.933e-08 &   1.7 & -1.0 &  1.6 &  0.0 &  0.1 & 0.171 & 0.3029 $\pm$ 0.0147 $\pm$ 0.0076\\
0.2736 &   4.47 & 0.102 & 1.312e-08 &   1.5 & -0.8 &  1.1 &  0.0 &  0.1 & 0.150 & 0.3045 $\pm$ 0.0102 $\pm$ 0.0060\\
0.2758 &   5.44 & 0.122 & 8.664e-09 &   1.3 & -0.5 &  0.8 &  0.0 &  0.1 & 0.129 & 0.3058 $\pm$ 0.0128 $\pm$ 0.0048\\
0.2803 &   6.85 & 0.151 & 5.503e-09 &   1.1 & -0.3 &  0.9 &  0.0 &  0.1 & 0.109 & 0.3200 $\pm$ 0.0139 $\pm$ 0.0048\\
0.2737 &   8.70 & 0.193 & 3.409e-09 &   0.9 & -0.2 &  1.1 &  0.1 &  0.1 & 0.093 & 0.3237 $\pm$ 0.0270 $\pm$ 0.0048\\
\hline
0.3166 &   4.62 & 0.091 & 1.009e-08 &   0.7 &  0.0 &  4.6 &  0.0 &  0.1 & 0.139 & 0.2884 $\pm$ 0.0123 $\pm$ 0.0134\\
0.3397 &   5.46 & 0.100 & 5.961e-09 &   0.6 &  0.1 &  3.6 &  0.0 &  0.1 & 0.120 & 0.2584 $\pm$ 0.0095 $\pm$ 0.0094\\
0.3511 &   6.87 & 0.122 & 3.805e-09 &   0.6 &  0.2 &  2.3 &  0.0 &  0.1 & 0.100 & 0.2759 $\pm$ 0.0103 $\pm$ 0.0065\\
0.3450 &   8.73 & 0.156 & 2.028e-09 &   0.5 &  0.3 &  2.1 &  0.0 &  0.1 & 0.085 & 0.2398 $\pm$ 0.0163 $\pm$ 0.0052\\
\hline
0.3973 &   5.73 & 0.090 & 3.735e-09 &  -2.4 &  3.2 & 11.0 &  0.0 &  0.1 & 0.111 & 0.2086 $\pm$ 0.0211 $\pm$ 0.0243\\
0.4425 &   6.97 & 0.099 & 2.164e-09 &  -2.0 &  2.9 &  9.1 &  0.0 &  0.1 & 0.093 & 0.2024 $\pm$ 0.0094 $\pm$ 0.0198\\
0.4653 &   8.75 & 0.117 & 1.151e-09 &  -1.7 &  2.7 &  6.8 &  0.0 &  0.1 & 0.078 & 0.1817 $\pm$ 0.0118 $\pm$ 0.0136\\
\hline
\end{tabular}

\caption{(a)
The differential scattering cross section, 
$d^2\sigma_p^{meas}/dxdQ^2$,
the value of $R$ as used in the
analysis (see also table 7), and the
structure function $F_2^p$ at 90 GeV 
incident muon energy, with the normalisation corrections of table~2.
Horizontal lines separate the $x$-bins.
The indicated five sources of systematic errors on $F_2$ are
the uncertainties on: $E$
the incident muon energy;
$E'$ the scattered muon energy; $AC$
the acceptance; $RC$ the
radiative correction; $RE$ the reconstruction efficiency.
The signs on $E$ and $E'$ correspond to an decrease in the
respective muon energy.
The various errors are not independent: $E$ and $E'$ 
are fully correlated between
the proton and the deuteron data, but
independent for the data taken at different energies;
$RC$ is fully correlated between all energies, but 
independent for the proton and the deuteron; whilst $AC$
and $RE$ should each be treated as fully correlated for all data
sets.
}
\end{table}

\setcounter{table}{2}
\begin{table}[t]
\centering

\tiny
\begin{tabular}{|c|c|c||c||c|r|c|c|c||c|r|}\hline
$x$ & $Q^2$ & $y$ &  $d^2\sigma_p^{meas}/dxdQ^2$
& $E$ & $E^{\prime}$ & $AC$ &  $RC$ & $RE$  & $R$ & $F_2^p\;\pm\Delta{F^{stat}_2}\;\pm\Delta{F^{syst}_2}$ \\
  & (${\rm GeV^2}$) & & $({\rm b\cdot GeV^{-2}})$ &  [\%]  &  [\%]  &  [\%]  &  [\%] &  [\%]  & & \\
 \hline
\hline
0.0087 &   1.18 & 0.626 & 3.938e-06 &   0.5 &  0.0 &  2.1 &  1.6 &  0.6 & 0.337 & 0.2987 $\pm$ 0.0090 $\pm$ 0.0082\\
\hline
0.0117 &   1.37 & 0.540 & 2.400e-06 &   0.7 & -0.2 &  1.6 &  1.1 &  0.4 & 0.246 & 0.3030 $\pm$ 0.0097 $\pm$ 0.0062\\
0.0129 &   1.70 & 0.604 & 1.410e-06 &   0.5 &  0.0 &  1.5 &  1.2 &  0.5 & 0.246 & 0.3200 $\pm$ 0.0071 $\pm$ 0.0066\\
\hline
0.0172 &   1.77 & 0.472 & 1.109e-06 &   0.7 & -0.2 &  1.2 &  0.7 &  0.4 & 0.190 & 0.3274 $\pm$ 0.0100 $\pm$ 0.0052\\
0.0178 &   2.34 & 0.599 & 6.064e-07 &   0.6 &  0.0 &  1.1 &  1.0 &  0.5 & 0.190 & 0.3519 $\pm$ 0.0072 $\pm$ 0.0057\\
\hline
0.0241 &   1.82 & 0.351 & 8.159e-07 &   0.9 & -0.4 &  0.7 &  0.3 &  0.3 & 0.099 & 0.3219 $\pm$ 0.0122 $\pm$ 0.0042\\
0.0245 &   2.50 & 0.471 & 4.142e-07 &   0.7 & -0.2 &  0.8 &  0.5 &  0.4 & 0.099 & 0.3440 $\pm$ 0.0054 $\pm$ 0.0043\\
0.0260 &   3.36 & 0.590 & 2.170e-07 &   0.6 &  0.0 &  1.0 &  1.2 &  0.5 & 0.099 & 0.3732 $\pm$ 0.0077 $\pm$ 0.0064\\
\hline
0.0348 &   2.51 & 0.332 & 3.497e-07 &   0.9 & -0.4 &  0.4 &  0.2 &  0.3 & 0.108 & 0.3733 $\pm$ 0.0094 $\pm$ 0.0043\\
0.0348 &   3.48 & 0.458 & 1.686e-07 &   0.7 & -0.2 &  0.6 &  0.7 &  0.4 & 0.108 & 0.3825 $\pm$ 0.0076 $\pm$ 0.0047\\
0.0354 &   4.41 & 0.568 & 9.558e-08 &   0.6 &  0.0 &  0.8 &  1.5 &  0.5 & 0.108 & 0.3825 $\pm$ 0.0103 $\pm$ 0.0070\\
0.0374 &   5.32 & 0.645 & 6.281e-08 &   0.5 &  0.0 &  1.0 &  2.0 &  0.6 & 0.108 & 0.4031 $\pm$ 0.0278 $\pm$ 0.0094\\
\hline
0.0497 &   2.54 & 0.237 & 2.342e-07 &   1.2 & -0.6 &  0.2 &  0.1 &  0.2 & 0.117 & 0.3394 $\pm$ 0.0087 $\pm$ 0.0048\\
0.0494 &   3.49 & 0.327 & 1.288e-07 &   0.9 & -0.4 &  0.2 &  0.3 &  0.3 & 0.117 & 0.3785 $\pm$ 0.0070 $\pm$ 0.0043\\
0.0495 &   4.47 & 0.416 & 7.061e-08 &   0.8 & -0.2 &  0.3 &  0.7 &  0.3 & 0.117 & 0.3681 $\pm$ 0.0073 $\pm$ 0.0043\\
0.0504 &   5.43 & 0.493 & 4.448e-08 &   0.7 & -0.1 &  0.5 &  1.0 &  0.4 & 0.117 & 0.3705 $\pm$ 0.0099 $\pm$ 0.0050\\
0.0524 &   6.61 & 0.576 & 3.014e-08 &   0.6 &  0.0 &  0.7 &  1.2 &  0.5 & 0.117 & 0.4086 $\pm$ 0.0162 $\pm$ 0.0065\\
\hline
0.0707 &   2.56 & 0.168 & 1.725e-07 &   1.5 & -0.9 &  0.3 &  0.1 &  0.2 & 0.113 & 0.3426 $\pm$ 0.0129 $\pm$ 0.0061\\
0.0702 &   3.50 & 0.229 & 9.223e-08 &   1.2 & -0.6 &  0.3 &  0.2 &  0.2 & 0.113 & 0.3584 $\pm$ 0.0089 $\pm$ 0.0051\\
0.0698 &   4.48 & 0.295 & 5.437e-08 &   1.0 & -0.4 &  0.3 &  0.3 &  0.2 & 0.113 & 0.3659 $\pm$ 0.0089 $\pm$ 0.0043\\
0.0700 &   5.47 & 0.357 & 3.656e-08 &   0.9 & -0.3 &  0.3 &  0.4 &  0.3 & 0.113 & 0.3867 $\pm$ 0.0106 $\pm$ 0.0040\\
0.0700 &   6.86 & 0.448 & 2.175e-08 &   0.7 & -0.1 &  0.4 &  0.7 &  0.4 & 0.113 & 0.3902 $\pm$ 0.0130 $\pm$ 0.0047\\
0.0718 &   8.62 & 0.544 & 1.460e-08 &   0.6 &  0.0 &  0.6 &  1.1 &  0.5 & 0.113 & 0.4555 $\pm$ 0.0308 $\pm$ 0.0068\\
\hline
0.0915 &   2.61 & 0.132 & 1.515e-07 &   1.8 & -1.2 &  0.2 &  0.0 &  0.2 & 0.096 & 0.3948 $\pm$ 0.0202 $\pm$ 0.0089\\
0.0909 &   3.50 & 0.177 & 7.095e-08 &   1.4 & -0.8 &  0.3 &  0.1 &  0.2 & 0.096 & 0.3430 $\pm$ 0.0109 $\pm$ 0.0058\\
0.0906 &   4.47 & 0.226 & 4.441e-08 &   1.2 & -0.6 &  0.4 &  0.2 &  0.2 & 0.096 & 0.3645 $\pm$ 0.0108 $\pm$ 0.0051\\
0.0904 &   5.48 & 0.278 & 2.948e-08 &   1.0 & -0.4 &  0.4 &  0.2 &  0.2 & 0.096 & 0.3807 $\pm$ 0.0120 $\pm$ 0.0047\\
0.0904 &   6.81 & 0.343 & 1.812e-08 &   0.9 & -0.3 &  0.5 &  0.4 &  0.3 & 0.096 & 0.3814 $\pm$ 0.0112 $\pm$ 0.0043\\
0.0905 &   8.77 & 0.440 & 1.099e-08 &   0.7 & -0.1 &  0.7 &  0.7 &  0.4 & 0.096 & 0.4158 $\pm$ 0.0221 $\pm$ 0.0054\\
\hline
0.1126 &   2.62 & 0.109 & 9.990e-08 &   2.2 & -1.5 &  0.2 &  0.0 &  0.2 & 0.043 & 0.3174 $\pm$ 0.0196 $\pm$ 0.0082\\
0.1120 &   3.48 & 0.143 & 6.469e-08 &   1.7 & -1.1 &  0.3 &  0.1 &  0.2 & 0.043 & 0.3721 $\pm$ 0.0165 $\pm$ 0.0078\\
0.1111 &   4.48 & 0.185 & 3.714e-08 &   1.4 & -0.8 &  0.4 &  0.1 &  0.2 & 0.043 & 0.3654 $\pm$ 0.0132 $\pm$ 0.0059\\
0.1109 &   5.49 & 0.227 & 2.441e-08 &   1.2 & -0.6 &  0.4 &  0.1 &  0.2 & 0.043 & 0.3718 $\pm$ 0.0134 $\pm$ 0.0050\\
0.1109 &   6.83 & 0.281 & 1.473e-08 &   1.0 & -0.4 &  0.6 &  0.2 &  0.2 & 0.043 & 0.3646 $\pm$ 0.0117 $\pm$ 0.0045\\
0.1106 &   8.87 & 0.365 & 9.308e-09 &   0.8 & -0.2 &  0.8 &  0.4 &  0.3 & 0.043 & 0.4145 $\pm$ 0.0209 $\pm$ 0.0053\\
0.1112 &  10.76 & 0.437 & 4.964e-09 &   0.7 & -0.1 &  1.1 &  0.6 &  0.4 & 0.043 & 0.3462 $\pm$ 0.0357 $\pm$ 0.0050\\
\hline
0.1435 &   3.48 & 0.113 & 4.974e-08 &   2.0 & -1.4 &  0.3 &  0.0 &  0.2 & 0.231 & 0.3610 $\pm$ 0.0134 $\pm$ 0.0088\\
0.1421 &   4.49 & 0.147 & 2.726e-08 &   1.6 & -1.0 &  0.2 &  0.0 &  0.2 & 0.198 & 0.3364 $\pm$ 0.0102 $\pm$ 0.0062\\
0.1411 &   5.48 & 0.179 & 1.738e-08 &   1.4 & -0.8 &  0.3 &  0.1 &  0.2 & 0.175 & 0.3258 $\pm$ 0.0101 $\pm$ 0.0051\\
0.1413 &   6.87 & 0.224 & 1.128e-08 &   1.2 & -0.5 &  0.5 &  0.1 &  0.2 & 0.151 & 0.3459 $\pm$ 0.0087 $\pm$ 0.0047\\
0.1407 &   8.90 & 0.291 & 6.779e-09 &   1.0 & -0.3 &  0.8 &  0.2 &  0.2 & 0.130 & 0.3682 $\pm$ 0.0147 $\pm$ 0.0050\\
0.1423 &  11.01 & 0.352 & 4.435e-09 &   0.8 & -0.2 &  1.1 &  0.3 &  0.3 & 0.115 & 0.3925 $\pm$ 0.0215 $\pm$ 0.0057\\
\hline
0.1821 &   3.62 & 0.092 & 3.152e-08 &   2.3 & -1.6 &  0.7 &  0.0 &  0.2 & 0.204 & 0.3116 $\pm$ 0.0180 $\pm$ 0.0091\\
0.1866 &   4.50 & 0.113 & 2.134e-08 &   1.9 & -1.2 &  0.3 &  0.0 &  0.2 & 0.175 & 0.3403 $\pm$ 0.0126 $\pm$ 0.0076\\
0.1845 &   5.49 & 0.137 & 1.336e-08 &   1.6 & -0.9 &  0.1 &  0.0 &  0.2 & 0.153 & 0.3198 $\pm$ 0.0119 $\pm$ 0.0060\\
0.1837 &   6.87 & 0.172 & 8.988e-09 &   1.3 & -0.7 &  0.4 &  0.1 &  0.2 & 0.132 & 0.3461 $\pm$ 0.0101 $\pm$ 0.0053\\
0.1813 &   8.94 & 0.227 & 4.992e-09 &   1.1 & -0.4 &  0.8 &  0.1 &  0.2 & 0.113 & 0.3361 $\pm$ 0.0159 $\pm$ 0.0047\\
0.1816 &  11.15 & 0.280 & 3.358e-09 &   0.9 & -0.3 &  1.1 &  0.2 &  0.2 & 0.099 & 0.3692 $\pm$ 0.0196 $\pm$ 0.0054\\
\hline
0.2292 &   4.56 & 0.093 & 1.562e-08 &   1.9 & -1.3 &  0.3 &  0.0 &  0.2 & 0.158 & 0.3112 $\pm$ 0.0149 $\pm$ 0.0072\\
0.2312 &   5.47 & 0.110 & 1.054e-08 &   1.7 & -1.0 &  0.2 &  0.0 &  0.2 & 0.139 & 0.3104 $\pm$ 0.0126 $\pm$ 0.0061\\
0.2311 &   6.92 & 0.139 & 6.614e-09 &   1.4 & -0.7 &  0.5 &  0.0 &  0.2 & 0.118 & 0.3188 $\pm$ 0.0099 $\pm$ 0.0051\\
0.2276 &   8.84 & 0.178 & 3.855e-09 &   1.1 & -0.5 &  0.8 &  0.1 &  0.2 & 0.101 & 0.3082 $\pm$ 0.0129 $\pm$ 0.0046\\
0.2276 &  11.24 & 0.226 & 2.722e-09 &   1.0 & -0.3 &  1.1 &  0.1 &  0.2 & 0.087 & 0.3664 $\pm$ 0.0184 $\pm$ 0.0054\\
\hline
0.2801 &   5.54 & 0.093 & 8.133e-09 &   1.5 & -0.8 &  0.8 &  0.0 &  0.2 & 0.127 & 0.2960 $\pm$ 0.0161 $\pm$ 0.0054\\
0.2853 &   6.93 & 0.113 & 4.854e-09 &   1.2 & -0.5 &  0.9 &  0.0 &  0.2 & 0.107 & 0.2866 $\pm$ 0.0106 $\pm$ 0.0049\\
0.2808 &   8.86 & 0.146 & 3.030e-09 &   1.0 & -0.3 &  1.1 &  0.0 &  0.2 & 0.091 & 0.2946 $\pm$ 0.0140 $\pm$ 0.0046\\
0.2790 &  11.27 & 0.186 & 1.738e-09 &   0.9 & -0.2 &  1.3 &  0.1 &  0.2 & 0.078 & 0.2814 $\pm$ 0.0164 $\pm$ 0.0046\\
0.2792 &  14.19 & 0.231 & 1.114e-09 &   0.8 &  0.0 &  1.5 &  0.1 &  0.2 & 0.068 & 0.2971 $\pm$ 0.0295 $\pm$ 0.0051\\
\hline
0.3411 &   7.08 & 0.097 & 3.646e-09 &   0.5 &  0.2 &  2.3 &  0.0 &  0.2 & 0.099 & 0.2675 $\pm$ 0.0104 $\pm$ 0.0063\\
0.3527 &   8.86 & 0.117 & 2.095e-09 &   0.5 &  0.3 &  2.1 &  0.0 &  0.2 & 0.083 & 0.2532 $\pm$ 0.0104 $\pm$ 0.0056\\
0.3528 &  11.33 & 0.149 & 1.152e-09 &   0.5 &  0.4 &  2.0 &  0.1 &  0.2 & 0.070 & 0.2340 $\pm$ 0.0117 $\pm$ 0.0048\\
0.3521 &  14.32 & 0.186 & 7.875e-10 &   0.5 &  0.4 &  1.9 &  0.1 &  0.2 & 0.060 & 0.2631 $\pm$ 0.0197 $\pm$ 0.0053\\
\hline
0.4334 &   9.02 & 0.097 & 1.050e-09 &  -2.5 &  3.4 &  6.8 &  0.0 &  0.2 & 0.077 & 0.1610 $\pm$ 0.0103 $\pm$ 0.0128\\
0.4626 &  11.36 & 0.115 & 6.386e-10 &  -2.0 &  3.0 &  4.7 &  0.0 &  0.2 & 0.064 & 0.1692 $\pm$ 0.0095 $\pm$ 0.0099\\
0.4722 &  14.39 & 0.141 & 3.458e-10 &  -1.5 &  2.6 &  2.7 &  0.0 &  0.2 & 0.054 & 0.1535 $\pm$ 0.0115 $\pm$ 0.0063\\
\hline
\end{tabular}

\caption{(b)
The differential scattering cross section,
$d^2\sigma_p^{meas}/dxdQ^2$,
the value of $R$ as used in the
analysis (see also table 7), and the
structure function $F_2^p$ at 120 GeV 
incident muon energy, with the normalisation corrections of table~2.
Horizontal lines separate the x-bins. See also caption (a).
 }
\end{table}

\setcounter{table}{2}
\begin{table}[t]
\centering

\tiny
\begin{tabular}{|c|c|c||c||c|r|c|c|c||c|r|}\hline
$x$ & $Q^2$ & $y$ &  $d^2\sigma_p^{meas}/dxdQ^2$
& $E$ & $E^{\prime}$ & $AC$ &  $RC$ & $RE$  & $R$ & $F_2^p\;\pm\Delta{F^{stat}_2}\;\pm\Delta{F^{syst}_2}$ \\
  & (${\rm GeV^2}$) & & $({\rm b\cdot GeV^{-2}})$ &  [\%]  &  [\%]  &  [\%]  &  [\%] &  [\%]  & & \\
 \hline
0.0035 &   0.83 & 0.654 & 1.785e-05 &   0.6 & -0.2 &  2.2 &  1.6 &  4.1 & 0.537 & 0.2890 $\pm$ 0.0107 $\pm$ 0.0144\\
0.0051 &   1.18 & 0.631 & 6.710e-06 &   0.5 &  0.0 &  1.7 &  1.6 &  3.8 & 0.537 & 0.3027 $\pm$ 0.0085 $\pm$ 0.0136\\
\hline
0.0075 &   1.32 & 0.486 & 4.363e-06 &   0.7 & -0.2 &  1.7 &  0.8 &  3.3 & 0.337 & 0.3158 $\pm$ 0.0083 $\pm$ 0.0123\\
0.0081 &   1.72 & 0.577 & 2.463e-06 &   0.6 & -0.1 &  1.2 &  1.1 &  3.2 & 0.337 & 0.3516 $\pm$ 0.0085 $\pm$ 0.0129\\
\hline
0.0122 &   1.76 & 0.395 & 1.798e-06 &   0.8 & -0.3 &  1.5 &  0.5 &  2.9 & 0.246 & 0.3445 $\pm$ 0.0095 $\pm$ 0.0118\\
0.0124 &   2.45 & 0.538 & 8.866e-07 &   0.6 & -0.1 &  1.1 &  0.8 &  2.6 & 0.246 & 0.3786 $\pm$ 0.0074 $\pm$ 0.0115\\
0.0138 &   3.25 & 0.640 & 4.107e-07 &   0.5 &  0.0 &  1.7 &  1.2 &  1.3 & 0.246 & 0.3684 $\pm$ 0.0183 $\pm$ 0.0093\\
\hline
0.0174 &   1.77 & 0.277 & 1.262e-06 &   1.1 & -0.5 &  1.7 &  0.2 &  2.6 & 0.190 & 0.3166 $\pm$ 0.0134 $\pm$ 0.0107\\
0.0174 &   2.48 & 0.388 & 6.767e-07 &   0.8 & -0.3 &  1.2 &  0.3 &  2.6 & 0.190 & 0.3675 $\pm$ 0.0090 $\pm$ 0.0110\\
0.0169 &   3.62 & 0.585 & 2.959e-07 &   0.6 & -0.1 &  1.5 &  1.2 &  1.1 & 0.190 & 0.3863 $\pm$ 0.0134 $\pm$ 0.0087\\
0.0185 &   4.34 & 0.638 & 1.935e-07 &   0.5 &  0.0 &  1.6 &  1.7 &  1.3 & 0.190 & 0.4076 $\pm$ 0.0163 $\pm$ 0.0112\\
\hline
0.0244 &   1.79 & 0.203 & 9.075e-07 &   1.4 & -0.8 &  1.9 &  0.1 &  2.4 & 0.099 & 0.3062 $\pm$ 0.0182 $\pm$ 0.0106\\
0.0246 &   2.50 & 0.279 & 4.911e-07 &   1.1 & -0.5 &  1.3 &  0.2 &  2.3 & 0.099 & 0.3494 $\pm$ 0.0081 $\pm$ 0.0101\\
0.0248 &   3.45 & 0.380 & 2.596e-07 &   0.8 & -0.3 &  0.9 &  0.4 &  2.3 & 0.099 & 0.3848 $\pm$ 0.0096 $\pm$ 0.0103\\
0.0240 &   4.54 & 0.520 & 1.431e-07 &   0.7 & -0.1 &  1.3 &  1.2 &  1.0 & 0.099 & 0.3946 $\pm$ 0.0113 $\pm$ 0.0084\\
0.0252 &   5.45 & 0.590 & 9.084e-08 &   0.6 &  0.0 &  1.5 &  1.6 &  1.1 & 0.099 & 0.3957 $\pm$ 0.0113 $\pm$ 0.0100\\
0.0277 &   6.51 & 0.636 & 5.973e-08 &   0.5 &  0.0 &  1.7 &  1.7 &  1.3 & 0.099 & 0.4177 $\pm$ 0.0159 $\pm$ 0.0118\\
\hline
0.0347 &   2.53 & 0.199 & 3.611e-07 &   1.4 & -0.8 &  1.3 &  0.1 &  2.1 & 0.108 & 0.3482 $\pm$ 0.0128 $\pm$ 0.0103\\
0.0349 &   3.46 & 0.271 & 1.924e-07 &   1.1 & -0.5 &  0.9 &  0.2 &  2.1 & 0.108 & 0.3703 $\pm$ 0.0115 $\pm$ 0.0095\\
0.0350 &   4.43 & 0.343 & 1.178e-07 &   0.9 & -0.3 &  0.8 &  0.4 &  2.1 & 0.108 & 0.3967 $\pm$ 0.0157 $\pm$ 0.0097\\
0.0346 &   5.50 & 0.434 & 7.039e-08 &   0.7 & -0.2 &  0.9 &  0.8 &  1.0 & 0.108 & 0.3885 $\pm$ 0.0135 $\pm$ 0.0069\\
0.0347 &   6.96 & 0.547 & 4.410e-08 &   0.6 & -0.1 &  1.2 &  1.1 &  1.1 & 0.108 & 0.4245 $\pm$ 0.0102 $\pm$ 0.0089\\
0.0368 &   8.65 & 0.635 & 2.361e-08 &   0.5 &  0.0 &  1.5 &  1.6 &  1.3 & 0.108 & 0.3915 $\pm$ 0.0161 $\pm$ 0.0100\\
\hline
0.0484 &   3.52 & 0.199 & 1.258e-07 &   1.4 & -0.8 &  1.0 &  0.1 &  1.8 & 0.117 & 0.3298 $\pm$ 0.0115 $\pm$ 0.0086\\
0.0493 &   4.46 & 0.248 & 8.407e-08 &   1.1 & -0.6 &  0.6 &  0.2 &  1.8 & 0.117 & 0.3764 $\pm$ 0.0122 $\pm$ 0.0086\\
0.0490 &   5.47 & 0.306 & 5.683e-08 &   1.0 & -0.4 &  0.4 &  0.3 &  1.3 & 0.117 & 0.3992 $\pm$ 0.0131 $\pm$ 0.0071\\
0.0489 &   6.99 & 0.393 & 3.122e-08 &   0.8 & -0.2 &  0.7 &  0.5 &  0.9 & 0.117 & 0.3851 $\pm$ 0.0089 $\pm$ 0.0058\\
0.0494 &   8.96 & 0.498 & 1.778e-08 &   0.7 & -0.1 &  0.9 &  0.9 &  1.0 & 0.117 & 0.3939 $\pm$ 0.0094 $\pm$ 0.0068\\
0.0516 &  11.12 & 0.586 & 1.133e-08 &   0.6 &  0.0 &  1.2 &  1.3 &  1.2 & 0.117 & 0.4282 $\pm$ 0.0118 $\pm$ 0.0094\\
\hline
0.0695 &   4.50 & 0.177 & 5.616e-08 &   1.5 & -0.8 &  0.6 &  0.1 &  1.6 & 0.113 & 0.3414 $\pm$ 0.0184 $\pm$ 0.0081\\
0.0690 &   5.48 & 0.216 & 3.943e-08 &   1.2 & -0.6 &  0.2 &  0.1 &  1.6 & 0.113 & 0.3644 $\pm$ 0.0192 $\pm$ 0.0078\\
0.0698 &   6.94 & 0.271 & 2.620e-08 &   1.0 & -0.4 &  0.3 &  0.2 &  1.0 & 0.113 & 0.4126 $\pm$ 0.0135 $\pm$ 0.0065\\
0.0695 &   8.99 & 0.353 & 1.397e-08 &   0.9 & -0.3 &  0.6 &  0.4 &  0.8 & 0.113 & 0.3942 $\pm$ 0.0120 $\pm$ 0.0054\\
0.0696 &  11.39 & 0.445 & 8.685e-09 &   0.7 & -0.1 &  0.8 &  0.7 &  0.9 & 0.113 & 0.4245 $\pm$ 0.0112 $\pm$ 0.0069\\
0.0707 &  14.46 & 0.553 & 4.527e-09 &   0.6 &  0.0 &  1.1 &  1.3 &  1.1 & 0.113 & 0.3920 $\pm$ 0.0141 $\pm$ 0.0082\\
\hline
0.0893 &   5.50 & 0.168 & 3.075e-08 &   1.5 & -0.9 &  0.4 &  0.1 &  1.7 & 0.096 & 0.3589 $\pm$ 0.0283 $\pm$ 0.0087\\
0.0897 &   6.96 & 0.211 & 2.080e-08 &   1.2 & -0.6 &  0.9 &  0.1 &  1.0 & 0.096 & 0.4048 $\pm$ 0.0169 $\pm$ 0.0082\\
0.0899 &   8.96 & 0.271 & 1.062e-08 &   1.0 & -0.4 &  1.3 &  0.2 &  0.9 & 0.096 & 0.3614 $\pm$ 0.0139 $\pm$ 0.0077\\
0.0899 &  11.45 & 0.346 & 6.432e-09 &   0.9 & -0.2 &  1.0 &  0.4 &  0.8 & 0.096 & 0.3804 $\pm$ 0.0121 $\pm$ 0.0061\\
0.0901 &  14.66 & 0.441 & 3.989e-09 &   0.7 & -0.1 &  1.2 &  0.7 &  0.9 & 0.096 & 0.4190 $\pm$ 0.0140 $\pm$ 0.0078\\
0.0907 &  19.02 & 0.566 & 1.981e-09 &   0.6 &  0.0 &  1.5 &  1.4 &  1.1 & 0.096 & 0.3856 $\pm$ 0.0242 $\pm$ 0.0093\\
\hline
0.1102 &   6.97 & 0.173 & 1.492e-08 &   1.4 & -0.8 &  0.5 &  0.1 &  1.0 & 0.043 & 0.3479 $\pm$ 0.0181 $\pm$ 0.0071\\
0.1101 &   8.99 & 0.222 & 8.715e-09 &   1.2 & -0.5 &  1.2 &  0.1 &  0.8 & 0.043 & 0.3522 $\pm$ 0.0161 $\pm$ 0.0072\\
0.1103 &  11.45 & 0.282 & 5.407e-09 &   1.0 & -0.4 &  1.1 &  0.2 &  0.8 & 0.043 & 0.3731 $\pm$ 0.0137 $\pm$ 0.0062\\
0.1100 &  14.76 & 0.364 & 3.054e-09 &   0.8 & -0.2 &  1.3 &  0.4 &  0.8 & 0.043 & 0.3734 $\pm$ 0.0136 $\pm$ 0.0068\\
0.1102 &  19.38 & 0.476 & 1.987e-09 &   0.7 & -0.1 &  1.7 &  0.7 &  1.0 & 0.043 & 0.4567 $\pm$ 0.0248 $\pm$ 0.0098\\
\hline
0.1395 &   7.15 & 0.141 & 1.148e-08 &   1.6 & -1.0 &  0.6 &  0.0 &  0.8 & 0.148 & 0.3502 $\pm$ 0.0195 $\pm$ 0.0078\\
0.1402 &   9.01 & 0.176 & 7.789e-09 &   1.4 & -0.7 &  0.8 &  0.1 &  0.7 & 0.129 & 0.3904 $\pm$ 0.0178 $\pm$ 0.0075\\
0.1398 &  11.43 & 0.223 & 4.431e-09 &   1.1 & -0.5 &  1.1 &  0.1 &  0.7 & 0.114 & 0.3713 $\pm$ 0.0115 $\pm$ 0.0068\\
0.1398 &  14.83 & 0.289 & 2.371e-09 &   0.9 & -0.3 &  1.4 &  0.2 &  0.8 & 0.101 & 0.3537 $\pm$ 0.0104 $\pm$ 0.0068\\
0.1400 &  19.32 & 0.374 & 1.330e-09 &   0.8 & -0.1 &  1.7 &  0.4 &  0.8 & 0.090 & 0.3616 $\pm$ 0.0128 $\pm$ 0.0077\\
0.1418 &  25.37 & 0.483 & 7.326e-10 &   0.6 &  0.0 &  2.1 &  0.8 &  1.0 & 0.080 & 0.3794 $\pm$ 0.0278 $\pm$ 0.0096\\
\hline
0.1789 &   7.38 & 0.114 & 7.087e-09 &   1.8 & -1.2 &  0.4 &  0.0 &  0.7 & 0.128 & 0.2916 $\pm$ 0.0243 $\pm$ 0.0067\\
0.1803 &   8.97 & 0.136 & 5.613e-09 &   1.6 & -0.9 &  0.8 &  0.0 &  0.7 & 0.113 & 0.3504 $\pm$ 0.0202 $\pm$ 0.0073\\
0.1818 &  11.44 & 0.172 & 3.256e-09 &   1.3 & -0.6 &  1.1 &  0.1 &  0.7 & 0.098 & 0.3437 $\pm$ 0.0133 $\pm$ 0.0066\\
0.1804 &  14.86 & 0.224 & 1.900e-09 &   1.1 & -0.4 &  1.4 &  0.1 &  0.7 & 0.086 & 0.3505 $\pm$ 0.0123 $\pm$ 0.0070\\
0.1804 &  19.42 & 0.292 & 1.065e-09 &   0.9 & -0.2 &  1.8 &  0.2 &  0.8 & 0.075 & 0.3552 $\pm$ 0.0131 $\pm$ 0.0076\\
0.1805 &  25.90 & 0.388 & 5.836e-10 &   0.7 & -0.1 &  2.1 &  0.5 &  0.9 & 0.067 & 0.3750 $\pm$ 0.0237 $\pm$ 0.0091\\
\hline
0.2226 &   9.15 & 0.113 & 4.115e-09 &   1.6 & -0.9 &  0.8 &  0.0 &  0.7 & 0.100 & 0.3263 $\pm$ 0.0233 $\pm$ 0.0067\\
0.2264 &  11.47 & 0.139 & 2.592e-09 &   1.3 & -0.6 &  1.1 &  0.0 &  0.7 & 0.086 & 0.3361 $\pm$ 0.0141 $\pm$ 0.0067\\
0.2263 &  14.90 & 0.180 & 1.363e-09 &   1.1 & -0.4 &  1.4 &  0.1 &  0.7 & 0.074 & 0.3083 $\pm$ 0.0114 $\pm$ 0.0061\\
0.2254 &  19.53 & 0.236 & 7.905e-10 &   0.9 & -0.2 &  1.7 &  0.1 &  0.7 & 0.065 & 0.3206 $\pm$ 0.0119 $\pm$ 0.0067\\
0.2258 &  26.32 & 0.316 & 4.192e-10 &   0.8 &  0.0 &  2.0 &  0.3 &  0.8 & 0.056 & 0.3309 $\pm$ 0.0196 $\pm$ 0.0076\\
\hline
0.2776 &  11.62 & 0.115 & 1.688e-09 &   1.1 & -0.4 &  1.3 &  0.0 &  0.7 & 0.077 & 0.2730 $\pm$ 0.0153 $\pm$ 0.0052\\
0.2755 &  14.82 & 0.147 & 1.078e-09 &   1.0 & -0.2 &  1.5 &  0.0 &  0.7 & 0.066 & 0.2885 $\pm$ 0.0133 $\pm$ 0.0056\\
0.2760 &  19.59 & 0.193 & 5.764e-10 &   0.8 & -0.1 &  1.7 &  0.1 &  0.7 & 0.057 & 0.2804 $\pm$ 0.0122 $\pm$ 0.0057\\
0.2764 &  26.49 & 0.260 & 2.869e-10 &   0.7 &  0.1 &  1.9 &  0.2 &  0.8 & 0.049 & 0.2705 $\pm$ 0.0184 $\pm$ 0.0059\\
0.2763 &  33.92 & 0.330 & 2.037e-10 &   0.6 &  0.1 &  2.1 &  0.4 &  0.8 & 0.043 & 0.3336 $\pm$ 0.0377 $\pm$ 0.0079\\
\hline
0.3421 &  15.02 & 0.121 & 6.364e-10 &   0.3 &  0.4 &  1.9 &  0.0 &  0.7 & 0.059 & 0.2152 $\pm$ 0.0091 $\pm$ 0.0045\\
0.3467 &  19.65 & 0.155 & 3.667e-10 &   0.4 &  0.5 &  1.8 &  0.1 &  0.7 & 0.050 & 0.2214 $\pm$ 0.0086 $\pm$ 0.0046\\
0.3473 &  26.20 & 0.205 & 1.848e-10 &   0.4 &  0.5 &  2.0 &  0.1 &  0.7 & 0.042 & 0.2070 $\pm$ 0.0112 $\pm$ 0.0046\\
0.3494 &  34.49 & 0.267 & 1.170e-10 &   0.4 &  0.5 &  2.3 &  0.2 &  0.8 & 0.037 & 0.2410 $\pm$ 0.0206 $\pm$ 0.0062\\
\hline
0.4554 &  20.05 & 0.121 & 1.789e-10 &  -2.2 &  3.2 &  1.5 &  0.0 &  0.7 & 0.044 & 0.1462 $\pm$ 0.0072 $\pm$ 0.0061\\
0.4745 &  26.22 & 0.152 & 8.466e-11 &  -1.5 &  2.7 &  1.8 &  0.1 &  0.7 & 0.037 & 0.1268 $\pm$ 0.0075 $\pm$ 0.0046\\
0.4769 &  34.80 & 0.200 & 3.899e-11 &  -1.0 &  2.2 &  3.0 &  0.1 &  0.7 & 0.031 & 0.1077 $\pm$ 0.0103 $\pm$ 0.0042\\
\hline
\end{tabular}

\caption{(c) 
The differential scattering cross section,
$d^2\sigma_p^{meas}/dxdQ^2$,
the value of $R$ as used in the
analysis (see also table 7), and the
structure function $F_2^p$ at 200 GeV 
incident muon energy, with the normalisation corrections of table~2.
Horizontal lines separate the x-bins. See also caption (a).
 }
\end{table}

\setcounter{table}{2}
\begin{table}[t]
\centering

\tiny
\begin{tabular}{|c|c|c||c||c|r|c|c|c||c|r|}\hline
$x$ & $Q^2$ & $y$ &  $d^2\sigma_p^{meas}/dxdQ^2$
& $E$ & $E^{\prime}$ & $AC$ &  $RC$ & $RE$  & $R$ & $F_2^p\;\pm\Delta{F^{stat}_2}\;\pm\Delta{F^{syst}_2}$ \\
  & (${\rm GeV^2}$) & & $({\rm b\cdot GeV^{-2}})$ &  [\%]  &  [\%]  &  [\%]  &  [\%] &  [\%]  & & \\
 \hline
0.0037 &   1.27 & 0.669 & 8.521e-06 &   0.6 & -0.1 &  2.5 &  1.7 &  3.7 & 0.537 & 0.3389 $\pm$ 0.0069 $\pm$ 0.0164\\
0.0050 &   1.71 & 0.672 & 3.769e-06 &   0.5 &  0.0 &  1.6 &  1.5 &  3.4 & 0.537 & 0.3606 $\pm$ 0.0069 $\pm$ 0.0147\\
0.0056 &   2.15 & 0.738 & 2.284e-06 &   0.4 &  0.1 &  0.9 &  1.6 &  3.2 & 0.537 & 0.3984 $\pm$ 0.0151 $\pm$ 0.0147\\
\hline
0.0073 &   1.81 & 0.487 & 2.638e-06 &   0.7 & -0.2 &  1.2 &  0.7 &  3.3 & 0.337 & 0.3484 $\pm$ 0.0091 $\pm$ 0.0127\\
0.0079 &   2.44 & 0.606 & 1.346e-06 &   0.6 & -0.1 &  1.5 &  1.0 &  3.0 & 0.337 & 0.3849 $\pm$ 0.0050 $\pm$ 0.0138\\
0.0090 &   3.37 & 0.733 & 6.304e-07 &   0.5 &  0.1 &  1.4 &  1.6 &  2.5 & 0.337 & 0.4113 $\pm$ 0.0080 $\pm$ 0.0140\\
\hline
0.0121 &   2.55 & 0.413 & 9.029e-07 &   0.8 & -0.3 &  1.0 &  0.4 &  2.9 & 0.246 & 0.3689 $\pm$ 0.0071 $\pm$ 0.0120\\
0.0123 &   3.47 & 0.552 & 4.692e-07 &   0.6 & -0.1 &  0.7 &  0.9 &  2.8 & 0.246 & 0.4012 $\pm$ 0.0062 $\pm$ 0.0123\\
0.0125 &   4.45 & 0.695 & 2.685e-07 &   0.5 &  0.0 &  1.5 &  2.1 &  2.1 & 0.246 & 0.4154 $\pm$ 0.0063 $\pm$ 0.0142\\
0.0139 &   5.38 & 0.754 & 1.761e-07 &   0.4 &  0.1 &  2.2 &  2.5 &  1.9 & 0.246 & 0.4469 $\pm$ 0.0112 $\pm$ 0.0174\\
\hline
0.0173 &   2.59 & 0.294 & 6.514e-07 &   1.0 & -0.5 &  1.2 &  0.2 &  2.7 & 0.190 & 0.3523 $\pm$ 0.0109 $\pm$ 0.0113\\
0.0173 &   3.49 & 0.393 & 3.537e-07 &   0.8 & -0.3 &  0.8 &  0.4 &  2.6 & 0.190 & 0.3804 $\pm$ 0.0079 $\pm$ 0.0111\\
0.0174 &   4.46 & 0.501 & 2.147e-07 &   0.7 & -0.1 &  0.4 &  1.1 &  2.6 & 0.190 & 0.4125 $\pm$ 0.0085 $\pm$ 0.0120\\
0.0168 &   5.60 & 0.658 & 1.326e-07 &   0.5 &  0.0 &  1.9 &  2.2 &  1.5 & 0.190 & 0.4242 $\pm$ 0.0105 $\pm$ 0.0139\\
0.0180 &   6.73 & 0.727 & 8.394e-08 &   0.5 &  0.1 &  2.1 &  2.3 &  1.7 & 0.190 & 0.4240 $\pm$ 0.0072 $\pm$ 0.0150\\
\hline
0.0245 &   3.49 & 0.280 & 2.693e-07 &   1.1 & -0.5 &  0.9 &  0.2 &  2.5 & 0.099 & 0.3716 $\pm$ 0.0084 $\pm$ 0.0110\\
0.0245 &   4.48 & 0.359 & 1.609e-07 &   0.9 & -0.3 &  0.6 &  0.5 &  2.4 & 0.099 & 0.3910 $\pm$ 0.0076 $\pm$ 0.0105\\
0.0246 &   5.47 & 0.435 & 1.063e-07 &   0.8 & -0.2 &  0.3 &  0.8 &  2.4 & 0.099 & 0.4107 $\pm$ 0.0082 $\pm$ 0.0109\\
0.0236 &   7.22 & 0.605 & 5.878e-08 &   0.6 &  0.0 &  1.7 &  1.4 &  1.3 & 0.099 & 0.4210 $\pm$ 0.0068 $\pm$ 0.0111\\
0.0253 &   8.88 & 0.684 & 3.691e-08 &   0.5 &  0.1 &  2.0 &  1.7 &  1.5 & 0.099 & 0.4441 $\pm$ 0.0064 $\pm$ 0.0136\\
0.0279 &  10.79 & 0.745 & 2.388e-08 &   0.4 &  0.1 &  2.1 &  2.1 &  1.9 & 0.099 & 0.4713 $\pm$ 0.0117 $\pm$ 0.0168\\
\hline
0.0347 &   4.48 & 0.253 & 1.307e-07 &   1.1 & -0.5 &  0.7 &  0.2 &  2.2 & 0.108 & 0.4139 $\pm$ 0.0109 $\pm$ 0.0110\\
0.0348 &   5.49 & 0.308 & 7.965e-08 &   1.0 & -0.4 &  0.5 &  0.3 &  2.2 & 0.108 & 0.3969 $\pm$ 0.0102 $\pm$ 0.0100\\
0.0348 &   6.86 & 0.383 & 4.857e-08 &   0.8 & -0.2 &  0.4 &  0.5 &  2.2 & 0.108 & 0.4027 $\pm$ 0.0091 $\pm$ 0.0097\\
0.0345 &   9.03 & 0.512 & 2.692e-08 &   0.6 & -0.1 &  1.3 &  0.9 &  1.3 & 0.108 & 0.4221 $\pm$ 0.0079 $\pm$ 0.0094\\
0.0348 &  11.44 & 0.641 & 1.610e-08 &   0.5 &  0.0 &  1.7 &  1.6 &  1.4 & 0.108 & 0.4406 $\pm$ 0.0063 $\pm$ 0.0123\\
0.0369 &  14.10 & 0.735 & 1.009e-08 &   0.5 &  0.1 &  2.0 &  2.2 &  1.7 & 0.108 & 0.4566 $\pm$ 0.0109 $\pm$ 0.0157\\
\hline
0.0469 &   5.53 & 0.232 & 6.014e-08 &   1.2 & -0.6 &  0.5 &  0.2 &  2.1 & 0.117 & 0.3872 $\pm$ 0.0110 $\pm$ 0.0099\\
0.0493 &   6.93 & 0.276 & 3.569e-08 &   1.0 & -0.4 &  0.4 &  0.2 &  2.0 & 0.117 & 0.3940 $\pm$ 0.0072 $\pm$ 0.0094\\
0.0490 &   8.99 & 0.361 & 2.018e-08 &   0.9 & -0.3 &  0.6 &  0.4 &  1.5 & 0.117 & 0.4008 $\pm$ 0.0076 $\pm$ 0.0078\\
0.0487 &  11.57 & 0.469 & 1.196e-08 &   0.7 & -0.1 &  1.2 &  0.8 &  1.0 & 0.117 & 0.4257 $\pm$ 0.0060 $\pm$ 0.0081\\
0.0494 &  14.87 & 0.589 & 6.673e-09 &   0.6 &  0.0 &  1.4 &  1.4 &  1.3 & 0.117 & 0.4317 $\pm$ 0.0053 $\pm$ 0.0106\\
0.0529 &  18.90 & 0.690 & 3.886e-09 &   0.5 &  0.1 &  1.7 &  2.1 &  1.6 & 0.117 & 0.4567 $\pm$ 0.0081 $\pm$ 0.0145\\
\hline
0.0663 &   7.32 & 0.217 & 2.366e-08 &   1.3 & -0.6 &  0.4 &  0.1 &  1.9 & 0.113 & 0.3753 $\pm$ 0.0137 $\pm$ 0.0091\\
0.0699 &   8.93 & 0.249 & 1.526e-08 &   1.1 & -0.5 &  0.5 &  0.2 &  1.9 & 0.113 & 0.3900 $\pm$ 0.0110 $\pm$ 0.0089\\
0.0696 &  11.47 & 0.323 & 9.131e-09 &   0.9 & -0.3 &  0.7 &  0.4 &  1.1 & 0.113 & 0.4086 $\pm$ 0.0085 $\pm$ 0.0070\\
0.0694 &  15.01 & 0.424 & 4.984e-09 &   0.7 & -0.1 &  1.1 &  0.7 &  0.9 & 0.113 & 0.4135 $\pm$ 0.0066 $\pm$ 0.0070\\
0.0699 &  19.65 & 0.547 & 2.720e-09 &   0.6 &  0.0 &  1.4 &  1.3 &  1.2 & 0.113 & 0.4269 $\pm$ 0.0061 $\pm$ 0.0098\\
0.0716 &  25.56 & 0.689 & 1.615e-09 &   0.5 &  0.1 &  1.7 &  2.2 &  1.5 & 0.113 & 0.4746 $\pm$ 0.0139 $\pm$ 0.0154\\
\hline
0.0863 &   9.35 & 0.213 & 1.039e-08 &   1.2 & -0.6 &  0.7 &  0.1 &  1.9 & 0.096 & 0.3505 $\pm$ 0.0188 $\pm$ 0.0085\\
0.0897 &  11.49 & 0.250 & 7.467e-09 &   1.1 & -0.4 &  0.9 &  0.2 &  1.2 & 0.096 & 0.4088 $\pm$ 0.0103 $\pm$ 0.0078\\
0.0897 &  14.97 & 0.326 & 4.005e-09 &   0.9 & -0.3 &  1.2 &  0.4 &  0.7 & 0.096 & 0.3964 $\pm$ 0.0084 $\pm$ 0.0069\\
0.0895 &  19.88 & 0.433 & 2.113e-09 &   0.7 & -0.1 &  1.5 &  0.7 &  0.9 & 0.096 & 0.4020 $\pm$ 0.0067 $\pm$ 0.0082\\
0.0903 &  26.07 & 0.558 & 1.126e-09 &   0.6 &  0.0 &  1.9 &  1.4 &  1.2 & 0.096 & 0.4068 $\pm$ 0.0084 $\pm$ 0.0108\\
\hline
0.1095 &  11.80 & 0.212 & 5.347e-09 &   1.2 & -0.6 &  1.1 &  0.1 &  1.0 & 0.043 & 0.3664 $\pm$ 0.0132 $\pm$ 0.0075\\
0.1097 &  14.93 & 0.265 & 3.404e-09 &   1.0 & -0.4 &  1.3 &  0.2 &  0.8 & 0.043 & 0.3908 $\pm$ 0.0103 $\pm$ 0.0075\\
0.1097 &  19.86 & 0.353 & 1.755e-09 &   0.8 & -0.2 &  1.7 &  0.4 &  0.7 & 0.043 & 0.3832 $\pm$ 0.0074 $\pm$ 0.0077\\
0.1100 &  26.38 & 0.465 & 8.933e-10 &   0.7 & -0.1 &  2.0 &  0.7 &  1.0 & 0.043 & 0.3760 $\pm$ 0.0078 $\pm$ 0.0091\\
0.1106 &  34.50 & 0.601 & 5.462e-10 &   0.6 &  0.1 &  2.3 &  1.4 &  1.3 & 0.043 & 0.4316 $\pm$ 0.0170 $\pm$ 0.0131\\
\hline
0.1391 &  11.77 & 0.167 & 4.294e-09 &   1.4 & -0.8 &  1.1 &  0.1 &  1.1 & 0.113 & 0.3620 $\pm$ 0.0219 $\pm$ 0.0079\\
0.1374 &  15.03 & 0.215 & 2.717e-09 &   1.2 & -0.5 &  1.4 &  0.1 &  0.7 & 0.101 & 0.3834 $\pm$ 0.0089 $\pm$ 0.0079\\
0.1389 &  19.89 & 0.280 & 1.389e-09 &   0.9 & -0.3 &  1.7 &  0.2 &  0.6 & 0.089 & 0.3664 $\pm$ 0.0060 $\pm$ 0.0078\\
0.1394 &  26.63 & 0.373 & 7.308e-10 &   0.8 & -0.1 &  2.1 &  0.4 &  0.8 & 0.080 & 0.3746 $\pm$ 0.0060 $\pm$ 0.0090\\
0.1401 &  35.24 & 0.489 & 3.687e-10 &   0.6 &  0.0 &  2.4 &  0.9 &  1.0 & 0.073 & 0.3639 $\pm$ 0.0100 $\pm$ 0.0103\\
\hline
0.1790 &  15.07 & 0.165 & 1.841e-09 &   1.3 & -0.6 &  1.4 &  0.1 &  0.5 & 0.085 & 0.3294 $\pm$ 0.0107 $\pm$ 0.0068\\
0.1796 &  19.99 & 0.218 & 1.015e-09 &   1.1 & -0.4 &  1.8 &  0.1 &  0.5 & 0.075 & 0.3349 $\pm$ 0.0070 $\pm$ 0.0072\\
0.1793 &  26.74 & 0.291 & 5.414e-10 &   0.9 & -0.2 &  2.1 &  0.2 &  0.6 & 0.066 & 0.3392 $\pm$ 0.0065 $\pm$ 0.0080\\
0.1800 &  35.13 & 0.378 & 3.040e-10 &   0.7 & -0.1 &  2.4 &  0.5 &  0.8 & 0.060 & 0.3548 $\pm$ 0.0083 $\pm$ 0.0096\\
0.1807 &  45.79 & 0.488 & 1.711e-10 &   0.6 &  0.1 &  2.7 &  0.9 &  1.0 & 0.055 & 0.3708 $\pm$ 0.0156 $\pm$ 0.0114\\
\hline
0.2245 &  15.14 & 0.133 & 1.440e-09 &   1.3 & -0.6 &  1.4 &  0.0 &  0.5 & 0.074 & 0.3210 $\pm$ 0.0114 $\pm$ 0.0067\\
0.2245 &  19.97 & 0.175 & 7.584e-10 &   1.1 & -0.4 &  1.7 &  0.1 &  0.5 & 0.064 & 0.3042 $\pm$ 0.0069 $\pm$ 0.0064\\
0.2244 &  26.77 & 0.233 & 4.171e-10 &   0.9 & -0.2 &  2.0 &  0.2 &  0.5 & 0.056 & 0.3154 $\pm$ 0.0064 $\pm$ 0.0071\\
0.2242 &  35.28 & 0.306 & 2.179e-10 &   0.8 &  0.0 &  2.3 &  0.3 &  0.6 & 0.050 & 0.3038 $\pm$ 0.0071 $\pm$ 0.0076\\
0.2250 &  46.62 & 0.401 & 1.199e-10 &   0.7 &  0.1 &  2.5 &  0.6 &  0.8 & 0.045 & 0.3167 $\pm$ 0.0121 $\pm$ 0.0090\\
\hline
0.2654 &  16.07 & 0.119 & 1.065e-09 &   1.1 & -0.3 &  1.5 &  0.0 &  0.1 & 0.065 & 0.3154 $\pm$ 0.0288 $\pm$ 0.0059\\
0.2744 &  19.99 & 0.143 & 5.872e-10 &   0.9 & -0.2 &  1.7 &  0.1 &  0.5 & 0.056 & 0.2840 $\pm$ 0.0079 $\pm$ 0.0058\\
0.2742 &  26.76 & 0.191 & 2.936e-10 &   0.8 &  0.0 &  1.9 &  0.1 &  0.5 & 0.049 & 0.2644 $\pm$ 0.0065 $\pm$ 0.0056\\
0.2749 &  35.44 & 0.251 & 1.683e-10 &   0.7 &  0.1 &  2.1 &  0.2 &  0.5 & 0.043 & 0.2800 $\pm$ 0.0074 $\pm$ 0.0064\\
0.2745 &  47.03 & 0.333 & 8.836e-11 &   0.6 &  0.2 &  2.3 &  0.4 &  0.6 & 0.038 & 0.2769 $\pm$ 0.0113 $\pm$ 0.0068\\
0.2772 &  59.81 & 0.413 & 3.442e-11 &   0.6 &  0.2 &  2.4 &  0.7 &  1.3 & 0.035 & 0.1879 $\pm$ 0.0288 $\pm$ 0.0054\\
\hline
0.3293 &  20.74 & 0.124 & 3.965e-10 &   0.3 &  0.5 &  1.8 &  0.0 &  0.4 & 0.050 & 0.2465 $\pm$ 0.0085 $\pm$ 0.0048\\
0.3447 &  26.74 & 0.153 & 1.998e-10 &   0.3 &  0.5 &  2.0 &  0.1 &  0.5 & 0.042 & 0.2216 $\pm$ 0.0049 $\pm$ 0.0047\\
0.3453 &  35.49 & 0.201 & 1.073e-10 &   0.3 &  0.5 &  2.3 &  0.1 &  0.5 & 0.037 & 0.2187 $\pm$ 0.0051 $\pm$ 0.0055\\
0.3461 &  46.63 & 0.262 & 6.040e-11 &   0.4 &  0.5 &  2.6 &  0.2 &  0.5 & 0.032 & 0.2242 $\pm$ 0.0066 $\pm$ 0.0062\\
0.3501 &  61.21 & 0.337 & 3.396e-11 &   0.4 &  0.5 &  2.8 &  0.4 &  1.2 & 0.029 & 0.2340 $\pm$ 0.0193 $\pm$ 0.0074\\
\hline
0.4331 &  27.70 & 0.125 & 1.111e-10 &  -2.4 &  3.5 &  1.8 &  0.0 &  0.5 & 0.037 & 0.1653 $\pm$ 0.0057 $\pm$ 0.0076\\
0.4652 &  35.55 & 0.150 & 4.990e-11 &  -1.7 &  2.9 &  3.0 &  0.1 &  0.5 & 0.031 & 0.1347 $\pm$ 0.0035 $\pm$ 0.0060\\
0.4701 &  46.57 & 0.194 & 2.514e-11 &  -1.2 &  2.4 &  3.9 &  0.1 &  0.4 & 0.027 & 0.1220 $\pm$ 0.0040 $\pm$ 0.0058\\
0.4790 &  62.34 & 0.254 & 1.315e-11 &  -0.7 &  1.9 &  4.6 &  0.3 &  1.1 & 0.023 & 0.1227 $\pm$ 0.0099 $\pm$ 0.0064\\
\hline
\end{tabular}

\caption{(d)
The differential scattering cross section,
$d^2\sigma_p^{meas}/dxdQ^2$,
the value of $R$ as used in the
analysis (see also table 7), and the
structure function $F_2^p$ at 280 GeV 
incident muon energy, with the normalisation corrections of table~2.
Horizontal lines separate the x-bins. See also caption (a).
 }
\end{table}


\setcounter{table}{3}
\begin{table}[t]
\centering

\tiny
\begin{tabular}{|c|c|c||c||c|r|c|c|c||c|r|}\hline
$x$ & $Q^2$ & $y$ &  $d^2\sigma_d^{meas}/dxdQ^2$
& $E$ & $E^{\prime}$ & $AC$ &  $RC$ & $RE$  & $R$ & $F_2^d\;\pm\Delta{F^{stat}_2}\;\pm\Delta{F^{syst}_2}$ \\
  & (${\rm GeV^2}$) & & $({\rm b\cdot GeV^{-2}})$ &  [\%]  &  [\%]  &  [\%]  &  [\%] &  [\%]  & & \\
 \hline
\hline
0.0078 &   0.80 & 0.635 & 8.019e-06 &   0.5 & -0.1 &  2.1 &  1.7 &  0.3 & 0.337 & 0.2667 $\pm$ 0.0027 $\pm$ 0.0074\\
0.0092 &   1.09 & 0.709 & 3.849e-06 &   0.4 &  0.1 &  1.0 &  2.2 &  0.5 & 0.337 & 0.2901 $\pm$ 0.0062 $\pm$ 0.0073\\
\hline
0.0120 &   0.90 & 0.462 & 5.052e-06 &   0.7 & -0.2 &  2.7 &  0.8 &  0.2 & 0.246 & 0.2694 $\pm$ 0.0054 $\pm$ 0.0078\\
0.0125 &   1.22 & 0.596 & 2.554e-06 &   0.6 & -0.1 &  1.7 &  1.4 &  0.3 & 0.246 & 0.2946 $\pm$ 0.0029 $\pm$ 0.0068\\
0.0139 &   1.62 & 0.703 & 1.286e-06 &   0.4 &  0.1 &  1.0 &  1.9 &  0.5 & 0.246 & 0.3102 $\pm$ 0.0074 $\pm$ 0.0069\\
\hline
0.0173 &   1.27 & 0.448 & 1.971e-06 &   0.7 & -0.2 &  2.0 &  0.7 &  0.2 & 0.190 & 0.2978 $\pm$ 0.0040 $\pm$ 0.0068\\
0.0176 &   1.72 & 0.594 & 1.046e-06 &   0.6 & -0.1 &  1.3 &  1.2 &  0.3 & 0.190 & 0.3304 $\pm$ 0.0041 $\pm$ 0.0063\\
0.0185 &   2.16 & 0.702 & 5.819e-07 &   0.5 &  0.0 &  1.0 &  1.5 &  0.4 & 0.190 & 0.3238 $\pm$ 0.0084 $\pm$ 0.0062\\
\hline
0.0246 &   1.27 & 0.320 & 1.541e-06 &   0.9 & -0.4 &  2.2 &  0.3 &  0.1 & 0.099 & 0.2967 $\pm$ 0.0049 $\pm$ 0.0072\\
0.0247 &   1.75 & 0.437 & 7.932e-07 &   0.7 & -0.2 &  1.4 &  0.5 &  0.2 & 0.099 & 0.3205 $\pm$ 0.0035 $\pm$ 0.0054\\
0.0254 &   2.36 & 0.566 & 4.123e-07 &   0.6 & -0.1 &  0.9 &  0.9 &  0.3 & 0.099 & 0.3409 $\pm$ 0.0034 $\pm$ 0.0049\\
0.0276 &   3.24 & 0.707 & 1.833e-07 &   0.5 &  0.1 &  0.9 &  1.7 &  0.5 & 0.099 & 0.3324 $\pm$ 0.0133 $\pm$ 0.0068\\
\hline
0.0348 &   1.30 & 0.231 & 1.163e-06 &   1.2 & -0.7 &  2.2 &  0.4 &  0.1 & 0.108 & 0.3057 $\pm$ 0.0077 $\pm$ 0.0082\\
0.0348 &   1.76 & 0.309 & 6.129e-07 &   1.0 & -0.4 &  1.3 &  0.3 &  0.1 & 0.108 & 0.3155 $\pm$ 0.0049 $\pm$ 0.0053\\
0.0349 &   2.45 & 0.428 & 3.062e-07 &   0.8 & -0.2 &  0.7 &  0.4 &  0.2 & 0.108 & 0.3405 $\pm$ 0.0036 $\pm$ 0.0039\\
0.0354 &   3.34 & 0.571 & 1.550e-07 &   0.6 &  0.0 &  0.5 &  1.1 &  0.3 & 0.108 & 0.3624 $\pm$ 0.0066 $\pm$ 0.0051\\
0.0364 &   4.27 & 0.705 & 8.467e-08 &   0.5 &  0.1 &  0.7 &  2.4 &  0.4 & 0.108 & 0.3541 $\pm$ 0.0205 $\pm$ 0.0089\\
\hline
0.0495 &   1.33 & 0.168 & 8.657e-07 &   1.6 & -1.0 &  2.6 &  0.4 &  0.1 & 0.117 & 0.3233 $\pm$ 0.0089 $\pm$ 0.0103\\
0.0493 &   1.76 & 0.221 & 4.663e-07 &   1.2 & -0.7 &  1.3 &  0.4 &  0.1 & 0.117 & 0.3189 $\pm$ 0.0050 $\pm$ 0.0063\\
0.0494 &   2.48 & 0.309 & 2.305e-07 &   1.0 & -0.4 &  0.5 &  0.3 &  0.1 & 0.117 & 0.3375 $\pm$ 0.0033 $\pm$ 0.0041\\
0.0501 &   3.41 & 0.416 & 1.157e-07 &   0.8 & -0.2 &  0.2 &  0.6 &  0.2 & 0.117 & 0.3574 $\pm$ 0.0042 $\pm$ 0.0036\\
0.0509 &   4.39 & 0.526 & 6.336e-08 &   0.6 & -0.1 &  0.3 &  1.2 &  0.3 & 0.117 & 0.3589 $\pm$ 0.0082 $\pm$ 0.0050\\
0.0526 &   5.31 & 0.609 & 4.373e-08 &   0.5 &  0.0 &  0.4 &  1.6 &  0.3 & 0.117 & 0.3957 $\pm$ 0.0198 $\pm$ 0.0069\\
\hline
0.0700 &   1.35 & 0.120 & 6.029e-07 &   2.0 & -1.4 &  3.5 &  0.5 &  0.1 & 0.113 & 0.3154 $\pm$ 0.0132 $\pm$ 0.0136\\
0.0699 &   1.77 & 0.156 & 3.485e-07 &   1.6 & -1.0 &  1.8 &  0.4 &  0.1 & 0.113 & 0.3229 $\pm$ 0.0073 $\pm$ 0.0088\\
0.0694 &   2.48 & 0.219 & 1.747e-07 &   1.2 & -0.6 &  0.5 &  0.4 &  0.1 & 0.113 & 0.3348 $\pm$ 0.0044 $\pm$ 0.0051\\
0.0696 &   3.45 & 0.303 & 8.975e-08 &   1.0 & -0.4 &  0.2 &  0.3 &  0.1 & 0.113 & 0.3608 $\pm$ 0.0050 $\pm$ 0.0039\\
0.0702 &   4.42 & 0.382 & 5.027e-08 &   0.8 & -0.2 &  0.4 &  0.5 &  0.1 & 0.113 & 0.3579 $\pm$ 0.0070 $\pm$ 0.0039\\
0.0703 &   5.38 & 0.463 & 3.236e-08 &   0.7 & -0.1 &  0.5 &  0.8 &  0.2 & 0.113 & 0.3654 $\pm$ 0.0122 $\pm$ 0.0043\\
0.0711 &   6.38 & 0.536 & 2.146e-08 &   0.6 &  0.0 &  0.5 &  1.1 &  0.3 & 0.113 & 0.3641 $\pm$ 0.0253 $\pm$ 0.0049\\
\hline
0.0902 &   1.38 & 0.095 & 4.694e-07 &   2.5 & -1.8 &  4.4 &  0.4 &  0.1 & 0.096 & 0.3267 $\pm$ 0.0195 $\pm$ 0.0176\\
0.0899 &   1.76 & 0.121 & 2.777e-07 &   2.0 & -1.4 &  2.6 &  0.4 &  0.1 & 0.096 & 0.3222 $\pm$ 0.0096 $\pm$ 0.0116\\
0.0897 &   2.47 & 0.169 & 1.414e-07 &   1.5 & -0.9 &  0.9 &  0.4 &  0.1 & 0.096 & 0.3358 $\pm$ 0.0057 $\pm$ 0.0065\\
0.0897 &   3.46 & 0.236 & 6.915e-08 &   1.1 & -0.5 &  0.2 &  0.3 &  0.1 & 0.096 & 0.3416 $\pm$ 0.0056 $\pm$ 0.0043\\
0.0900 &   4.44 & 0.299 & 4.154e-08 &   0.9 & -0.3 &  0.5 &  0.3 &  0.1 & 0.096 & 0.3577 $\pm$ 0.0075 $\pm$ 0.0040\\
0.0903 &   5.44 & 0.366 & 2.770e-08 &   0.8 & -0.2 &  0.7 &  0.4 &  0.1 & 0.096 & 0.3801 $\pm$ 0.0117 $\pm$ 0.0046\\
0.0897 &   6.50 & 0.436 & 1.708e-08 &   0.7 & -0.1 &  0.8 &  0.6 &  0.2 & 0.096 & 0.3527 $\pm$ 0.0173 $\pm$ 0.0045\\
\hline
0.1090 &   1.78 & 0.101 & 2.283e-07 &   2.3 & -1.7 &  3.2 &  0.4 &  0.1 & 0.043 & 0.3220 $\pm$ 0.0117 $\pm$ 0.0141\\
0.1096 &   2.49 & 0.140 & 1.162e-07 &   1.7 & -1.1 &  1.4 &  0.4 &  0.1 & 0.043 & 0.3346 $\pm$ 0.0069 $\pm$ 0.0083\\
0.1098 &   3.48 & 0.193 & 5.685e-08 &   1.3 & -0.7 &  0.3 &  0.3 &  0.1 & 0.043 & 0.3353 $\pm$ 0.0065 $\pm$ 0.0054\\
0.1100 &   4.45 & 0.246 & 3.308e-08 &   1.1 & -0.5 &  0.4 &  0.3 &  0.1 & 0.043 & 0.3360 $\pm$ 0.0077 $\pm$ 0.0042\\
0.1105 &   5.44 & 0.299 & 2.084e-08 &   0.9 & -0.3 &  0.7 &  0.2 &  0.1 & 0.043 & 0.3314 $\pm$ 0.0105 $\pm$ 0.0042\\
0.1096 &   6.60 & 0.363 & 1.380e-08 &   0.8 & -0.2 &  0.9 &  0.3 &  0.1 & 0.043 & 0.3379 $\pm$ 0.0144 $\pm$ 0.0043\\
\hline
0.1282 &   1.86 & 0.090 & 1.926e-07 &   2.7 & -2.0 &  3.5 &  0.4 &  0.1 & 0.296 & 0.3479 $\pm$ 0.0198 $\pm$ 0.0168\\
0.1383 &   2.49 & 0.112 & 9.229e-08 &   2.0 & -1.4 &  2.0 &  0.4 &  0.1 & 0.271 & 0.3298 $\pm$ 0.0060 $\pm$ 0.0105\\
0.1396 &   3.48 & 0.154 & 4.525e-08 &   1.5 & -0.9 &  0.7 &  0.3 &  0.1 & 0.233 & 0.3328 $\pm$ 0.0057 $\pm$ 0.0064\\
0.1394 &   4.45 & 0.196 & 2.633e-08 &   1.2 & -0.6 &  0.3 &  0.3 &  0.1 & 0.201 & 0.3272 $\pm$ 0.0062 $\pm$ 0.0047\\
0.1398 &   5.46 & 0.239 & 1.749e-08 &   1.1 & -0.4 &  0.5 &  0.3 &  0.1 & 0.176 & 0.3411 $\pm$ 0.0085 $\pm$ 0.0043\\
0.1407 &   6.69 & 0.289 & 1.092e-08 &   0.9 & -0.3 &  0.8 &  0.2 &  0.1 & 0.154 & 0.3363 $\pm$ 0.0097 $\pm$ 0.0042\\
\hline
0.1737 &   2.67 & 0.095 & 6.374e-08 &   2.2 & -1.5 &  1.7 &  0.4 &  0.1 & 0.247 & 0.3277 $\pm$ 0.0091 $\pm$ 0.0103\\
0.1803 &   3.46 & 0.119 & 3.339e-08 &   1.7 & -1.0 &  1.1 &  0.3 &  0.1 & 0.211 & 0.3069 $\pm$ 0.0066 $\pm$ 0.0071\\
0.1799 &   4.46 & 0.152 & 2.048e-08 &   1.4 & -0.7 &  0.6 &  0.3 &  0.1 & 0.179 & 0.3197 $\pm$ 0.0076 $\pm$ 0.0054\\
0.1806 &   5.46 & 0.185 & 1.276e-08 &   1.2 & -0.5 &  0.4 &  0.3 &  0.1 & 0.156 & 0.3088 $\pm$ 0.0088 $\pm$ 0.0045\\
0.1801 &   6.77 & 0.229 & 8.304e-09 &   1.0 & -0.4 &  0.5 &  0.2 &  0.1 & 0.135 & 0.3203 $\pm$ 0.0097 $\pm$ 0.0039\\
0.1769 &   8.50 & 0.289 & 4.918e-09 &   0.9 & -0.2 &  0.8 &  0.2 &  0.1 & 0.118 & 0.3092 $\pm$ 0.0232 $\pm$ 0.0037\\
\hline
0.2048 &   2.90 & 0.087 & 4.151e-08 &   2.1 & -1.4 &  0.9 &  0.3 &  0.1 & 0.223 & 0.2970 $\pm$ 0.0263 $\pm$ 0.0081\\
0.2203 &   3.51 & 0.099 & 2.567e-08 &   1.8 & -1.1 &  0.8 &  0.3 &  0.1 & 0.193 & 0.2924 $\pm$ 0.0070 $\pm$ 0.0065\\
0.2255 &   4.47 & 0.123 & 1.458e-08 &   1.4 & -0.7 &  0.7 &  0.3 &  0.1 & 0.162 & 0.2820 $\pm$ 0.0070 $\pm$ 0.0050\\
0.2259 &   5.47 & 0.149 & 9.905e-09 &   1.2 & -0.5 &  0.6 &  0.3 &  0.1 & 0.140 & 0.2937 $\pm$ 0.0090 $\pm$ 0.0043\\
0.2273 &   6.81 & 0.184 & 5.838e-09 &   1.0 & -0.3 &  0.6 &  0.3 &  0.1 & 0.120 & 0.2789 $\pm$ 0.0086 $\pm$ 0.0034\\
0.2211 &   8.64 & 0.236 & 3.680e-09 &   0.9 & -0.2 &  0.6 &  0.2 &  0.1 & 0.104 & 0.2874 $\pm$ 0.0182 $\pm$ 0.0032\\
\hline
0.2643 &   3.75 & 0.088 & 1.722e-08 &   1.4 & -0.7 &  1.7 &  0.3 &  0.1 & 0.172 & 0.2690 $\pm$ 0.0118 $\pm$ 0.0062\\
0.2723 &   4.47 & 0.102 & 1.153e-08 &   1.3 & -0.5 &  1.3 &  0.3 &  0.1 & 0.150 & 0.2672 $\pm$ 0.0079 $\pm$ 0.0052\\
0.2746 &   5.44 & 0.123 & 7.263e-09 &   1.1 & -0.3 &  1.0 &  0.3 &  0.1 & 0.130 & 0.2562 $\pm$ 0.0091 $\pm$ 0.0039\\
0.2792 &   6.84 & 0.151 & 4.327e-09 &   0.9 & -0.2 &  0.7 &  0.3 &  0.1 & 0.109 & 0.2512 $\pm$ 0.0090 $\pm$ 0.0030\\
0.2729 &   8.70 & 0.194 & 2.684e-09 &   0.8 &  0.0 &  0.6 &  0.2 &  0.1 & 0.094 & 0.2550 $\pm$ 0.0167 $\pm$ 0.0028\\
\hline
0.3145 &   4.61 & 0.091 & 8.036e-09 &   0.4 &  0.3 &  4.4 &  0.3 &  0.1 & 0.140 & 0.2286 $\pm$ 0.0088 $\pm$ 0.0100\\
0.3373 &   5.46 & 0.101 & 5.253e-09 &   0.4 &  0.4 &  3.3 &  0.2 &  0.1 & 0.121 & 0.2267 $\pm$ 0.0071 $\pm$ 0.0077\\
0.3485 &   6.87 & 0.123 & 3.044e-09 &   0.4 &  0.4 &  2.1 &  0.2 &  0.1 & 0.101 & 0.2195 $\pm$ 0.0068 $\pm$ 0.0047\\
0.3432 &   8.73 & 0.157 & 1.758e-09 &   0.4 &  0.5 &  1.2 &  0.2 &  0.1 & 0.085 & 0.2074 $\pm$ 0.0108 $\pm$ 0.0028\\
\hline
0.3931 &   5.72 & 0.091 & 3.036e-09 &  -2.9 &  3.7 &  9.4 &  0.2 &  0.1 & 0.112 & 0.1680 $\pm$ 0.0147 $\pm$ 0.0177\\
0.4379 &   6.96 & 0.100 & 1.731e-09 &  -2.5 &  3.4 &  7.0 &  0.2 &  0.1 & 0.094 & 0.1603 $\pm$ 0.0061 $\pm$ 0.0132\\
0.4603 &   8.74 & 0.118 & 8.371e-10 &  -1.9 &  3.0 &  4.4 &  0.2 &  0.1 & 0.078 & 0.1310 $\pm$ 0.0070 $\pm$ 0.0074\\
\hline
\end{tabular}

\caption{(a)
The differential scattering cross section,
$d^2\sigma_d^{meas}/dxdQ^2$,
the value of $R$ as used in the
analysis (see also table 7), and
structure function $F_2^d$ at 90 GeV 
incident muon energy, with the normalisation corrections of table~2.
Horizontal lines separate the $x$-bins.
The indicated five sources of systematic errors on $F_2$ are
the uncertainties on: $E$
the incident muon energy;
$E'$ the scattered muon energy; $AC$
the acceptance; $RC$ the
radiative correction; $RE$ the reconstruction efficiency.
The signs on $E$ and $E'$ correspond to an decrease in the
respective muon energy.
The various errors are not independent: $E$ and $E'$ 
are fully correlated between
the proton and the deuteron data, but
independent for the data taken at different energies;
$RC$ is fully correlated between all energies, but 
independent for the proton and the deuteron; whilst $AC$
and $RE$ should each be treated as fully correlated for all data
sets.
}
\end{table}

\setcounter{table}{3}
\begin{table}[t]
\centering

\tiny
\begin{tabular}{|c|c|c||c||c|r|c|c|c||c|r|}\hline
$x$ & $Q^2$ & $y$ &  $d^2\sigma_d^{meas}/dxdQ^2$
& $E$ & $E^{\prime}$ & $AC$ &  $RC$ & $RE$  & $R$ & $F_2^d\;\pm\Delta{F^{stat}_2}\;\pm\Delta{F^{syst}_2}$ \\
  & (${\rm GeV^2}$) & & $({\rm b\cdot GeV^{-2}})$ &  [\%]  &  [\%]  &  [\%]  &  [\%] &  [\%]  & & \\
 \hline
\hline
0.0087 &   1.19 & 0.626 & 3.651e-06 &   0.5 &  0.0 &  1.0 &  1.6 &  0.6 & 0.337 & 0.2894 $\pm$ 0.0070 $\pm$ 0.0058\\
\hline
0.0117 &   1.37 & 0.539 & 2.248e-06 &   0.6 & -0.2 &  1.7 &  1.1 &  0.4 & 0.246 & 0.2907 $\pm$ 0.0077 $\pm$ 0.0061\\
0.0129 &   1.70 & 0.605 & 1.392e-06 &   0.5 &  0.0 &  1.0 &  1.2 &  0.5 & 0.246 & 0.3247 $\pm$ 0.0056 $\pm$ 0.0057\\
\hline
0.0172 &   1.77 & 0.472 & 1.095e-06 &   0.7 & -0.2 &  1.3 &  0.7 &  0.4 & 0.190 & 0.3279 $\pm$ 0.0082 $\pm$ 0.0056\\
0.0178 &   2.34 & 0.599 & 5.760e-07 &   0.6 &  0.0 &  1.0 &  1.0 &  0.5 & 0.190 & 0.3428 $\pm$ 0.0054 $\pm$ 0.0054\\
\hline
0.0241 &   1.82 & 0.351 & 8.001e-07 &   0.9 & -0.4 &  1.4 &  0.3 &  0.3 & 0.099 & 0.3184 $\pm$ 0.0102 $\pm$ 0.0056\\
0.0245 &   2.50 & 0.472 & 3.977e-07 &   0.7 & -0.2 &  0.9 &  0.5 &  0.4 & 0.099 & 0.3351 $\pm$ 0.0042 $\pm$ 0.0046\\
0.0260 &   3.36 & 0.591 & 2.031e-07 &   0.6 &  0.0 &  0.9 &  1.2 &  0.5 & 0.099 & 0.3560 $\pm$ 0.0056 $\pm$ 0.0058\\
\hline
0.0347 &   2.51 & 0.332 & 3.240e-07 &   0.9 & -0.4 &  0.7 &  0.3 &  0.3 & 0.108 & 0.3486 $\pm$ 0.0075 $\pm$ 0.0046\\
0.0348 &   3.48 & 0.458 & 1.557e-07 &   0.7 & -0.2 &  0.5 &  0.7 &  0.4 & 0.108 & 0.3579 $\pm$ 0.0056 $\pm$ 0.0043\\
0.0354 &   4.41 & 0.568 & 8.904e-08 &   0.6 &  0.0 &  0.7 &  1.4 &  0.5 & 0.108 & 0.3619 $\pm$ 0.0072 $\pm$ 0.0063\\
0.0374 &   5.32 & 0.645 & 6.119e-08 &   0.5 &  0.0 &  0.8 &  1.9 &  0.6 & 0.108 & 0.4004 $\pm$ 0.0195 $\pm$ 0.0087\\
\hline
0.0496 &   2.54 & 0.237 & 2.194e-07 &   1.2 & -0.6 &  0.5 &  0.4 &  0.2 & 0.117 & 0.3200 $\pm$ 0.0072 $\pm$ 0.0047\\
0.0494 &   3.49 & 0.327 & 1.196e-07 &   0.9 & -0.3 &  0.2 &  0.3 &  0.3 & 0.117 & 0.3544 $\pm$ 0.0054 $\pm$ 0.0039\\
0.0495 &   4.47 & 0.416 & 6.821e-08 &   0.8 & -0.2 &  0.3 &  0.7 &  0.3 & 0.117 & 0.3589 $\pm$ 0.0055 $\pm$ 0.0039\\
0.0505 &   5.43 & 0.492 & 4.383e-08 &   0.7 & -0.1 &  0.4 &  0.9 &  0.4 & 0.117 & 0.3689 $\pm$ 0.0072 $\pm$ 0.0047\\
0.0524 &   6.61 & 0.576 & 2.839e-08 &   0.6 &  0.0 &  0.6 &  1.2 &  0.5 & 0.117 & 0.3901 $\pm$ 0.0111 $\pm$ 0.0061\\
\hline
0.0707 &   2.56 & 0.168 & 1.655e-07 &   1.5 & -0.9 &  0.5 &  0.4 &  0.2 & 0.113 & 0.3307 $\pm$ 0.0112 $\pm$ 0.0064\\
0.0702 &   3.50 & 0.229 & 8.718e-08 &   1.2 & -0.6 &  0.2 &  0.3 &  0.2 & 0.113 & 0.3408 $\pm$ 0.0072 $\pm$ 0.0046\\
0.0698 &   4.48 & 0.295 & 5.082e-08 &   1.0 & -0.4 &  0.4 &  0.3 &  0.2 & 0.113 & 0.3442 $\pm$ 0.0068 $\pm$ 0.0040\\
0.0700 &   5.47 & 0.357 & 3.384e-08 &   0.8 & -0.2 &  0.5 &  0.4 &  0.3 & 0.113 & 0.3606 $\pm$ 0.0078 $\pm$ 0.0042\\
0.0700 &   6.86 & 0.448 & 2.042e-08 &   0.7 & -0.1 &  0.5 &  0.7 &  0.4 & 0.113 & 0.3691 $\pm$ 0.0091 $\pm$ 0.0043\\
0.0718 &   8.61 & 0.543 & 1.262e-08 &   0.6 &  0.0 &  0.5 &  1.1 &  0.5 & 0.113 & 0.3976 $\pm$ 0.0199 $\pm$ 0.0056\\
\hline
0.0915 &   2.61 & 0.132 & 1.368e-07 &   1.8 & -1.2 &  0.9 &  0.4 &  0.2 & 0.096 & 0.3587 $\pm$ 0.0171 $\pm$ 0.0087\\
0.0909 &   3.50 & 0.177 & 6.689e-08 &   1.4 & -0.8 &  0.2 &  0.3 &  0.2 & 0.096 & 0.3253 $\pm$ 0.0090 $\pm$ 0.0055\\
0.0906 &   4.47 & 0.226 & 4.225e-08 &   1.2 & -0.5 &  0.5 &  0.3 &  0.2 & 0.096 & 0.3487 $\pm$ 0.0086 $\pm$ 0.0047\\
0.0904 &   5.48 & 0.278 & 2.705e-08 &   1.0 & -0.4 &  0.7 &  0.2 &  0.2 & 0.096 & 0.3513 $\pm$ 0.0089 $\pm$ 0.0044\\
0.0903 &   6.81 & 0.344 & 1.705e-08 &   0.8 & -0.2 &  0.8 &  0.3 &  0.3 & 0.096 & 0.3618 $\pm$ 0.0081 $\pm$ 0.0044\\
0.0905 &   8.77 & 0.440 & 1.005e-08 &   0.7 & -0.1 &  0.8 &  0.6 &  0.4 & 0.096 & 0.3827 $\pm$ 0.0149 $\pm$ 0.0049\\
\hline
0.1125 &   2.62 & 0.109 & 9.641e-08 &   2.1 & -1.4 &  1.4 &  0.4 &  0.2 & 0.043 & 0.3078 $\pm$ 0.0176 $\pm$ 0.0088\\
0.1120 &   3.48 & 0.143 & 6.068e-08 &   1.6 & -1.0 &  0.3 &  0.4 &  0.2 & 0.043 & 0.3508 $\pm$ 0.0140 $\pm$ 0.0067\\
0.1110 &   4.49 & 0.185 & 3.429e-08 &   1.3 & -0.7 &  0.4 &  0.3 &  0.2 & 0.043 & 0.3390 $\pm$ 0.0106 $\pm$ 0.0054\\
0.1108 &   5.49 & 0.227 & 2.240e-08 &   1.1 & -0.5 &  0.7 &  0.3 &  0.2 & 0.043 & 0.3428 $\pm$ 0.0102 $\pm$ 0.0050\\
0.1108 &   6.84 & 0.282 & 1.399e-08 &   1.0 & -0.3 &  0.9 &  0.2 &  0.2 & 0.043 & 0.3487 $\pm$ 0.0087 $\pm$ 0.0047\\
0.1105 &   8.86 & 0.365 & 7.705e-09 &   0.8 & -0.2 &  1.0 &  0.3 &  0.3 & 0.043 & 0.3448 $\pm$ 0.0134 $\pm$ 0.0047\\
0.1112 &  10.75 & 0.437 & 5.054e-09 &   0.7 & -0.1 &  1.0 &  0.5 &  0.4 & 0.043 & 0.3544 $\pm$ 0.0251 $\pm$ 0.0050\\
\hline
0.1432 &   3.48 & 0.113 & 4.394e-08 &   1.9 & -1.2 &  0.7 &  0.4 &  0.2 & 0.231 & 0.3197 $\pm$ 0.0109 $\pm$ 0.0077\\
0.1419 &   4.49 & 0.147 & 2.508e-08 &   1.5 & -0.9 &  0.3 &  0.4 &  0.2 & 0.199 & 0.3105 $\pm$ 0.0082 $\pm$ 0.0054\\
0.1409 &   5.48 & 0.179 & 1.689e-08 &   1.3 & -0.6 &  0.5 &  0.3 &  0.2 & 0.175 & 0.3177 $\pm$ 0.0081 $\pm$ 0.0049\\
0.1411 &   6.87 & 0.224 & 1.032e-08 &   1.1 & -0.4 &  0.8 &  0.3 &  0.2 & 0.151 & 0.3180 $\pm$ 0.0064 $\pm$ 0.0045\\
0.1405 &   8.90 & 0.291 & 6.161e-09 &   0.9 & -0.3 &  1.0 &  0.2 &  0.2 & 0.130 & 0.3356 $\pm$ 0.0101 $\pm$ 0.0047\\
0.1421 &  11.00 & 0.353 & 3.766e-09 &   0.8 & -0.1 &  1.2 &  0.3 &  0.3 & 0.115 & 0.3345 $\pm$ 0.0138 $\pm$ 0.0049\\
\hline
0.1814 &   3.62 & 0.093 & 3.050e-08 &   2.1 & -1.4 &  1.1 &  0.3 &  0.2 & 0.205 & 0.3014 $\pm$ 0.0158 $\pm$ 0.0085\\
0.1859 &   4.50 & 0.113 & 1.861e-08 &   1.7 & -1.0 &  0.6 &  0.3 &  0.2 & 0.175 & 0.2968 $\pm$ 0.0099 $\pm$ 0.0062\\
0.1840 &   5.49 & 0.138 & 1.216e-08 &   1.5 & -0.8 &  0.4 &  0.3 &  0.2 & 0.154 & 0.2913 $\pm$ 0.0092 $\pm$ 0.0050\\
0.1833 &   6.87 & 0.173 & 7.810e-09 &   1.2 & -0.5 &  0.5 &  0.3 &  0.2 & 0.132 & 0.3012 $\pm$ 0.0073 $\pm$ 0.0045\\
0.1810 &   8.93 & 0.227 & 4.410e-09 &   1.0 & -0.3 &  0.8 &  0.2 &  0.2 & 0.113 & 0.2973 $\pm$ 0.0109 $\pm$ 0.0039\\
0.1813 &  11.15 & 0.281 & 3.103e-09 &   0.9 & -0.2 &  1.0 &  0.2 &  0.2 & 0.099 & 0.3418 $\pm$ 0.0135 $\pm$ 0.0046\\
\hline
0.2282 &   4.55 & 0.093 & 1.362e-08 &   1.7 & -1.0 &  0.7 &  0.3 &  0.2 & 0.159 & 0.2707 $\pm$ 0.0118 $\pm$ 0.0058\\
0.2304 &   5.47 & 0.111 & 9.247e-09 &   1.5 & -0.8 &  0.6 &  0.3 &  0.2 & 0.139 & 0.2722 $\pm$ 0.0097 $\pm$ 0.0049\\
0.2304 &   6.92 & 0.139 & 5.745e-09 &   1.2 & -0.5 &  0.6 &  0.3 &  0.2 & 0.118 & 0.2769 $\pm$ 0.0073 $\pm$ 0.0041\\
0.2270 &   8.83 & 0.179 & 3.482e-09 &   1.0 & -0.3 &  0.6 &  0.3 &  0.2 & 0.101 & 0.2783 $\pm$ 0.0092 $\pm$ 0.0036\\
0.2271 &  11.23 & 0.226 & 2.258e-09 &   0.9 & -0.2 &  0.7 &  0.2 &  0.2 & 0.087 & 0.3041 $\pm$ 0.0119 $\pm$ 0.0035\\
\hline
0.2795 &   5.54 & 0.093 & 7.133e-09 &   1.2 & -0.5 &  1.0 &  0.3 &  0.2 & 0.127 & 0.2598 $\pm$ 0.0125 $\pm$ 0.0044\\
0.2849 &   6.93 & 0.114 & 4.042e-09 &   1.0 & -0.3 &  0.7 &  0.3 &  0.2 & 0.107 & 0.2389 $\pm$ 0.0075 $\pm$ 0.0034\\
0.2801 &   8.85 & 0.146 & 2.491e-09 &   0.9 & -0.1 &  0.6 &  0.3 &  0.2 & 0.091 & 0.2423 $\pm$ 0.0095 $\pm$ 0.0029\\
0.2783 &  11.26 & 0.186 & 1.453e-09 &   0.8 &  0.0 &  0.7 &  0.2 &  0.2 & 0.078 & 0.2350 $\pm$ 0.0106 $\pm$ 0.0026\\
0.2788 &  14.18 & 0.231 & 8.263e-10 &   0.7 &  0.1 &  0.8 &  0.2 &  0.2 & 0.068 & 0.2204 $\pm$ 0.0173 $\pm$ 0.0024\\
\hline
0.3421 &   7.08 & 0.097 & 2.936e-09 &   0.3 &  0.5 &  2.1 &  0.2 &  0.2 & 0.099 & 0.2168 $\pm$ 0.0073 $\pm$ 0.0047\\
0.3535 &   8.86 & 0.117 & 1.641e-09 &   0.3 &  0.5 &  1.2 &  0.2 &  0.2 & 0.083 & 0.1993 $\pm$ 0.0068 $\pm$ 0.0027\\
0.3526 &  11.32 & 0.149 & 9.001e-10 &   0.3 &  0.5 &  0.7 &  0.2 &  0.2 & 0.070 & 0.1829 $\pm$ 0.0073 $\pm$ 0.0019\\
0.3516 &  14.31 & 0.187 & 5.527e-10 &   0.3 &  0.5 &  0.9 &  0.2 &  0.2 & 0.060 & 0.1846 $\pm$ 0.0112 $\pm$ 0.0020\\
\hline
0.4383 &   9.03 & 0.096 & 8.531e-10 &  -2.9 &  3.8 &  4.4 &  0.2 &  0.2 & 0.077 & 0.1330 $\pm$ 0.0068 $\pm$ 0.0087\\
0.4698 &  11.36 & 0.114 & 4.403e-10 &  -2.3 &  3.3 &  2.5 &  0.2 &  0.2 & 0.064 & 0.1189 $\pm$ 0.0054 $\pm$ 0.0057\\
0.4780 &  14.38 & 0.140 & 2.700e-10 &  -1.7 &  2.8 &  1.5 &  0.1 &  0.2 & 0.054 & 0.1215 $\pm$ 0.0069 $\pm$ 0.0045\\
\hline
\end{tabular}

\caption{(b)
The differential scattering cross section, 
$d^2\sigma_d^{meas}/dxdQ^2$,
the value of $R$ as used in the
analysis (see also table 7), and the
structure function $F_2^d$ at 120 GeV 
incident muon energy, with the normalisation corrections of table~2.
Horizontal lines separate the x-bins. See also caption (a).
 }
\end{table}

\setcounter{table}{3}
\begin{table}[t]
\centering

\tiny
\begin{tabular}{|c|c|c||c||c|r|c|c|c||c|r|}\hline
$x$ & $Q^2$ & $y$ &  $d^2\sigma_d^{meas}/dxdQ^2$
& $E$ & $E^{\prime}$ & $AC$ &  $RC$ & $RE$  & $R$ & $F_2^d\;\pm\Delta{F^{stat}_2}\;\pm\Delta{F^{syst}_2}$ \\
  & (${\rm GeV^2}$) & & $({\rm b\cdot GeV^{-2}})$ &  [\%]  &  [\%]  &  [\%]  &  [\%] &  [\%]  & & \\
 \hline
0.0035 &   0.83 & 0.654 & 1.619e-05 &   0.6 & -0.2 &  5.0 &  1.2 &  4.1 & 0.537 & 0.2756 $\pm$ 0.0078 $\pm$ 0.0181\\
0.0051 &   1.18 & 0.633 & 6.441e-06 &   0.5 &  0.0 &  3.6 &  1.1 &  3.8 & 0.537 & 0.3032 $\pm$ 0.0064 $\pm$ 0.0161\\
\hline
0.0075 &   1.32 & 0.486 & 4.059e-06 &   0.7 & -0.2 &  2.2 &  0.4 &  3.3 & 0.337 & 0.2996 $\pm$ 0.0062 $\pm$ 0.0121\\
0.0081 &   1.72 & 0.578 & 2.297e-06 &   0.6 & -0.1 &  2.8 &  0.2 &  3.2 & 0.337 & 0.3364 $\pm$ 0.0062 $\pm$ 0.0145\\
\hline
0.0122 &   1.76 & 0.395 & 1.698e-06 &   0.8 & -0.3 &  1.6 &  0.1 &  2.9 & 0.246 & 0.3295 $\pm$ 0.0073 $\pm$ 0.0113\\
0.0124 &   2.45 & 0.537 & 8.446e-07 &   0.6 & -0.1 &  1.9 &  0.4 &  2.6 & 0.246 & 0.3662 $\pm$ 0.0055 $\pm$ 0.0122\\
0.0138 &   3.25 & 0.640 & 3.951e-07 &   0.5 &  0.0 &  1.1 &  1.3 &  1.3 & 0.246 & 0.3632 $\pm$ 0.0146 $\pm$ 0.0079\\
\hline
0.0175 &   1.77 & 0.277 & 1.195e-06 &   1.0 & -0.5 &  1.6 &  0.1 &  2.6 & 0.190 & 0.3026 $\pm$ 0.0107 $\pm$ 0.0099\\
0.0174 &   2.48 & 0.388 & 6.429e-07 &   0.8 & -0.3 &  1.5 &  0.0 &  2.6 & 0.190 & 0.3532 $\pm$ 0.0068 $\pm$ 0.0110\\
0.0169 &   3.62 & 0.586 & 2.822e-07 &   0.6 & -0.1 &  1.1 &  1.2 &  1.1 & 0.190 & 0.3759 $\pm$ 0.0106 $\pm$ 0.0076\\
0.0185 &   4.34 & 0.638 & 1.836e-07 &   0.5 &  0.0 &  1.2 &  1.8 &  1.3 & 0.190 & 0.3948 $\pm$ 0.0125 $\pm$ 0.0100\\
\hline
0.0244 &   1.79 & 0.203 & 8.997e-07 &   1.4 & -0.8 &  1.9 &  0.2 &  2.4 & 0.099 & 0.3058 $\pm$ 0.0154 $\pm$ 0.0105\\
0.0247 &   2.50 & 0.279 & 4.645e-07 &   1.1 & -0.5 &  1.6 &  0.1 &  2.3 & 0.099 & 0.3334 $\pm$ 0.0063 $\pm$ 0.0102\\
0.0248 &   3.45 & 0.380 & 2.480e-07 &   0.8 & -0.3 &  1.6 &  0.1 &  2.3 & 0.099 & 0.3709 $\pm$ 0.0071 $\pm$ 0.0108\\
0.0239 &   4.54 & 0.520 & 1.380e-07 &   0.7 & -0.1 &  1.0 &  1.1 &  1.0 & 0.099 & 0.3855 $\pm$ 0.0090 $\pm$ 0.0074\\
0.0252 &   5.45 & 0.590 & 8.728e-08 &   0.6 &  0.0 &  1.2 &  1.5 &  1.1 & 0.099 & 0.3864 $\pm$ 0.0086 $\pm$ 0.0089\\
0.0277 &   6.51 & 0.637 & 5.517e-08 &   0.5 &  0.0 &  1.4 &  1.7 &  1.3 & 0.099 & 0.3930 $\pm$ 0.0115 $\pm$ 0.0101\\
\hline
0.0347 &   2.53 & 0.199 & 3.382e-07 &   1.4 & -0.8 &  1.8 &  0.2 &  2.1 & 0.108 & 0.3290 $\pm$ 0.0102 $\pm$ 0.0103\\
0.0349 &   3.46 & 0.270 & 1.888e-07 &   1.1 & -0.5 &  1.6 &  0.1 &  2.1 & 0.108 & 0.3657 $\pm$ 0.0090 $\pm$ 0.0104\\
0.0350 &   4.43 & 0.343 & 1.089e-07 &   0.9 & -0.3 &  1.6 &  0.2 &  2.1 & 0.108 & 0.3692 $\pm$ 0.0112 $\pm$ 0.0102\\
0.0345 &   5.49 & 0.433 & 6.676e-08 &   0.7 & -0.2 &  0.9 &  0.7 &  1.1 & 0.108 & 0.3701 $\pm$ 0.0104 $\pm$ 0.0066\\
0.0347 &   6.96 & 0.546 & 4.035e-08 &   0.6 & -0.1 &  1.1 &  1.1 &  1.1 & 0.108 & 0.3931 $\pm$ 0.0075 $\pm$ 0.0077\\
0.0368 &   8.66 & 0.636 & 2.208e-08 &   0.5 &  0.0 &  1.3 &  1.5 &  1.3 & 0.108 & 0.3725 $\pm$ 0.0114 $\pm$ 0.0090\\
\hline
0.0486 &   3.52 & 0.198 & 1.285e-07 &   1.4 & -0.8 &  1.7 &  0.1 &  1.8 & 0.117 & 0.3398 $\pm$ 0.0094 $\pm$ 0.0101\\
0.0494 &   4.46 & 0.248 & 8.038e-08 &   1.1 & -0.5 &  1.6 &  0.0 &  1.8 & 0.117 & 0.3619 $\pm$ 0.0092 $\pm$ 0.0099\\
0.0491 &   5.46 & 0.305 & 5.267e-08 &   1.0 & -0.4 &  1.1 &  0.2 &  1.4 & 0.117 & 0.3717 $\pm$ 0.0099 $\pm$ 0.0078\\
0.0490 &   6.98 & 0.391 & 3.073e-08 &   0.8 & -0.2 &  0.7 &  0.5 &  1.0 & 0.117 & 0.3808 $\pm$ 0.0071 $\pm$ 0.0058\\
0.0494 &   8.96 & 0.498 & 1.726e-08 &   0.7 & -0.1 &  0.9 &  0.8 &  1.0 & 0.117 & 0.3865 $\pm$ 0.0070 $\pm$ 0.0066\\
0.0516 &  11.12 & 0.586 & 1.032e-08 &   0.6 &  0.0 &  1.2 &  1.3 &  1.2 & 0.117 & 0.3954 $\pm$ 0.0081 $\pm$ 0.0087\\
\hline
0.0697 &   4.50 & 0.177 & 5.410e-08 &   1.4 & -0.8 &  1.7 &  0.1 &  1.6 & 0.113 & 0.3316 $\pm$ 0.0142 $\pm$ 0.0094\\
0.0691 &   5.47 & 0.215 & 3.825e-08 &   1.2 & -0.6 &  1.5 &  0.1 &  1.6 & 0.113 & 0.3558 $\pm$ 0.0145 $\pm$ 0.0093\\
0.0699 &   6.92 & 0.270 & 2.403e-08 &   1.0 & -0.4 &  0.8 &  0.1 &  1.1 & 0.113 & 0.3788 $\pm$ 0.0104 $\pm$ 0.0066\\
0.0695 &   8.99 & 0.353 & 1.282e-08 &   0.8 & -0.2 &  0.5 &  0.4 &  0.8 & 0.113 & 0.3638 $\pm$ 0.0090 $\pm$ 0.0051\\
0.0696 &  11.39 & 0.445 & 7.847e-09 &   0.7 & -0.1 &  0.8 &  0.7 &  0.9 & 0.113 & 0.3868 $\pm$ 0.0079 $\pm$ 0.0061\\
0.0707 &  14.46 & 0.553 & 4.450e-09 &   0.6 &  0.0 &  1.1 &  1.2 &  1.1 & 0.113 & 0.3892 $\pm$ 0.0100 $\pm$ 0.0080\\
\hline
0.0895 &   5.50 & 0.168 & 2.954e-08 &   1.4 & -0.8 &  1.4 &  0.1 &  1.7 & 0.096 & 0.3473 $\pm$ 0.0215 $\pm$ 0.0096\\
0.0901 &   6.95 & 0.210 & 1.853e-08 &   1.2 & -0.6 &  1.4 &  0.2 &  1.1 & 0.096 & 0.3629 $\pm$ 0.0128 $\pm$ 0.0083\\
0.0900 &   8.96 & 0.271 & 1.090e-08 &   1.0 & -0.4 &  0.8 &  0.2 &  0.9 & 0.096 & 0.3735 $\pm$ 0.0114 $\pm$ 0.0059\\
0.0899 &  11.45 & 0.346 & 6.142e-09 &   0.8 & -0.2 &  0.8 &  0.4 &  0.8 & 0.096 & 0.3655 $\pm$ 0.0091 $\pm$ 0.0055\\
0.0900 &  14.68 & 0.442 & 3.556e-09 &   0.7 & -0.1 &  0.8 &  0.7 &  0.9 & 0.096 & 0.3770 $\pm$ 0.0096 $\pm$ 0.0060\\
0.0907 &  19.01 & 0.566 & 1.918e-09 &   0.6 &  0.0 &  1.2 &  1.3 &  1.1 & 0.096 & 0.3765 $\pm$ 0.0162 $\pm$ 0.0081\\
\hline
0.1102 &   6.96 & 0.173 & 1.415e-08 &   1.3 & -0.7 &  1.0 &  0.2 &  1.1 & 0.043 & 0.3308 $\pm$ 0.0145 $\pm$ 0.0072\\
0.1101 &   8.99 & 0.222 & 8.313e-09 &   1.1 & -0.5 &  1.4 &  0.3 &  0.8 & 0.043 & 0.3374 $\pm$ 0.0128 $\pm$ 0.0072\\
0.1103 &  11.45 & 0.282 & 5.087e-09 &   0.9 & -0.3 &  1.0 &  0.2 &  0.8 & 0.043 & 0.3527 $\pm$ 0.0104 $\pm$ 0.0059\\
0.1100 &  14.77 & 0.364 & 2.838e-09 &   0.8 & -0.1 &  1.0 &  0.4 &  0.8 & 0.043 & 0.3494 $\pm$ 0.0097 $\pm$ 0.0056\\
0.1102 &  19.37 & 0.475 & 1.629e-09 &   0.7 &  0.0 &  1.0 &  0.7 &  1.0 & 0.043 & 0.3768 $\pm$ 0.0153 $\pm$ 0.0063\\
\hline
0.1388 &   7.16 & 0.142 & 1.073e-08 &   1.5 & -0.9 &  0.9 &  0.3 &  0.9 & 0.149 & 0.3291 $\pm$ 0.0162 $\pm$ 0.0073\\
0.1400 &   9.01 & 0.176 & 6.979e-09 &   1.3 & -0.6 &  1.0 &  0.3 &  0.7 & 0.129 & 0.3508 $\pm$ 0.0141 $\pm$ 0.0066\\
0.1395 &  11.43 & 0.224 & 4.009e-09 &   1.1 & -0.4 &  1.2 &  0.3 &  0.7 & 0.114 & 0.3370 $\pm$ 0.0087 $\pm$ 0.0060\\
0.1396 &  14.84 & 0.290 & 2.233e-09 &   0.9 & -0.2 &  1.3 &  0.2 &  0.8 & 0.101 & 0.3348 $\pm$ 0.0076 $\pm$ 0.0059\\
0.1398 &  19.32 & 0.375 & 1.248e-09 &   0.7 & -0.1 &  1.4 &  0.4 &  0.8 & 0.090 & 0.3413 $\pm$ 0.0088 $\pm$ 0.0063\\
0.1417 &  25.36 & 0.483 & 6.323e-10 &   0.6 &  0.0 &  1.4 &  0.8 &  1.0 & 0.080 & 0.3290 $\pm$ 0.0173 $\pm$ 0.0065\\
\hline
0.1785 &   7.38 & 0.114 & 6.502e-09 &   1.7 & -1.0 &  0.5 &  0.3 &  0.7 & 0.128 & 0.2678 $\pm$ 0.0201 $\pm$ 0.0056\\
0.1798 &   8.97 & 0.137 & 4.896e-09 &   1.4 & -0.7 &  0.8 &  0.3 &  0.7 & 0.113 & 0.3060 $\pm$ 0.0158 $\pm$ 0.0059\\
0.1815 &  11.44 & 0.172 & 2.709e-09 &   1.2 & -0.5 &  1.0 &  0.3 &  0.7 & 0.098 & 0.2867 $\pm$ 0.0096 $\pm$ 0.0050\\
0.1802 &  14.86 & 0.225 & 1.678e-09 &   1.0 & -0.3 &  1.3 &  0.2 &  0.7 & 0.086 & 0.3107 $\pm$ 0.0089 $\pm$ 0.0056\\
0.1802 &  19.43 & 0.293 & 9.090e-10 &   0.8 & -0.1 &  1.5 &  0.2 &  0.8 & 0.075 & 0.3047 $\pm$ 0.0087 $\pm$ 0.0058\\
0.1804 &  25.90 & 0.388 & 4.681e-10 &   0.7 &  0.0 &  1.7 &  0.5 &  0.9 & 0.067 & 0.3020 $\pm$ 0.0144 $\pm$ 0.0063\\
\hline
0.2225 &   9.15 & 0.113 & 3.494e-09 &   1.4 & -0.7 &  0.6 &  0.3 &  0.7 & 0.100 & 0.2781 $\pm$ 0.0180 $\pm$ 0.0051\\
0.2261 &  11.47 & 0.139 & 2.107e-09 &   1.2 & -0.4 &  0.7 &  0.3 &  0.7 & 0.086 & 0.2741 $\pm$ 0.0101 $\pm$ 0.0045\\
0.2262 &  14.90 & 0.180 & 1.132e-09 &   1.0 & -0.2 &  1.0 &  0.3 &  0.7 & 0.074 & 0.2569 $\pm$ 0.0079 $\pm$ 0.0041\\
0.2253 &  19.56 & 0.236 & 6.436e-10 &   0.8 & -0.1 &  1.3 &  0.2 &  0.7 & 0.065 & 0.2627 $\pm$ 0.0078 $\pm$ 0.0045\\
0.2257 &  26.32 & 0.317 & 3.587e-10 &   0.7 &  0.0 &  1.7 &  0.3 &  0.8 & 0.056 & 0.2840 $\pm$ 0.0123 $\pm$ 0.0057\\
\hline
0.2769 &  11.62 & 0.115 & 1.501e-09 &   0.9 & -0.2 &  0.7 &  0.3 &  0.7 & 0.077 & 0.2428 $\pm$ 0.0117 $\pm$ 0.0033\\
0.2756 &  14.82 & 0.147 & 9.132e-10 &   0.8 &  0.0 &  0.8 &  0.2 &  0.7 & 0.066 & 0.2453 $\pm$ 0.0094 $\pm$ 0.0034\\
0.2759 &  19.60 & 0.193 & 4.836e-10 &   0.7 &  0.1 &  0.9 &  0.2 &  0.7 & 0.057 & 0.2363 $\pm$ 0.0082 $\pm$ 0.0032\\
0.2763 &  26.49 & 0.261 & 2.413e-10 &   0.6 &  0.2 &  1.2 &  0.2 &  0.8 & 0.049 & 0.2283 $\pm$ 0.0114 $\pm$ 0.0036\\
0.2761 &  33.94 & 0.330 & 1.283e-10 &   0.6 &  0.2 &  1.8 &  0.3 &  0.8 & 0.043 & 0.2110 $\pm$ 0.0197 $\pm$ 0.0044\\
\hline
0.3410 &  15.01 & 0.121 & 5.132e-10 &   0.1 &  0.7 &  0.9 &  0.2 &  0.7 & 0.059 & 0.1733 $\pm$ 0.0062 $\pm$ 0.0023\\
0.3460 &  19.65 & 0.156 & 3.019e-10 &   0.2 &  0.7 &  1.2 &  0.2 &  0.7 & 0.050 & 0.1825 $\pm$ 0.0057 $\pm$ 0.0028\\
0.3469 &  26.24 & 0.206 & 1.575e-10 &   0.3 &  0.6 &  1.4 &  0.1 &  0.7 & 0.042 & 0.1774 $\pm$ 0.0072 $\pm$ 0.0030\\
0.3493 &  34.49 & 0.267 & 7.797e-11 &   0.3 &  0.6 &  1.5 &  0.2 &  0.8 & 0.037 & 0.1610 $\pm$ 0.0110 $\pm$ 0.0030\\
\hline
0.4516 &  20.02 & 0.122 & 1.422e-10 &  -2.5 &  3.6 &  2.0 &  0.2 &  0.7 & 0.044 & 0.1152 $\pm$ 0.0047 $\pm$ 0.0055\\
0.4710 &  26.12 & 0.152 & 6.470e-11 &  -1.8 &  2.9 &  2.8 &  0.1 &  0.7 & 0.037 & 0.0957 $\pm$ 0.0042 $\pm$ 0.0043\\
0.4737 &  34.79 & 0.201 & 3.526e-11 &  -1.2 &  2.4 &  3.3 &  0.1 &  0.7 & 0.031 & 0.0969 $\pm$ 0.0065 $\pm$ 0.0042\\
\hline
\end{tabular}

\caption{(c) 
The differential scattering cross section, 
$d^2\sigma_d^{meas}/dxdQ^2$,
the value of $R$ as used in the
analysis (see also table 7), and the
structure function $F_2^d$ at 200 GeV 
incident muon energy, with the normalisation corrections of table~2.
Horizontal lines separate the x-bins. See also caption (a).
 }
\end{table}

\setcounter{table}{3}
\begin{table}[t]
\centering

\tiny
\begin{tabular}{|c|c|c||c||c|r|c|c|c||c|r|}\hline
$x$ & $Q^2$ & $y$ &  $d^2\sigma_d^{meas}/dxdQ^2$
& $E$ & $E^{\prime}$ & $AC$ &  $RC$ & $RE$  & $R$ & $F_2^d\;\pm\Delta{F^{stat}_2}\;\pm\Delta{F^{syst}_2}$ \\
  & (${\rm GeV^2}$) & & $({\rm b\cdot GeV^{-2}})$ &  [\%]  &  [\%]  &  [\%]  &  [\%] &  [\%]  & & \\
 \hline
0.0037 &   1.27 & 0.670 & 8.092e-06 &   0.6 & -0.1 &  1.8 &  1.8 &  3.7 & 0.537 & 0.3363 $\pm$ 0.0055 $\pm$ 0.0153\\
0.0050 &   1.71 & 0.673 & 3.544e-06 &   0.5 &  0.0 &  1.4 &  1.6 &  3.4 & 0.537 & 0.3531 $\pm$ 0.0053 $\pm$ 0.0142\\
0.0056 &   2.15 & 0.738 & 2.090e-06 &   0.4 &  0.1 &  1.0 &  1.7 &  3.3 & 0.537 & 0.3841 $\pm$ 0.0111 $\pm$ 0.0146\\
\hline
0.0073 &   1.81 & 0.488 & 2.508e-06 &   0.7 & -0.2 &  1.0 &  0.7 &  3.3 & 0.337 & 0.3370 $\pm$ 0.0073 $\pm$ 0.0120\\
0.0079 &   2.44 & 0.606 & 1.246e-06 &   0.6 & -0.1 &  1.5 &  1.0 &  3.0 & 0.337 & 0.3655 $\pm$ 0.0038 $\pm$ 0.0131\\
0.0090 &   3.37 & 0.729 & 5.855e-07 &   0.4 &  0.1 &  1.3 &  1.7 &  2.5 & 0.337 & 0.3963 $\pm$ 0.0060 $\pm$ 0.0134\\
\hline
0.0121 &   2.55 & 0.413 & 8.675e-07 &   0.8 & -0.3 &  0.9 &  0.4 &  2.9 & 0.246 & 0.3585 $\pm$ 0.0058 $\pm$ 0.0115\\
0.0123 &   3.47 & 0.552 & 4.498e-07 &   0.6 & -0.1 &  1.1 &  0.9 &  2.8 & 0.246 & 0.3912 $\pm$ 0.0048 $\pm$ 0.0123\\
0.0126 &   4.45 & 0.693 & 2.518e-07 &   0.5 &  0.0 &  1.2 &  2.1 &  2.1 & 0.246 & 0.4014 $\pm$ 0.0047 $\pm$ 0.0133\\
0.0139 &   5.38 & 0.753 & 1.639e-07 &   0.4 &  0.1 &  1.3 &  2.6 &  1.9 & 0.246 & 0.4314 $\pm$ 0.0084 $\pm$ 0.0153\\
\hline
0.0173 &   2.59 & 0.294 & 6.274e-07 &   1.0 & -0.5 &  1.2 &  0.4 &  2.7 & 0.190 & 0.3427 $\pm$ 0.0092 $\pm$ 0.0110\\
0.0173 &   3.49 & 0.393 & 3.388e-07 &   0.8 & -0.3 &  1.0 &  0.4 &  2.6 & 0.190 & 0.3674 $\pm$ 0.0063 $\pm$ 0.0109\\
0.0174 &   4.46 & 0.501 & 2.054e-07 &   0.7 & -0.1 &  0.9 &  1.0 &  2.5 & 0.190 & 0.3991 $\pm$ 0.0064 $\pm$ 0.0118\\
0.0168 &   5.60 & 0.658 & 1.239e-07 &   0.5 &  0.0 &  1.3 &  2.0 &  1.5 & 0.190 & 0.4051 $\pm$ 0.0079 $\pm$ 0.0115\\
0.0180 &   6.73 & 0.727 & 8.140e-08 &   0.5 &  0.1 &  1.4 &  2.2 &  1.7 & 0.190 & 0.4233 $\pm$ 0.0055 $\pm$ 0.0134\\
\hline
0.0245 &   3.49 & 0.280 & 2.613e-07 &   1.0 & -0.5 &  1.3 &  0.3 &  2.5 & 0.099 & 0.3630 $\pm$ 0.0071 $\pm$ 0.0113\\
0.0245 &   4.48 & 0.359 & 1.586e-07 &   0.9 & -0.3 &  1.2 &  0.4 &  2.4 & 0.099 & 0.3883 $\pm$ 0.0061 $\pm$ 0.0111\\
0.0246 &   5.47 & 0.435 & 1.028e-07 &   0.7 & -0.2 &  1.1 &  0.7 &  2.4 & 0.099 & 0.4009 $\pm$ 0.0063 $\pm$ 0.0113\\
0.0235 &   7.22 & 0.605 & 5.557e-08 &   0.6 &  0.0 &  1.4 &  1.4 &  1.3 & 0.099 & 0.4043 $\pm$ 0.0051 $\pm$ 0.0098\\
0.0253 &   8.88 & 0.684 & 3.509e-08 &   0.5 &  0.1 &  1.5 &  1.7 &  1.6 & 0.099 & 0.4314 $\pm$ 0.0047 $\pm$ 0.0123\\
0.0279 &  10.78 & 0.744 & 2.188e-08 &   0.4 &  0.1 &  1.7 &  2.1 &  1.9 & 0.099 & 0.4436 $\pm$ 0.0083 $\pm$ 0.0148\\
\hline
0.0347 &   4.48 & 0.253 & 1.217e-07 &   1.1 & -0.5 &  1.4 &  0.3 &  2.3 & 0.108 & 0.3877 $\pm$ 0.0088 $\pm$ 0.0113\\
0.0348 &   5.49 & 0.308 & 7.567e-08 &   1.0 & -0.4 &  1.3 &  0.3 &  2.2 & 0.108 & 0.3794 $\pm$ 0.0080 $\pm$ 0.0105\\
0.0348 &   6.86 & 0.383 & 4.809e-08 &   0.8 & -0.2 &  1.2 &  0.5 &  2.2 & 0.108 & 0.4019 $\pm$ 0.0070 $\pm$ 0.0105\\
0.0345 &   9.01 & 0.511 & 2.583e-08 &   0.6 & -0.1 &  1.3 &  0.9 &  1.3 & 0.108 & 0.4080 $\pm$ 0.0059 $\pm$ 0.0089\\
0.0348 &  11.44 & 0.641 & 1.520e-08 &   0.5 &  0.0 &  1.6 &  1.5 &  1.4 & 0.108 & 0.4227 $\pm$ 0.0046 $\pm$ 0.0113\\
0.0369 &  14.09 & 0.735 & 9.379e-09 &   0.5 &  0.1 &  1.8 &  2.2 &  1.7 & 0.108 & 0.4346 $\pm$ 0.0076 $\pm$ 0.0145\\
\hline
0.0469 &   5.53 & 0.232 & 5.708e-08 &   1.2 & -0.6 &  1.5 &  0.3 &  2.1 & 0.117 & 0.3696 $\pm$ 0.0089 $\pm$ 0.0107\\
0.0493 &   6.93 & 0.276 & 3.474e-08 &   1.0 & -0.4 &  1.4 &  0.2 &  2.0 & 0.117 & 0.3858 $\pm$ 0.0057 $\pm$ 0.0105\\
0.0491 &   8.99 & 0.360 & 1.948e-08 &   0.9 & -0.3 &  1.1 &  0.3 &  1.5 & 0.117 & 0.3888 $\pm$ 0.0058 $\pm$ 0.0083\\
0.0487 &  11.57 & 0.468 & 1.135e-08 &   0.7 & -0.1 &  1.2 &  0.7 &  1.0 & 0.117 & 0.4067 $\pm$ 0.0045 $\pm$ 0.0078\\
0.0494 &  14.87 & 0.590 & 6.312e-09 &   0.6 &  0.0 &  1.5 &  1.3 &  1.3 & 0.117 & 0.4133 $\pm$ 0.0038 $\pm$ 0.0103\\
0.0529 &  18.90 & 0.690 & 3.549e-09 &   0.5 &  0.1 &  1.9 &  2.0 &  1.6 & 0.117 & 0.4244 $\pm$ 0.0055 $\pm$ 0.0137\\
\hline
0.0663 &   7.32 & 0.217 & 2.352e-08 &   1.2 & -0.6 &  1.4 &  0.3 &  1.9 & 0.113 & 0.3749 $\pm$ 0.0113 $\pm$ 0.0103\\
0.0698 &   8.93 & 0.250 & 1.454e-08 &   1.1 & -0.4 &  1.3 &  0.3 &  1.9 & 0.113 & 0.3738 $\pm$ 0.0085 $\pm$ 0.0096\\
0.0696 &  11.45 & 0.322 & 8.538e-09 &   0.9 & -0.3 &  0.9 &  0.3 &  1.1 & 0.113 & 0.3832 $\pm$ 0.0064 $\pm$ 0.0068\\
0.0694 &  15.01 & 0.423 & 4.747e-09 &   0.7 & -0.1 &  1.1 &  0.6 &  0.9 & 0.113 & 0.3964 $\pm$ 0.0049 $\pm$ 0.0070\\
0.0699 &  19.64 & 0.547 & 2.475e-09 &   0.6 &  0.0 &  1.5 &  1.2 &  1.2 & 0.113 & 0.3920 $\pm$ 0.0042 $\pm$ 0.0091\\
0.0716 &  25.55 & 0.689 & 1.420e-09 &   0.5 &  0.1 &  1.9 &  2.2 &  1.5 & 0.113 & 0.4238 $\pm$ 0.0090 $\pm$ 0.0141\\
\hline
0.0863 &   9.34 & 0.213 & 1.000e-08 &   1.2 & -0.6 &  1.3 &  0.3 &  1.9 & 0.096 & 0.3390 $\pm$ 0.0147 $\pm$ 0.0090\\
0.0896 &  11.48 & 0.250 & 6.718e-09 &   1.0 & -0.4 &  1.0 &  0.2 &  1.2 & 0.096 & 0.3689 $\pm$ 0.0077 $\pm$ 0.0071\\
0.0897 &  14.95 & 0.326 & 3.694e-09 &   0.9 & -0.2 &  0.8 &  0.3 &  0.8 & 0.096 & 0.3668 $\pm$ 0.0063 $\pm$ 0.0056\\
0.0895 &  19.88 & 0.433 & 1.954e-09 &   0.7 & -0.1 &  1.2 &  0.7 &  0.9 & 0.096 & 0.3742 $\pm$ 0.0047 $\pm$ 0.0066\\
0.0903 &  26.06 & 0.558 & 1.062e-09 &   0.6 &  0.0 &  1.6 &  1.3 &  1.2 & 0.096 & 0.3865 $\pm$ 0.0058 $\pm$ 0.0094\\
\hline
0.1094 &  11.80 & 0.212 & 4.982e-09 &   1.1 & -0.5 &  1.0 &  0.3 &  1.1 & 0.043 & 0.3427 $\pm$ 0.0102 $\pm$ 0.0070\\
0.1097 &  14.90 & 0.264 & 3.208e-09 &   1.0 & -0.3 &  1.0 &  0.2 &  0.8 & 0.043 & 0.3688 $\pm$ 0.0078 $\pm$ 0.0062\\
0.1097 &  19.86 & 0.353 & 1.639e-09 &   0.8 & -0.2 &  1.0 &  0.4 &  0.8 & 0.043 & 0.3599 $\pm$ 0.0054 $\pm$ 0.0054\\
0.1100 &  26.36 & 0.464 & 8.460e-10 &   0.7 &  0.0 &  1.3 &  0.7 &  1.0 & 0.043 & 0.3580 $\pm$ 0.0055 $\pm$ 0.0067\\
0.1106 &  34.48 & 0.601 & 4.834e-10 &   0.6 &  0.1 &  1.7 &  1.4 &  1.3 & 0.043 & 0.3849 $\pm$ 0.0110 $\pm$ 0.0099\\
\hline
0.1390 &  11.77 & 0.167 & 3.463e-09 &   1.3 & -0.6 &  1.2 &  0.3 &  1.1 & 0.113 & 0.2929 $\pm$ 0.0151 $\pm$ 0.0064\\
0.1372 &  15.03 & 0.215 & 2.354e-09 &   1.1 & -0.4 &  1.2 &  0.3 &  0.8 & 0.101 & 0.3332 $\pm$ 0.0065 $\pm$ 0.0063\\
0.1388 &  19.88 & 0.280 & 1.243e-09 &   0.9 & -0.2 &  1.4 &  0.2 &  0.6 & 0.089 & 0.3291 $\pm$ 0.0043 $\pm$ 0.0058\\
0.1393 &  26.62 & 0.373 & 6.462e-10 &   0.7 & -0.1 &  1.4 &  0.4 &  0.8 & 0.080 & 0.3328 $\pm$ 0.0041 $\pm$ 0.0059\\
0.1400 &  35.22 & 0.489 & 3.446e-10 &   0.6 &  0.0 &  1.4 &  0.8 &  1.0 & 0.073 & 0.3419 $\pm$ 0.0067 $\pm$ 0.0071\\
\hline
0.1790 &  15.07 & 0.165 & 1.705e-09 &   1.2 & -0.5 &  1.3 &  0.3 &  0.5 & 0.085 & 0.3063 $\pm$ 0.0084 $\pm$ 0.0058\\
0.1795 &  19.99 & 0.218 & 8.991e-10 &   1.0 & -0.3 &  1.5 &  0.2 &  0.6 & 0.075 & 0.2978 $\pm$ 0.0050 $\pm$ 0.0057\\
0.1791 &  26.74 & 0.291 & 4.814e-10 &   0.8 & -0.1 &  1.7 &  0.2 &  0.6 & 0.066 & 0.3028 $\pm$ 0.0045 $\pm$ 0.0061\\
0.1799 &  35.13 & 0.378 & 2.574e-10 &   0.7 &  0.0 &  1.9 &  0.4 &  0.8 & 0.060 & 0.3017 $\pm$ 0.0053 $\pm$ 0.0067\\
0.1807 &  45.74 & 0.488 & 1.473e-10 &   0.6 &  0.1 &  2.1 &  0.8 &  1.0 & 0.055 & 0.3201 $\pm$ 0.0099 $\pm$ 0.0081\\
\hline
0.2240 &  15.14 & 0.133 & 1.192e-09 &   1.2 & -0.4 &  1.0 &  0.3 &  0.5 & 0.074 & 0.2661 $\pm$ 0.0083 $\pm$ 0.0045\\
0.2242 &  19.97 & 0.175 & 6.788e-10 &   1.0 & -0.2 &  1.3 &  0.3 &  0.5 & 0.064 & 0.2729 $\pm$ 0.0051 $\pm$ 0.0047\\
0.2241 &  26.76 & 0.233 & 3.529e-10 &   0.8 & -0.1 &  1.7 &  0.2 &  0.6 & 0.056 & 0.2674 $\pm$ 0.0043 $\pm$ 0.0052\\
0.2239 &  35.30 & 0.306 & 1.864e-10 &   0.7 &  0.0 &  2.0 &  0.3 &  0.6 & 0.050 & 0.2609 $\pm$ 0.0047 $\pm$ 0.0058\\
0.2249 &  46.57 & 0.401 & 9.883e-11 &   0.6 &  0.1 &  2.4 &  0.5 &  0.8 & 0.045 & 0.2614 $\pm$ 0.0075 $\pm$ 0.0071\\
\hline
0.2646 &  16.06 & 0.119 & 9.405e-10 &   0.9 & -0.1 &  0.8 &  0.0 &  0.1 & 0.065 & 0.2782 $\pm$ 0.0232 $\pm$ 0.0033\\
0.2740 &  19.99 & 0.143 & 4.845e-10 &   0.8 &  0.0 &  0.9 &  0.2 &  0.6 & 0.056 & 0.2346 $\pm$ 0.0055 $\pm$ 0.0031\\
0.2739 &  26.76 & 0.191 & 2.540e-10 &   0.7 &  0.1 &  1.2 &  0.2 &  0.5 & 0.049 & 0.2292 $\pm$ 0.0045 $\pm$ 0.0035\\
0.2745 &  35.42 & 0.251 & 1.398e-10 &   0.6 &  0.2 &  1.8 &  0.2 &  0.6 & 0.043 & 0.2328 $\pm$ 0.0048 $\pm$ 0.0046\\
0.2744 &  46.98 & 0.332 & 7.511e-11 &   0.6 &  0.3 &  2.4 &  0.4 &  0.7 & 0.038 & 0.2354 $\pm$ 0.0071 $\pm$ 0.0060\\
0.2769 &  59.79 & 0.413 & 3.105e-11 &   0.5 &  0.3 &  2.9 &  0.6 &  1.3 & 0.035 & 0.1698 $\pm$ 0.0183 $\pm$ 0.0057\\
\hline
0.3289 &  20.73 & 0.124 & 3.185e-10 &   0.0 &  0.8 &  1.2 &  0.2 &  0.5 & 0.050 & 0.1981 $\pm$ 0.0058 $\pm$ 0.0031\\
0.3440 &  26.73 & 0.153 & 1.567e-10 &   0.1 &  0.7 &  1.4 &  0.2 &  0.5 & 0.042 & 0.1739 $\pm$ 0.0032 $\pm$ 0.0029\\
0.3447 &  35.48 & 0.202 & 8.717e-11 &   0.2 &  0.7 &  1.5 &  0.1 &  0.5 & 0.037 & 0.1778 $\pm$ 0.0032 $\pm$ 0.0031\\
0.3455 &  46.61 & 0.262 & 4.561e-11 &   0.3 &  0.6 &  1.6 &  0.2 &  0.5 & 0.032 & 0.1694 $\pm$ 0.0039 $\pm$ 0.0032\\
0.3491 &  61.17 & 0.338 & 2.314e-11 &   0.3 &  0.6 &  2.6 &  0.4 &  1.2 & 0.029 & 0.1592 $\pm$ 0.0106 $\pm$ 0.0047\\
\hline
0.4334 &  27.70 & 0.125 & 8.355e-11 &  -2.7 &  3.8 &  2.8 &  0.2 &  0.5 & 0.037 & 0.1246 $\pm$ 0.0035 $\pm$ 0.0068\\
0.4657 &  35.56 & 0.150 & 3.690e-11 &  -1.9 &  3.1 &  3.3 &  0.1 &  0.5 & 0.031 & 0.0999 $\pm$ 0.0021 $\pm$ 0.0050\\
0.4702 &  46.58 & 0.194 & 1.972e-11 &  -1.3 &  2.5 &  3.6 &  0.1 &  0.5 & 0.027 & 0.0960 $\pm$ 0.0024 $\pm$ 0.0044\\
0.4772 &  62.24 & 0.254 & 8.989e-12 &  -0.8 &  2.1 &  3.7 &  0.2 &  1.1 & 0.023 & 0.0835 $\pm$ 0.0054 $\pm$ 0.0037\\
\hline
\end{tabular}

\caption{(d)
The differential scattering cross section, 
$d^2\sigma_d^{meas}/dxdQ^2$,
the value of $R$ as used in the
analysis (see also table 7), and the
structure function $F_2^d$ at 280 GeV 
incident muon energy, with the normalisation corrections of table~2.
Horizontal lines separate the x-bins. See also caption (a).
 }
\end{table}